\documentclass[twocolumn,floatfix]{aastex62}
\usepackage{amsmath,amssymb,amstext}

\usepackage{aas_macros}
\usepackage{natbib}
\usepackage{tikz}
\usetikzlibrary{shapes,arrows,intersections}
\shorttitle{TINYMO}
\shortauthors{Riedel et al.}

\begin{document}

\title{The Solar Neighborhood. XLIII: Discovery of New Nearby Stars with $\mu < 0.18\arcsec$ yr$^{-1}$ (TINYMO sample)}
\author{Adric~R.~Riedel}
\affil{Space Telescope Science Institute, Baltimore, MD 21218}
\email{riedel@stsci.edu}

\author{Michele~L.~Silverstein}
\affil{Physics and Astronomy Department, Georgia State University, Atlanta, GA 30302}

\author{Todd~J.~Henry}
\affil{RECONS Institute, Chambersburg, PA}

\author{Wei-Chun Jao}
\affil{Physics and Astronomy Department, Georgia State University, Atlanta, GA 30302}

\author{Jennifer~G.~Winters}
\affil{Harvard-Smithsonian Center for Astrophysics, Cambridge, MA 02138}

\author{John~P.~Subasavage}
\affil{US Naval Observatory Flagstaff Station, Flagstaff, AZ 86005}

\author{Lison~Malo}
\affil{Canada-France-Hawaii Telescope, Kamuela, HI 96743}


\author{Nigel~C.~Hambly}

\affil{Institute for Astronomy, University of Edinburgh, Blackford Hill, Edinburgh, EH9 3HJ, Scotland, UK}

\begin{abstract}

We have conducted a novel search of most of the southern sky for nearby red dwarfs having low proper motions, with specific emphasis on those with $\mu$ $<$ 0\farcs18 yr$^{-1}$, the lower cutoff of Luyten's classic proper motion catalog.  We used a tightly constrained search of the SuperCOSMOS database and a suite of photometric distance relations for photographic $BRI$ and 2MASS $JHK_s$ magnitudes to estimate distances to more than fourteen million red dwarf candidates. Here we discuss 29 stars in 26 systems estimated to be within 25 parsecs, all of which have $\mu$ $<$0\farcs18 yr$^{-1}$, which we have investigated using milliarcsecond astrometry, $VRI$ photometry, and low-resolution spectroscopy. In total, we present the first parallaxes of 20 star systems, nine of which are within 25 parsecs. We have additionally identified eight young M dwarfs, of which two are new members of the nearby young moving groups, and 72 new giants, including two new carbon stars. We also present the entire catalog of 1215 sources we have identified by this means.

\end{abstract}

\keywords{astrometry --- solar neighborhood --- stars: distances ---
stars: low mass --- stars: statistics --- surveys}


\section{Introduction}
\label{sec:introduction}

The Solar Neighborhood is the best laboratory for studying the Galaxy in which we live. The optimal place to make a volume-limited study of stars is nearby, where the very faintest stellar and substellar objects are easiest to detect and measure. Nearby binary systems are excellent targets for dynamical mass determination; they are resolvable with smaller orbits and shorter orbital periods than their more distant counterparts. Planetary-mass objects are brighter and have larger angular separations for the same linear separation when they are closer.

Most surveys to reveal the Sun's nearest neighbors focus on detecting stars exhibiting high proper motions, $\mu$. Such surveys identify two categories of stars --- disk stars that are close enough that their modest Galactic orbital motion yields an apparent angular motion above the search threshold, and more distant stars with much higher intrinsic motions, e.g., subdwarfs and halo stars.  This property of large proper motion has served nearby star research well from the very beginning, forming at least part of the decisions of \citet{Bessel1838} and \citet{Henderson1839} to observe 61 Cygni and Alpha Centauri (respectively) for the first parallaxes. These searches have continued on into the present day, encompassing everything from historical efforts like the Luyten Half Second catalog \citep{Luyten1957,Luyten1979b} to recent efforts like LSPM-North \citep{Lepine2005a} and the Research Consortium On Nearby Stars (RECONS) group's own work (e.g. \citealt{Henry2004,Subasavage2005a,Subasavage2005b,Finch2007,Boyd2011a, Boyd2011b}). These searches have yielded thousands of stars that are candidates for stars within 25 parsecs (pc), the horizon adopted by the Catalog of Nearby Stars \citep{Gliese1991} and NStars \citep{Backman2001} compendia. 

Nearly all known nearby stars have high proper motions. An analysis of the current RECONS 10 pc sample\footnote{See \cite{Henry2006} for discussion of the definition of a RECONS 10 pc system, and \url{www.recons.org} for updated statistics.} to explore the realm of low $\mu$ nearby stars is revealing. Of the 259 systems (not including the Sun) within 10 pc as of 2012 JAN 01, 133 (52\%) have $\mu \ge 1.00\arcsec yr^{-1}$, 88 (35\%) have $1.00\arcsec yr^{-1} > \mu \ge 0.50\arcsec yr^{-1}$, and 32 (13\%) have $0.50\arcsec yr^{-1} > \mu \ge 0.18\arcsec yr^{-1}$. Only two stars, less than 1\% of the total sample, have $\mu <$ 0.18\arcsec yr$^{-1}$: GJ 566 AB (spectral type G8V, $V =$ 4.67 \citep{Hoeg2000}, $\mu = 0.169\arcsec yr^{-1}$), and LSPM~J0330+5413 (an M dwarf with $V \sim$ 16, $\mu = 0.150\arcsec yr^{-1}$, \citealt{Lepine2005a}).

There are reasons to suspect that a small but significant population of nearby, very low proper motion stars have been overlooked. The limits of the proper motion samples set above are based on historical precedent. 
In particular, the value of 0.18\arcsec~yr$^{-1}$ as the lowest interesting proper motion, used by RECONS' other survey samples (e.g. \citealt{Winters2017}) as its lower limit, comes from the influential surveys of Luyten (Luyten Palomar, Luyten Bruce, Luyten Two Tenths, New Luyten Two Tenths) and Giclas (Southern survey). Those studies were themselves influenced by the work of the Royal Greenwich Observatory, particularly \citet{Thackeray1917} and \citet{Dyson1917}, the latter of which suggests that Greenwich set their 0.2\arcsec~yr$^{-1}$ limit based on calculations that suggested only one-eighth of all nearby stars ($<$20 pc) should have lower proper motions. Thus, from the outset, it was understood that some population of nearby stars would be overlooked.

In this paper, we present a survey of those very low proper motion stars, along with astrometric, photometric, and spectroscopic follow-up observations for selected high priority stars and other additional targets of interest from the Cerro Tololo Inter-american Observatory Parallax Investigation (CTIOPI) parallax program. In Section \ref{sec:expectations} we lay out the background work done on these ``TINYMO'' systems that have $\mu < 0.18\arcsec yr^{-1}$. In Section \ref{sec:TINYMO} we discuss the design and methodology of the TINYMO survey itself. In Section \ref{sec:followup}, we discuss further target characterization, which results in the final catalog of targets presented in Section \ref{sec:catalog}. We then discuss the results of the observational followup of our targets in Section \ref{sec:results}, and discuss the implications of the TINYMO survey in detail in Section \ref{sec:discussion}.

\section{Expected Distribution of TINYMOs}
\label{sec:expectations}


A few surveys have delved into searches for stars with smaller proper motions, most notably Wroblewski-Torres-Costa \citep[][and subsequent]{Wroblewski1989} (0.15\arcsec~yr$^{-1}$), the LSPM survey \citep{Lepine2005a} (0.15\arcsec~yr$^{-1}$), the `Meet the Cool Neighbors' group \citep{Reid2007} (limit 0.11\arcsec~yr$^{-1}$ northern hemisphere, 0.28\arcsec~yr$^{-1}$ southern hemisphere), and \citet{Deacon2007} (0.1\arcsec~yr$^{-1}$), and \citet{Deacon2009} (0.08\arcsec~yr$^{-1}$). Apart from the anticipated but currently unreleased Lepine SUPERBLINK catalogs (0.04\arcsec~yr$^{-1}$ and larger), no efforts are searching for stars with proper motions smaller than 0.1\arcsec~yr$^{-1}$, or down to truly zero proper motions. These comprehensive searches have not been done, because without the telltale marker of motion on photographic plates (or more recently, CCD images), the investigator looking for nearby stars is inundated by huge numbers of candidates that come pouring out of automated searches (UCAC, \citealt{Zacharias2013}; PPMXL, \citealt{Roeser2011}).

\subsection{Incompleteness of the 25 parsec sample}
\label{sec:simulation}

\begin{figure}
\centering
\includegraphics[angle=0,width=0.5\textwidth]{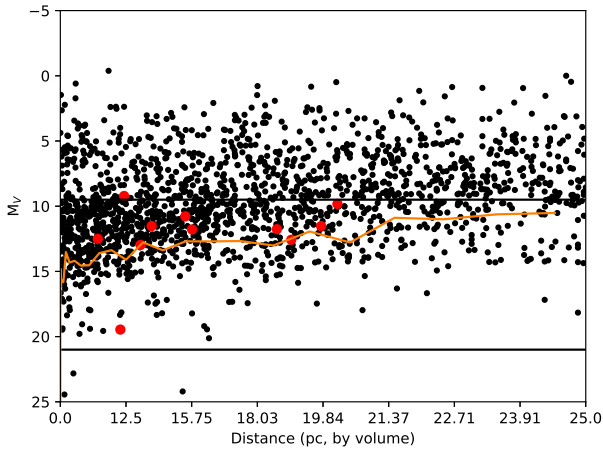}
\caption{Absolute V magnitude versus Distance for stars within 25 pc, from the RECONS 25 pc database (Jao et al. in prep). The distance is given in equal-volume-elements to properly represent stellar density. The density is essentially uniform for stars brighter than M$_V=$9, suggesting completeness for A,F,G, and K stars; 90\% of stars lie above the orange line, demonstrating completeness decreases at larger distances. Red dots represent the eleven systems within 25 pc with new parallaxes in this paper.}
\label{fig:RECX25completion}
\end{figure}

How many stars do we expect to find within 25 pc at very low proper motions? For the purposes of this work we have made an estimation, using a simulation of the Solar Neighborhood, accounting for spatial and velocity distributions. The spatial distribution within 25 pc is assumed to be uniform because the volume density of K stars (and hotter) in the RECONS 25 pc sample (Jao et al. in prep) is essentially uniform (Figure \ref{fig:RECX25completion}). The decreasing spatial distribution of M dwarfs is assumed to be the result of luminosity-related incompleteness. Accordingly, we assume the overall stellar density matches that of the nearest 5 pc (52 systems in 5 pc, or 0.099 systems pc$^{-3}$) and expect 6500 systems within 25 pc (Note that Regulus is the sole known B star within 25 parsecs, and not within that 5 pc radius). The velocity distribution of the Solar Neighborhood is modeled according to the spectral type of the stars, as given in \citet{Aumer2009}. The hottest stars have the lowest dispersions around the Local Standard of Rest, and cooler stars have increasingly large velocity dispersions up until the Parenago Discontinuity around $B-V$=$+$0.9, where the average stellar population has had uniform amounts of disk heating. Additional kinematic parameters for subdwarfs and white dwarfs are sourced from \citet{Gizis1997} and \citet{Mihalas1981}, respectively.

\begin{figure}
\centering
\includegraphics[angle=0,width=0.5\textwidth]{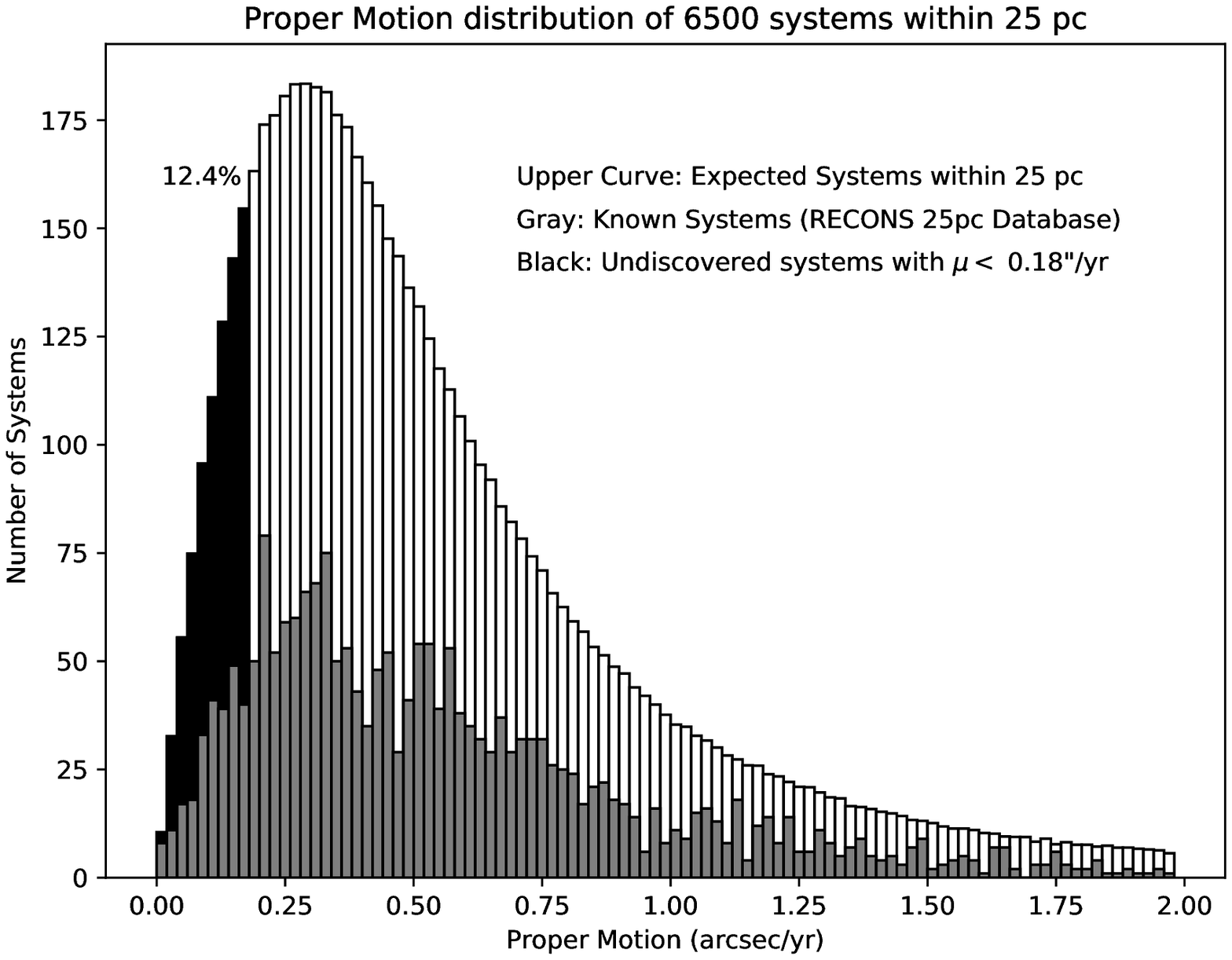}
\caption{A simulated proper motion distribution using \citet{Aumer2009}, \citet{Gizis1997} and \citet{Mihalas1981} velocity distributions, and a Cumulative Mass Function (CMF) from RECONS data. The top black and white plotted histogram assumes 6500 star systems within 25 pc, based on the assumption that 52 systems within 5 pc is a representative spatial density. Systems moving slower than 0.18\arcsec~yr$^{-1}$ are noted in black; the gray curve represents actual proper motions
from the RECONS 25 pc sample (Jao et al. in preparation).}
\label{fig:Proper_Motion_Distribution}
\end{figure}

To tie the spherical and velocity distributions together, we used the color distribution of spectral types in the RECONS 25 pc database (together with the assumption that all of the missing star systems would be K, M, L, or T dwarfs with the same velocity dispersion) and generated a cumulative luminosity distribution (Table \ref{tab:syntheticParameters}) out of which a random number generator can provide appropriately distributed stars of different spectral types, luminosities, and dispersions. These randomly generated stars were placed in a uniform spatial distribution with a radius of 25 pc. Str\"{o}mberg's asymmetric drift equation ($<V>=\frac{U^2}{k},k=74\pm5$; \citealt{Aumer2009}) was added to the stars, and the UVW velocity of the Sun relative to the local standard of rest (U=11.10, V=12.24, W=7.25 km s$^{-1}$, \citealt{Schoenrich2010}) was subtracted. We then derived the observational properties (RA, DEC, proper motion, radial velocity) from these synthetic stars. The distribution of proper motions, as derived from 10 million synthetic stars, is shown in Figure \ref{fig:Proper_Motion_Distribution}.

\begin{deluxetable}{llcccl}
\setlength{\tabcolsep}{0.02in}
\tablewidth{0pt}
\tabletypesize{\tiny}
\tablecaption{Parameters for synthetic 25 pc sample\label{tab:syntheticParameters}}
\tablehead{
\colhead{$V-K_s$} &
\colhead{Cum. Frac.} &
\colhead{$\sigma$U} &
\colhead{$\sigma$V} &
\colhead{$\sigma$W} &
\colhead{Note}\\
\colhead{} &
\colhead{} &
\colhead{km s$^{-1}$} &
\colhead{km s$^{-1}$} &
\colhead{km s$^{-1}$} & 
\colhead{} 
}
\startdata
-1    & 0.0000   &   8    &    8    &   5   & B systems (Regulus= 1/6375) \\
 0    & 0.00016  &  14    &    9    &   4.5 & A systems (4/408) \\
 1    & 0.0098   &  22    &   14    &  10   & F systems (6/408) \\
 2    & 0.0245   &  38    &   26    &  20   & G systems (20/408)\\
 3    & 0.0735   &  37    &   26    &  19   & K systems (44/408)\\
 3.8  & 0.1814   &  37    &   26    &  19   & M0-3 systems\\
 5    & 0.3500   &  37    &   26    &  19   & M3-5 systems\\
 6    & 0.5000   &  37    &   26    &  19   & M5-7 systems\\
 8    & 0.7200   &  37    &   26    &  19   & M7-9.5 systems\\
10    & 0.8100   &  37    &   26    &  19   & L,T systems \\
20    & 0.91186  &  37    &   26    &  19   & Transition \tablenotemark{a}\\
-1    & 0.91187  & 177    &  100    &  82   & Subdwarfs \tablenotemark{b}\\
20    & 0.92336  & 177    &  100    &  82   & Transition \tablenotemark{a}\\
-1    & 0.92337  &  50    &   30    &  20   & White dwarfs \tablenotemark{c}\\
 0    & 0.9500   &  50    &   30    &  20   & White dwarfs \\
2.7   & 1.0000   &  50    &   30    &  20   & White dwarfs \\
\enddata
\tablenotetext{a}{These are not real; they are a computational necessity included to separate the ``sequences'' and prevent interpolation from making many oddly-distributed stars from a continuous function.}
\tablenotetext{b}{\citet{Gizis1997}}
\tablenotetext{c}{\citet{Mihalas1981}}
\tablecomments{The cumulative luminosity function (CMF) distribution of stars is in three sequences - main sequence, subdwarfs, and white dwarfs - used to randomly generate a proportional and representative 25 pc sample.}
\end{deluxetable}

As can be seen in Figure \ref{fig:Proper_Motion_Distribution}, the RECONS 25 pc sample is incomplete for all proper motions, but particularly incomplete for proper motions less than 0.5\arcsec~yr$^{-1}$, with peaks in both around 0.3\arcsec~yr$^{-1}$. Overall, 12.4\% of all stars within 25 pc should be moving at speeds slower than 0.18\arcsec~yr$^{-1}$, which is in line with other estimates (\citealt{Reid2007}, for example, find 11\%). 

There are potential improvements to this simulation: Neither giants nor young stars, nor any of the local kinematic streams (as seen in \citealt{Skuljan1999}, \citealt{Nordstroem2004}) were included in this analysis. An additional possible improvement would be to model stars as bursts of star formation with a cluster mass, IMF distribution, and age-associated velocity dispersion. 

\section{The TINYMO Survey}
\label{sec:TINYMO}

To create a more detailed picture of the Solar Neighborhood, we have carried out a search of the southern sky for stars with tiny proper motions, less than $0.18\arcsec yr^{-1}$, dubbed TINYMO. This is a regime of proper motions that has not been explored in a rigorous way. The discoveries reported here complement previous SuperCOSMOS-RECONS (SCR) searches of the southern sky \citep{Henry2004, Hambly2004, Subasavage2005a, Subasavage2005b,Finch2007,Boyd2011a,Boyd2011b}; in particular, the last three revealed 6007 new proper systems with $0.40\arcsec yr^{-1} > \mu \ge 0.18\arcsec ^{-1}$ between declinations $-90^\circ$ and $-00^\circ$, and $R_{59F}$ $=<$ 18.0.

Those previous searches used proper motion cuts to identify potential nearby stars, followed by photometric estimates of distance to pick the most promising nearby young objects for astrometric and spectroscopic follow-up through the Cerro Tololo Inter-american Observatory Parallax Investigation (CTIOPI). The search discussed here is almost entirely the opposite of those searches, starting with a rough proper motion limit and then using photometry to select the promising nearby stars. Photometry is rarely the primary method of identifying nearby stars (one of the rare other examples is \citealt{Cruz2007}, which identified late-type red dwarfs and brown dwarfs by their extreme colors) because of the enormous contamination of distant giants and other non-nearby sources. 

Despite this challenge, a photometric search is the only way to reliably identify genuinely nearby tiny proper motion stars. Even allowing for the practical limitations of proper motion measurements (particularly those of compiled catalogs, which have uncertainties introduced by source/scanning resolution and optical defects), below a certain level (See Section \ref{sec:pm_lower_limit}), even distant background stars have some non-zero proper motion, because they too are in orbit around the Galactic center.

In this paper we estimate distances {\it en masse} for millions of sources, then target those with the smallest distances for further consideration. The list of selected nearby red dwarf candidates is sequentially winnowed with quality and color cuts until only the most promising targets remain, and these are investigated individually for available data in the literature and targeted in observational programs (See Figure \ref{fig:flowchart}).

\onecolumngrid
\begin{center}
  \footnotesize
  \begin{tikzpicture}[auto,
    block/.style ={rectangle, draw=black, thick, fill=white,
      text width=10em, text centered,
      minimum height=3em},
    decision/.style ={rectangle, draw=black, thick, fill=white,
      text width=18em, text centered, rounded corners, minimum height=2em, inner sep=2pt},
    line/.style ={draw, thick, -latex', shorten >=0pt}]
    \label{fig:flowchart}
    \matrix [column sep=5mm,row sep=3mm] {
    \node [block]  (one) {SuperCOSMOS (1.9 billion sources) -- See Section \ref{sec:1_superc}};    & & \\
    \node [decision]  (two) {Position and photometric data cuts -- See Section \ref{sec:2_sql}}; & & \\
    \node [block]  (three) {14 million sources};                                                 & & \\
    \node [decision]  (four) {Photometric Distance cut -- See Section \ref{sec:3_photsift}};     & & \\
    \node [block]  (five) {88586 sources};                                                       & & \\
    \node [decision]  (six) {Color-color cuts -- See Section \ref{sec:colorcuts}};               & & \\
    \node [block]  (seven) {1077 sources};                                                       & & \\
    \node [block]  (eight) {1154 sources};    & \node [decision] (eightalt) {Old color boxes (77 additional sources)};     & \\
    \node [block]  (nine) {\bf{1215 sources} (Section \ref{sec:catalog})}; & \node [decision] (ninealt) {Manually identified sources (61 additional sources)}; & \\
    \node [decision] (ten) {Quality cuts -- See Section \ref{sec:5_quality}};                    & & \\
    \node [block] (eleven) {651 non-giant and non-suspected-giant sources (Samples 1 and 2)}; & & \node [block] (elevenc) {Giants -- Sections \ref{sec:giants},\ref{sec:carbon}}; \\
    \node [decision] (twelve) {Within 15 pc -- See Section \ref{sec:followup}};                  & & \\
    \node [block] (thirteen) {115 sources};                                                      & & \\
    \node [decision] (14) {...and X-ray bright or already in CTIOPI -- See Section \ref{sec:followup}}; & & \\
    \node [block] (15) {187 sources};                                                            & & \\
    \node [decision] (16) {Astrometry, Photometry, Spectroscopy -- Section \ref{sec:followup}};  & & \node [block] (16c) {Additional non-survey CTIOPI targets}; \\
    & \node [block] (17) {48 Systems \citep{Riedel2014}}; & \\
    & \node [block] (18) {77 Systems \citep{Riedel2017a}}; & \\
    & \node [block] (19) {26 Systems (This paper)}; & \\
    };
    \begin{scope}[every path/.style=line]
      \path (one)   -- (two);
      \path (two)   -- (three);
      \path (three)   -- (four);
      \path (four)   -- (five);
      \path (five)   -- (six);
      \path (six)   -- (seven);
      \path (seven)   -- (eight);
      \path (eight)   -- (nine);
      \path (nine)   -- (ten);
      \path (ten)   -- (eleven);
      \path (eleven)   -- (twelve);
      \path (twelve)   -- (thirteen);
      \path (thirteen)   -- (14);
      \path (14)   -- (15);
      \path (15)   -- (16);
      \path (16)   -- (17);
      \path (16)   -- (18);
      \path (16)   -- (19);
      \path (eightalt)       -- (eight);
      \path (ninealt)       -- (nine);
      \path (ten)       -- (elevenc);
      \path (16c)       -- (17);
      \path (16c)       -- (18);
      \path (16c)       -- (19);
    \end{scope}
  \end{tikzpicture}
\end{center}
\twocolumngrid

\subsection{SuperCOSMOS (1.9 billion sources)}

SuperCOSMOS was a machine that scanned glass photographic plates for more than a decade at the Royal Observatory in Edinburgh (ROE), Scotland. The SuperCOSMOS Science Archive (SSA) database \citep{Hambly2001a} is built from scans the machine made of primarily Palomar Observatory Sky Survey (POSS) and Science and Engineering Research Council (SERC) sky survey plates.  The survey covers the entire sky  at four different epochs and in four different passbands, deriving positions, proper motions, and (up to) four-color photometry for 1.9 billion sources. SuperCOSMOS magnitude limits vary by field but are generally equivalent to $B$=22, $R$=20, $I$=19 in the plate photographic magnitude system of \citep[e.g.][]{Bessell1986}. 2MASS $JHK_s$ photometry has been cross-matched to sources where available. SuperCOSMOS is {\it not} a source of absolute positions or proper motions, though attempts were made to force the mean Galaxy proper motions (field by field) to zero in fields where galaxies were available \citep{Hambly2001c}.  The overall reference frame was shifted to ICRS via cross-matching with 2MASS (which is linked to TYCHO-2).
\label{sec:1_superc}

Of interest to TINYMO, the plates were aligned by cross-matching stars out to distances of 6\arcsec~(in a spiral search pattern) between two plates. This matching constraint actually provides a variable {\it upper} limit on measurable proper motions. For the southern hemisphere where epoch spreads are 30-40 years, the maximum proper motion detectable is around 0.2-0.3\arcsec~yr$^{-1}$, above which an object would move more than 6\arcsec~in that time. This represents a tradeoff: a few higher proper motion stars -- perhaps a hundred thousand out of two billion -- will be identified as multiple transient objects. Previous RECONS proper motion searches have been carried out using additional software designed to match up otherwise unmatched sources in the SSA. Other surveys using the SuperCOSMOS Database (and their own special software) include \citet{Scholz2002} (and subsequent), the Liverpool-Edinburgh High Proper Motion Survey \citep{Pokorny2003}, and the Southern Infrared Proper Motion Survey \citep{Deacon2005a}.

For the purposes of TINYMO, the main catalog is sufficient, provided we limit ourselves to sources identified on all four plates. The catalog contains proper motions up to 0.3\arcsec yr$^{-1}$ for sources of interest, except in regions north of DEC = $-$18\arcdeg, where far older POSS-I E red plates were used. In those areas, the larger epoch spread means that the highest proper motion that can be reliably extracted from the 6\arcsec~crossmatch is roughly 0.12\arcsec~yr$^{-1}$; it is also incomplete for a 25 square degree region around RA=16h, DEC=$-$12\arcdeg where POSS-I E field 1038 is missing (and thus no four-color detections are possible).
 
\subsection{SQL Query (14 million sources)}
\label{sec:2_sql}

The initial sift of the TINYMO survey was an SQL query, meant to identify meaningful targets in the Southern hemisphere. To avoid overloading the server, the queries were conducted in tiles of RA and DEC. The selection criteria were as follows: 

\subsubsection{Location Cuts}
\begin{itemize}
\item Regions in the Southern hemisphere.

\item More than 20 degrees from the Galactic Center.

\item More than 10 degrees from the Galactic Plane.
\end{itemize}

\begin{figure*}
\centering
\includegraphics[angle=0,width=\textwidth]{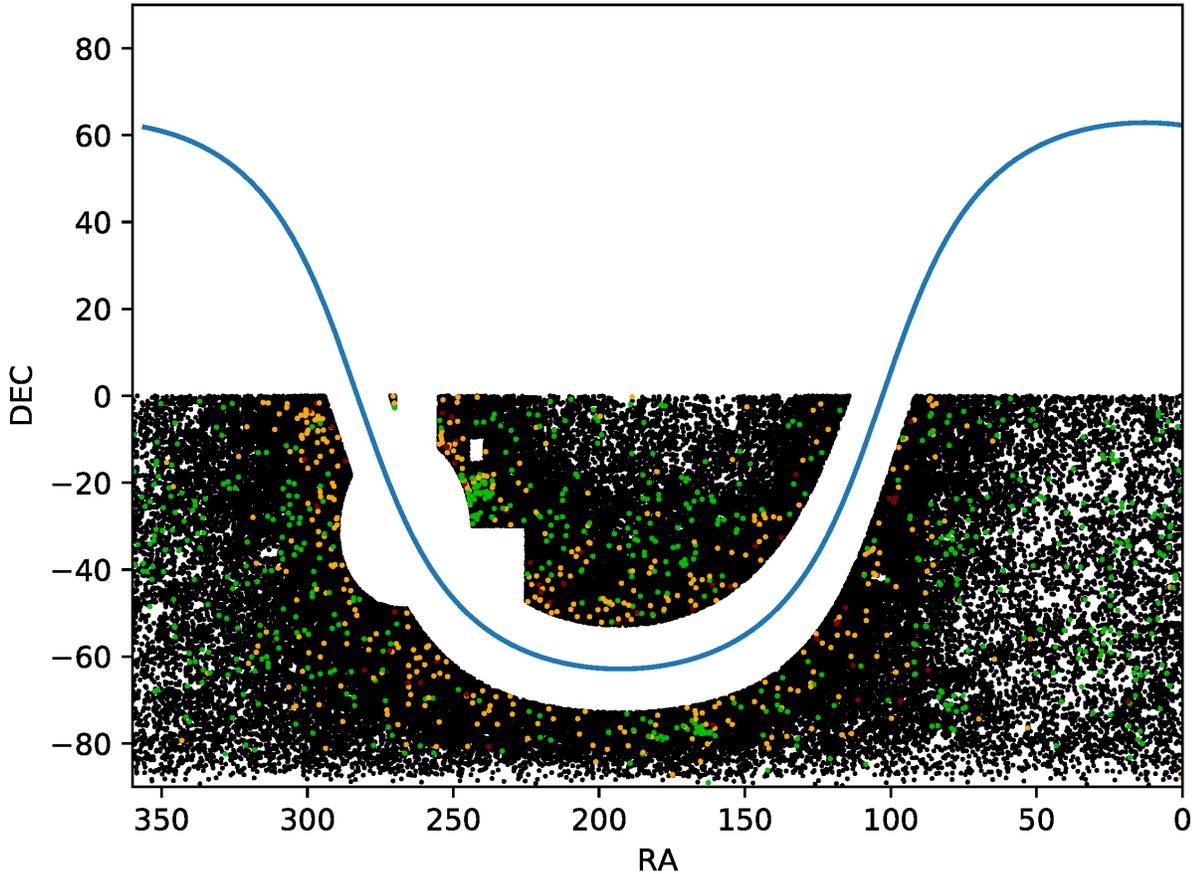}
\caption{Sky Map of the TINYMO survey extraction centered on RA=0h, DEC=0\arcdeg, with a 10$^\circ$ band around the Galactic equator and a 20$^\circ$ region around the Galactic Center removed, along with two regions near the North Galactic Spur.  The colored points (green = high-probability M dwarfs, yellow = lower probability M dwarfs, red = probable giants) are stars selected by later phases of the survey processing (see Fig \ref{fig:TINYMO_boxes}), demonstrating the crowding of sources around the Galactic bulge. The resulting sky coverage is roughly 16,000 square degrees, or 39.3\% of the total sky, and contains just under 14 million targets that satisfy our photometric quality criteria.}
\label{fig:TINYMO_sky_area}
\end{figure*}

These positional cuts were designed to limit the survey to the Southern hemisphere, and remove extremely dense areas (full of highly reddened stars that would contaminate the sample) from consideration. After the fact, additional cuts were made to the extracted data to remove regions near the North Galactic Spur: 15h $\leq$ RA $\leq$ 16h, $-$30 $\leq$ DEC $\leq$ $+$00; 15h $\leq$ RA $\leq$ 16h, $-$60 $\leq$ DEC $\leq$ $-$30; 17h $\leq$ RA $\leq$ 18h, $-$30 $\leq$ DEC $\leq$ $+$00. Within those regions there were as many stars with apparent photometric distances within 25 pc, all most likely giants at far greater distances, as there were in the rest of the sample (see Figure \ref{fig:TINYMO_sky_area}). The Large Magellanic Cloud and Small Magellanic Cloud regions were not removed. This cut defines our coverage of 16214 square degrees, or 39.3\% of the sky.

\subsubsection{Plate Detection Cuts}
\begin{itemize}
\item Detected on all four plates.
\end{itemize}

This criterion sets an upper proper motion limit as described above, as well as limits on color -- the star could not be so red it did not appear in the $B_J$ plate, which cuts out a number of cool and faint stars. As mentioned earlier, this also cut out a small region of sky (roughly 16h $\leq$ RA $\leq$ 16:20, $-$15 $\leq$ DEC $\leq$ $-$10) where there is no $R_1$ plate.

\subsubsection{Quality Cuts}
\begin{itemize}
\item Internal quality measure $>$ 128 on all plates.

\item Ellipticity less than 0.2 on all plates.
\end{itemize}

These cuts removed a large number of extragalactic sources, unresolved binaries, and spurious sources including plate defects.

\subsubsection{Luminosity Cuts}
\begin{itemize}
\item Brighter than $R_2$=16.5.
\end{itemize}

The $R_2$ magnitude limit allows for detection of stars with $B_J$=21 and $B_J-R_2$ colors as red as 4.5 (a brown dwarf), and matches the Giclas surveys \citep{Giclas1979} as well as previous SCR proper motion surveys, not including \citet{Boyd2011b}

\subsubsection{Offset cuts}
\begin{itemize}
\item Detected in 2MASS within 5\arcsec~of the weighted mean plate position.
\end{itemize}

The mean plate position recorded by SuperCOSMOS is weighted by the positional accuracy of each of the detections; the epoch of this effective plate position is usually around 1985, while the mean epoch of 2MASS is around 2000.  Thus, any star moving slower than $\mu<$0.333\arcsec~yr$^{-1}$ (less than $\sim$5\arcsec~motion over 15 years) will be matched to its 2MASS entry. This, in concert with the four-plate detection requirement, makes the most stringent cut.

The photometric limits chosen influence the kinds of stars we expect to find. The limit of 2MASS is effectively $JHK\approx15$. SuperCOSMOS contains sources as faint as $B_J=21$, so with a magnitude cutoff of $R_2=16.5$, the limiting magnitudes for M dwarfs are all therefore set by the $R_2$ filter. The magnitudes of an M0V star ($M_{B_J}=10$, $B_J-R_2=2.3$, $B_J-K=4.5$) corresponding to our cutoff at $R_2=16.5$ are $B_J=18.8$, $R_2=16.5$, and $K=14.3$.  This implies a limiting distance of 630 pc. For an M9.0V star ($M_{B_J}=20.4$, $B_J-R_2=3.0$, $B_J-K=10.2$) the magnitude limit is $B_J=19.5$, $R_2=16.5$, and $K=9.3$, which implies a limiting distance of 6.6 pc. Within 25 pc we should be able to detect every M dwarf bluer than $B_J-R_2=2.6$ (M7V).

Ultimately, the search identified just short of 14 million stars in the covered 16214 square degree region seen in Figure \ref{fig:TINYMO_sky_area}.

\subsection{Photometric Sift (88,586 sources)}
\label{sec:3_photsift}

The next phase of the search for low proper motion nearby stars was the computation of photometric distance estimates \citep{Hambly2004} for all stars. This method uses the plate $BR_2I$ and 2MASS $JHK$ colors to produce up to 11 distance estimates (out of a total possible 15 colors; $B-R_2$, $J-H$, $J-K$, and $H-K$ do not provide useful discriminants for red dwarfs) that are then combined into a weighted mean with a typical uncertainty of 26\%. These color-magnitude relationships, described by fourth-order fits to the main sequence, are only valid for K and M dwarfs, which removes all hotter stars from our consideration. We expect that no stars hotter than K remain undiscovered within 25 pc thanks to the work of {\it HIPPARCOS}. Of the 14 million point sources from the first step, slightly fewer than 89,000 (see Figure \ref{fig:TINYMO_sky_area}) were estimated to be within 25 pc by those relations.

As there are only roughly 6,500 systems expected within 25 pc (Section \ref{sec:simulation}), the $\approx$89,000 figure suggests massive contamination. This is as expected: apart from subdwarfs and (theoretically) stars with unresolved white dwarf companions, contaminants with the colors of main-sequence stars are much brighter objects that will land in a magnitude-limited survey such as ours, and include:
\begin{itemize}
\item Giants that mimic main sequence colors or were caught at fortuitous times in their light curves; particularly Mira variables due to their intrinsic luminosity
\item Metal-rich stars just beyond 25 pc
\item Unresolved multiple stars, where there is extra luminosity and therefore a smaller expected distance.
\item Pre-main-sequence stars, where the extra luminosity is due to the enlarged radius of the gravitationally contracting protostar
\item Reddened (and extincted) objects in molecular cloud regions
\item Redshifted Active Galactic Nuclei
\end{itemize}
The 11 plate relations were calibrated to colors typical of K and M main sequence stars; if a star has unusual colors outside the valid color ranges, it is less likely to be a main sequence star. We therefore flagged all objects with fewer than 9 valid distance relations (out of 11 total). It should be noted that this limit is different from that used in other publications in this series, where as few as 7 relations were accepted to accommodate the possibility that a single $Bj$ or $R2$ filter magnitude might be erroneous.


\subsection{Color-Color cuts (1154 sources)}
\label{sec:colorcuts}

To identify specifically main sequence stars, we applied a color-color cut, in $J-K$ versus $v-K$ space, where $v$ is an estimated $V$ magnitude formed by taking the average of $B$ and $R_2$.

\begin{figure*}
\centering
\includegraphics[angle=0,width=\textwidth]{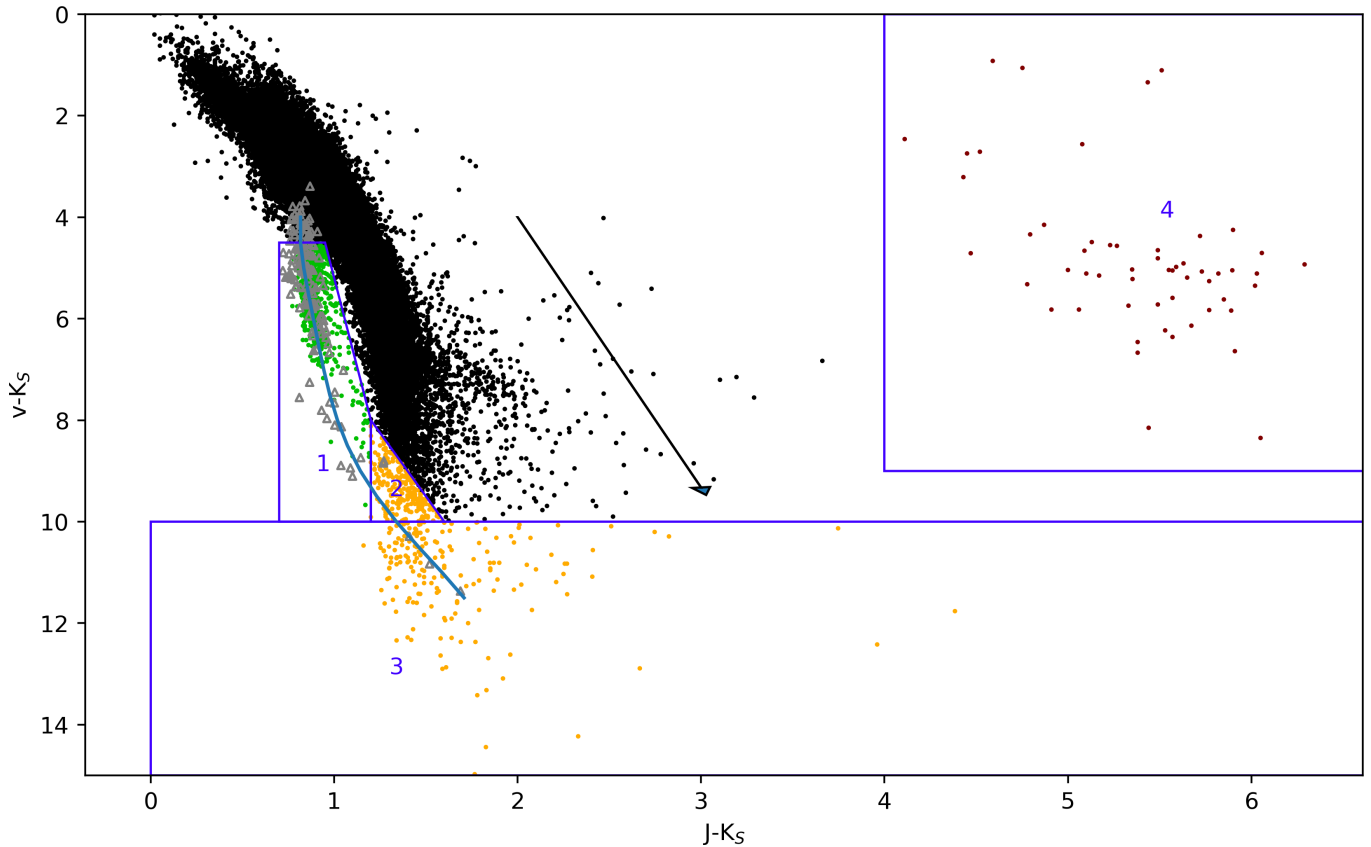}
\caption{Using the color-color regions shown on this
$J-K$ vs $v-K$ diagram, we separate the 88,586 stars in Figure \ref{fig:TINYMO_sky_area} into giants (black), dwarfs (green), and interesting overlap regions (yellow). The curve is a fifth-order fit to the main
sequence (as determined by the RECONS 10 pc sample, plotted as gray
triangles). The cluster of red points beyond $J-K$=4 (``flyers'') were
later revealed to be accidental cross-matches to spurious 2MASS
entries. The selection boxes from Table \ref{tab:TINYMO_regions} and Section \ref{sec:colorcuts} are shown here. The appropriate interstellar reddening vector from
\citet{Fitzpatrick1999} (assuming $v,J,K_s$=Johnson $V,J,K$)
is also shown. Colors are the same as on Figure \ref{fig:TINYMO_sky_area}.}
\label{fig:TINYMO_boxes}
\end{figure*}

There are, among $BR_2IJHK$ color combinations, two particular colors in which M dwarfs are distinguishable from red giants: $J-H$ and $J-K$ (Figure \ref{fig:TINYMO_boxes}). In these colors (and only these), mid-M dwarfs are bluer than mid-M giants of the same $v-K$ color. This property does not appear in any other combination of colors, including $H$-$K$, but it shows up when $J-H$ or $J-K$ is plotted against any other color. This behavior is most likely due to gravity-sensitive absorption features in all three bands: The $J$ band feature decreases in strength as gravity increases, and the $H$ and $K$ band features both increase in strength as gravity increases. \citet{Allers2007} identifies a number of potentially gravity-sensitive features that may fit those requirements: VO and TiO weaken with increasing gravity (and are predominantly found in the J band); CO (which dominates in the $K$ band), K I, and Na I all strengthen with increasing gravity. This would explain why dwarfs are bluer in $J-H$ and $J-K$ (increased $J$ flux, decreased $H$ or $K$ flux) and yet there is no effect on $H-K$ (correlated loss of flux). This behavior does not appear in other Johnson/Kron-Cousins/2MASS filter combinations, though $R$ and $I$ are also dominated by TiO; it may have to do with the rate at which the band strengths change.

Plotting the $J-K$ versus $v-K$ combination of colors (Figure \ref{fig:TINYMO_boxes}) demonstrates a region of color-color space where M dwarfs are distinguishable from M giants entirely by photometric colors. To take advantage of that property, we have created four selection regions to separate out the data, as shown in Figure \ref{fig:TINYMO_boxes} and Table \ref{tab:TINYMO_regions}.

{\it Region 1: Red Dwarf Candidates}

In the Red Dwarf region, the main sequence is clearly separated from the giants, as seen in Figure \ref{fig:TINYMO_boxes}. This is the most reliable region for nearby star detections using our search technique and encompasses spectral types M3.0V through M9.0V.  The region has been drawn with a blue cutoff of $v-K = 4.50$ to avoid the broad sequence of giants with bluer $v-K$ seen above the dwarfs, although some nearby stars should be found in this region. Some giants will still bleed into Region 1 near the $J-K = 0.95$, $v-K = 4.50$ corner; the amount of contamination varies from field to field and appears to be related to the Galactic latitude of the region and its particular reddening. The red edge of the sample is cut at $J-K = 1.2$, beyond which the giant and brown dwarf colors overlap.

{\it Region 2: Giants' Tail}

The tail of the giant sequence crosses the dwarf sequence in Region 2, but we have retained these targets because one or more may be a very nearby late-type red dwarf or brown dwarf.

{\it Region 3: Very Red Candidates}

Region 3 includes extremely red objects ($v-K$ $>$ 10) that are also likely giants or highly reddened distant stars, but could be interesting unusually red sources.

{\it Region 4: Flyers}

A small group of extremely red sources (called ``Flyers'') were found to have $v-K$ = 0--7 and $J-K$ = 4--7. Investigation showed that all were bright targets on the SuperCOSMOS plates, with erroneous matches to 2MASS sources. All checked objects were later determined to be giants falling within the giant locus once their photometry was corrected.

The result of accepting only the objects in these four regions was a reduction of 88,586 candidates to 1077 promising nearby objects. An early selection attempt used different boxes which included 77 stars\footnote{None of the 77 objects show signs of being main sequence stars.} not in the final set of boxes (see Figure \ref{fig:flowchart}), which we retain in our final catalog for bookkeeping reasons. This brings the total to 1154 stars. 

\begin{deluxetable}{lccl}
\tablecaption{The TINYMO Color Selection Regions\label{tab:TINYMO_regions}}
\tablehead{
\colhead{}     &
\multicolumn{2}{c}{Vertices}&
\colhead{}     \\
\colhead{Box}  &
\colhead{$J-K$} &
\colhead{$v-K$} &
\colhead{Purpose} }
\startdata
1 & 0.7  &  4.5 & Main Sequence \\
  & 0.95 &  4.5 & \\
  & 1.2  &  8.0 & \\
  & 1.2  & 10.0 & \\
  & 0.7  & 10.0 & \\
\hline  
2 & 1.2  &  8.0 & Brown dwarfs \\
  & 1.6  & 10.0 & \\
  & 1.2  & 10.0 & \\
\hline
3 & 0.0  & 10.0 & Very red dwarfs \\
  & 7.0  & 10.0 & \\
  & 7.0  & 15.0 & \\
  & 0.0  & 15.0 & \\
\hline
4 & 4.0  & 0.0 & ``Flyers'' \\
  & 7.0  & 0.0 & \\
  & 7.0  & 9.0 & \\
  & 4.0  & 9.0 & 
\enddata
\end{deluxetable}

\subsection{Further Quality Cuts}
\label{sec:5_quality}
  
 In the fourth and final phase of the winnowing, more quality cuts were made to improve the nearby star recovery rate: 
\begin{enumerate}
\item visual inspection (``blinking'') of SuperCOSMOS plate scans in the Aladin Skyview Desktop applet with the 2MASS Point Source Catalog loaded as an overlay, to ensure detected stars were a.) real objects, b.) moving - proper motions larger than 0.08\arcsec~yr$^{-1}$ were identifiable under visual examination, c.) matched to the proper 2MASS point (mistakes in the 2MASS identification account for the ``flyers'' mentioned in Section \ref{sec:colorcuts}). At this point, an additional 61 proper motion objects (generally companions) were non-exhaustively identified by eye to bring the total candidate list to 1215 objects.
\item comparisons of the two R band SuperCOSMOS magnitudes for consistency --- values differing by more than 1.00 magnitude were likely variable giants and were discarded. This is admittedly imperfect: low-amplitude Mira variables or Miras caught at two similar points in their lightcurve will not be flagged by their $R_1-R_2$ magnitudes, while stars with bad $R_1$ or $R_2$ photometry will be unfairly excluded.
\item elimination of sources with $J-K$ $\ge$ 2.00, which are presumed giants or stars with poor $JHK$ magnitudes that corrupt the distance estimates
\item searches of the SIMBAD database to determine whether or not sources are previously documented nearby stars, giants, Mira stars, carbon stars, and/or pulsating or variable stars
\item searches of the Rosat All-Sky Survey (RASS) Bright Source \citep{Voges1999} and Faint Source \citep{Voges2000} catalogs for stars with X-ray detections, which are most likely dwarfs or young stars.
\end{enumerate}

We thus arrive at a sample of 651 stars that pass all tests, while the remaining 564 of the 1215 objects were flagged for any number of the above quality reasons.

We developed five classifications for stars based on the above quality cuts, which we use to reclassify the sources identified in the four color boxes defined above, and which we will refer to from this point on:
\begin{enumerate}
\item X-ray: Stars that had X-ray counterparts in the RASS-BSC and RASS-FSC catalogs were highest priority, as they were most likely to be nearby dwarfs
\item Good: Stars that passed all quality cuts but did not have X-ray detections
\item Probable: Stars that failed either the $R_1-R_2$ test, had fewer than 9 valid photometric plate distance relations, or $J-K$ $\ge$ 2.00, but were not already know to be giants (as of 2012, \citealt{Riedel2012}).
\item Giants: Stars known to be giants according to the General Catalog of Variable Stars \citep[][in VizieR as b/GCVS]{Samus2010}, the Catalog of Galactic Carbon Stars \citep{Alksnis2001}, or SIMBAD.
\item Flyers: Stars from Region 4 of the color-color boxes, the older boxes, or spuriously identified by-eye.
\end{enumerate}

Proper motions from the survey ranged from 0.000\arcsec~yr$^{-1}$ to 0.397\arcsec~yr$^{-1}$; additional targets found by eye were found to be moving as fast as  0.444\arcsec~yr$^{-1}$. Overall, 1016 of the stars found in the survey were moving slower than 0.18\arcsec~yr$^{-1}$

In practice, all but one of the flagged stars in the ``Probable'' group were revealed to be giants after a literature search or low-resolution spectroscopy (Section \ref{sec:spectroscopy}). The one potential nearby star is SCR~1931-1757 (19:31:39.88 -17:57:36.0, $\mu=$0.028 $P.A.$=188.2$^{\circ}$), a spectroscopically confirmed M2.0Ve star with all 11 valid plate relations and $R_1-R_2=-3.03$ (SuperCOSMOS colors are apparently erroneous); its predicted distance was too far (17.67 pc by the average of 12 CCD distance estimates) to earn astrometric follow-up (Section \ref{sec:followup}).

\section{Follow-up Observations}
\label{sec:followup}

Given limited observing resources, it was decided to define a higher-priority sample of stars for follow-up. This sample included the 115 tiny proper motion ($<$0.18\arcsec~yr$^{-1}$) candidates with an estimated distance within 15 pc that had not already been identified as giants in the literature (Regions 1,2, and 3 (if they had more than 9 valid plate relations) of Figure \ref{fig:TINYMO_boxes}), plus all 55 of the targets within 25 pc found to be X-ray bright (Section \ref{sec:literature}). Additional tiny proper motion targets from the survey that were already on the observing programs were folded into our observational list, bringing it to 187 total targets of interest.

For the purposes of providing a larger selection of tiny proper motion objects for analysis in this paper, we added an additional 12 targets from the CTIOPI program that were not found in the TINYMO survey. These additional 12 targets do not appear in the master catalog (Section \ref{sec:catalog}) or discussion thereof, and are marked as such in Tables where they do appear. Their astrometry, photometry, and spectroscopy (where applicable) were obtained in the same way as our survey followup described below.

Analysis of some stars found in the TINYMO sample also appears in \citet{Riedel2014} and \citet{Riedel2017a}, and objects with proper motions higher than 0.18\arcsec~yr$^{-1}$ were folded into the study published in \citet{Winters2017}.

\subsection{Literature Search}
\label{sec:literature}
 
There are useful bodies of work in the literature that can be used to further characterize the remaining stars of interest. Apart from SIMBAD, the General Catalog of Variable Stars \citep[][in VizieR as b/GCVS]{Samus2010} maintains a list of all known variable stars and can be used to identify Mira variables, Carbon stars, and other semi-regular and irregular giant stars. The Catalog of Galactic Carbon Stars \citep{Alksnis2001} also furnished some Carbon star identifications. Finally, the entire list was checked against the VizieR versions of the LSPM \citep{Lepine2005a} and NLTT \citep{Luyten1979b} catalogs to identify previously known proper motion objects. Identifications from these catalogs appear in the catalog (Table \ref{tab:tinymo}).

We searched the ROSAT \citep{Voges1999,Voges2000} catalog for cross-matches to our objects, as giants are not generally expected to be strong X-ray emitters (I. Song, Priv. Comm.). \citet{Voges1999} defines the 90\% limit on detections as being sources within 25\arcsec~of the optical source, with less than 25\% uncertainty on the count rate; those guidelines were followed when identifying X-ray sources prioritized for photometry, spectroscopy, and astrometry. Most of these X-ray bright objects were identified as objects of interest by \citet{Riaz2006}. Because the ROSAT observations were carried out in the early 1990s, we applied our proper motions to move the targets back to their epoch 1991 positions using the SuperCOSMOS proper motions before carrying out the X-ray search.

\subsection{Photometry}
\label{sec:photometry}

Through the existing CTIOPI program (operating since 1999 on the CTIO 0.9m, \citealt{Jao2005,Henry2006}) we have obtained Johnson-Kron-Cousins $VRI$ photometry \citep{Jao2003,Winters2015} for all 187 targets. Target fields are observed in each filter on photometric nights and then transformed to Johnson-Kron-Cousins $VRI$ through the use of standards from \citet{Landolt1992} and \citet{Landolt2007}. Stars were observed on at least two nights to check for consistent $VRI$ photometry.

The faintest star in our sample is 2MASS 0936-2610B, with $V$=19.92. The brightest star is the unresolved binary \object[GJ 2122]{GJ 2122AB} (V=9.68), which is a well-known M1.0V star.

Accurate $VRI$ photometric data increase our confidence that the candidate stars are truly nearby dwarfs because (a) the derived distance estimate uncertainties drop from 26\% for the plate photometry (and 2MASS) based-$BR_2IJHK$ relations to 15\% for the CCD photometry (and 2MASS) based-$VRIJHK$ relations \citep{Henry2004}, and (b) many giants can be eliminated from the candidate pool based on photometric variability revealed by comparing their new CCD Kron-Cousins $R$ magnitudes to existing SuperCOSMOS plate $R_1$ and $R_2$.

The vast majority of this photometry can be found in Table \ref{tab:tinymo}. Photometry for all the astrometric targets reported in this paper (including the tiny proper motion systems not found as part of the TINYMO survey) is given in Table \ref{tab:photometry}.

\clearpage
\begin{rotatetable*}
\begin{deluxetable*}{llrrrclcccrrrccccl}
\setlength{\tabcolsep}{0.01in}
\tablewidth{0pt}
\tabletypesize{\tiny}
\tablecaption{Photometric Results for 26 Selected Star Systems}
\label{tab:photometry}
\tablehead{\colhead{}              &
	   \colhead{Alternate}     &
	   \colhead{}              &
       \colhead{}              &
	   \colhead{}              &
	   \colhead{No. of abs.}   &
	   \colhead{}              &
	   \colhead{$\sigma$}      &
	   \colhead{No. of rel.}   &
	   \colhead{No. of}        &
	   \colhead{$J$}           &
	   \colhead{$H$}           &
	   \colhead{$K_s$}         &
	   \colhead{spectral}      &
	   \colhead{}              &
	   \colhead{phot}          &
	   \colhead{No. of}        &
	   \colhead{}              \\
        \colhead{Name}          &
        \colhead{Name}          &
	   \colhead{$V_{J}$}       &
 	   \colhead{$R_{KC}$}      &
	   \colhead{$I_{KC}$}      &
	   \colhead{Nights}        &
	   \colhead{$\pi$ filter}  &
	   \colhead{(mag)}         &
 	   \colhead{Nights}        &
 	   \colhead{Frames}        &
	   \colhead{(2MASS)}       &
	   \colhead{(2MASS)}       &
	   \colhead{(2MASS)}       &
	   \colhead{type}          &
	   \colhead{ref\tablenotemark{b}}&
	   \colhead{dist}          &
	   \colhead{Relations}     &
	   \colhead{Notes}         \\
           \colhead{(1)}           &         
           \colhead{(2)}           &
           \colhead{(3)}           &
           \colhead{(4)}           &
           \colhead{(5)}           &
           \colhead{(6)}           &
           \colhead{(7)}           &
           \colhead{(8)}           &
           \colhead{(9)}           &
           \colhead{(10)}          &
           \colhead{(11)}          &
           \colhead{(12)}          &
           \colhead{(13)}          &
           \colhead{(14)}          &
           \colhead{(15)}          &
           \colhead{(16)}          &
           \colhead{(17)}          &
           \colhead{(18)}          }
\startdata
\hline
NLTT01261     & DY Psc          & 19.88$\pm$0.03 & 17.46$\pm$0.07 & 15.12$\pm$0.04 & 2 & I & .0076 &  7 &  29 & 11.99$\pm$0.04 & 11.08$\pm$0.02 & 10.54$\pm$0.02 & M9.5V  & (1) & 11.35$\pm$2.00 &  8 & \\
GIC0050       & GR 50           & 13.97$\pm$0.05 & 12.68$\pm$0.03 & 11.00$\pm$0.03 & 2 & R & .0130 & 16 &  85 &  9.28$\pm$0.02 &  8.62$\pm$0.03 &  8.35$\pm$0.02 & M3.0Ve & (2) & 11.73$\pm$2.04 & 12 & * \\
2MA0112+1703  & GU Psc          & 14.14$\pm$0.04 & 13.01$\pm$0.04 & 11.61$\pm$0.03 & 2 & I & .0167 &  7 &  36 & 10.21$\pm$0.02 &  9.60$\pm$0.02 &  9.35$\pm$0.02 & M3     & (3) & 31.24$\pm$4.95 & 12 & \\
2MA0123-6921  &                 & 19.12$\pm$0.19 & 17.22$\pm$0.04 & 14.91$\pm$0.03 & 2 & I & .0139 & 13 &  53 & 12.32$\pm$0.02 & 11.71$\pm$0.03 & 11.32$\pm$0.03 & M8     & (4) & 20.42$\pm$3.22 & 12 & \\
SCR0128-1458  &                 & 13.60$\pm$0.04 & 12.33$\pm$0.03 & 10.67$\pm$0.03 & 3 & V & .0099 & 13 &  70 &  9.06$\pm$0.02 &  8.56$\pm$0.06 &  8.20$\pm$0.03 & M3.0Ve & (2) & 12.77$\pm$2.00 & 12 & * \\
BAR161-012    & Barta 161 12    & 13.42$\pm$0.03 & 12.19$\pm$0.03 & 10.57$\pm$0.03 & 2 & R & .0510 & 13 &  70 &  8.96$\pm$0.02 &  8.39$\pm$0.03 &  8.08$\pm$0.03 & M3.0Ve & (2) & 12.34$\pm$1.95 & 12 & * \\
SCR0143-0602  & RBS 237         & 13.01$\pm$0.03 & 11.80$\pm$0.03 & 10.25$\pm$0.03 & 2 & V & .0368 & 12 &  61 &  8.77$\pm$0.02 &  8.17$\pm$0.03 &  7.91$\pm$0.02 & M4.0Ve & (2) & 13.33$\pm$2.05 & 12 & * \\
SIP0152-6329  &                 & 15.41$\pm$0.05 & 13.93$\pm$0.03 & 12.01$\pm$0.03 & 2 & R & .0141 & 11 &  56 & 10.17$\pm$0.02 &  9.60$\pm$0.02 &  9.26$\pm$0.02 & M4.5Ve & (2) & 13.69$\pm$2.15 & 12 & * \\
SCR0222-6022  & RBS 309         & 13.36$\pm$0.05 & 12.12$\pm$0.04 & 10.52$\pm$0.04 & 3 & V & .0408 & 11 &  52 &  8.99$\pm$0.02 &  8.39$\pm$0.04 &  8.10$\pm$0.03 & M3.0Ve & (2) & 13.25$\pm$2.03 & 12 & * \\
2MA0236-5203  & EXO 0235.2-5216 & 12.06$\pm$0.03 & 11.02$\pm$0.03 &  9.75$\pm$0.03 & 2 & V & .0397 & 13 &  60 &  8.42$\pm$0.02 &  7.76$\pm$0.02 &  7.50$\pm$0.03 & M2.5Ve & (2) & 15.61$\pm$2.88 & 12 & \\ 
2MA0254-5108A & GSC 08057-00342 & 12.08$\pm$0.04 & 11.06$\pm$0.04 &  9.87$\pm$0.03 & 3 & V & .0514 & 14 &  70 &  8.67$\pm$0.03 &  8.07$\pm$0.06 &  7.79$\pm$0.03 & M2.0Ve & (2) & 21.45$\pm$3.40 & 12 & \\
2MA0254-5108B &                 & 17.56$\pm$0.06 & 15.95$\pm$0.09 & 13.90$\pm$0.04 & 3 & V & .0458 & 14 &  70 & 12.07$\pm$0.02 & 11.49$\pm$0.02 & 11.19$\pm$0.02 &        &     & 30.48$\pm$5.94 & 12 & \\
SCR0336-2619  &                 & 16.33$\pm$0.04 & 14.76$\pm$0.03 & 12.72$\pm$0.03 & 2 & I & .0117 & 13 &  68 & 10.68$\pm$0.02 & 10.13$\pm$0.02 &  9.76$\pm$0.02 & M4.5Ve & (2) & 14.19$\pm$2.21 & 12 & * \\
RX0413-0139   &                 & 13.96$\pm$0.03 & 12.66$\pm$0.03 & 10.97$\pm$0.03 & 2 & V & .0345 & 14 &  58 &  9.38$\pm$0.02 &  8.76$\pm$0.03 &  8.50$\pm$0.02 & K5.0Ve & (2) & 13.97$\pm$2.17 & 12 & \\
2MA0446-1116AB & RBS 584        & 12.25$\pm$0.05 & 11.05$\pm$0.03 &  9.57$\pm$0.04 & 2 & V & .0162 & 14 &  71 &  8.14$\pm$0.02 &  7.56$\pm$0.03 &  7.29$\pm$0.02 & M4.9V  & (5) & 11.07$\pm$1.74 & 12 & \\
HD271076      &                 & 11.35$\pm$0.03 & 10.33$\pm$0.03 &  9.11$\pm$0.03 & 3 & V & .0076 &  9 &  40 &  7.89$\pm$0.03 &  7.32$\pm$0.03 &  7.05$\pm$0.02 & M2.0V  & (2) & 15.06$\pm$2.32 & 12 & * \\
SCR0533-4257AB & RBS 661        & 12.58$\pm$0.05 & 11.27$\pm$0.04 &  9.59$\pm$0.03 & 3 & R & .0138 & 28 & 141 &  8.00$\pm$0.03 &  7.40$\pm$0.03 &  7.12$\pm$0.03 & M4.0VeJ& (2) &  7.41$\pm$1.15 & 12 & * \\
LP780-032     &                 & 12.77$\pm$0.03 & 11.55$\pm$0.04 &  9.99$\pm$0.03 & 3 & V & .0111 & 23 & 111 &  8.51$\pm$0.02 &  7.91$\pm$0.03 &  7.65$\pm$0.02 & M4.0Ve & (2) & 11.67$\pm$1.80 & 12 & \\
2MA0936-2610AC&                 & 13.12$\pm$0.03 & 11.87$\pm$0.03 & 10.32$\pm$0.03 & 3 & V & .0088 &  9 &  46 &  8.86$\pm$0.03 &  8.29$\pm$0.05 &  7.96$\pm$0.02 & M4.0Ve & (2) & 13.57$\pm$2.20 & 12 & * \\
2MA0936-2610B &                 & 19.92$\pm$0.29 & 17.46$\pm$0.08 & 14.98$\pm$0.02 & 2 & V &\ldots&\ldots&\ldots&12.27$\pm$0.02 &11.61$\pm$0.02 & 11.21$\pm$0.02 & \ldots &\ldots& 17.03$\pm$2.73 & 9 & * \\
SIP1110-3731ABC&TWA 3ABC        & 12.06$\pm$0.03 & 10.82$\pm$0.03 &  9.20$\pm$0.03 & 3 & V &\ldots& 13 &  63 &  7.65$\pm$0.02 &  7.04$\pm$0.03 &  6.77$\pm$0.02 & M4.0VeJ& (2) &  6.99$\pm$1.07 & 12 & * \\
STEPH0164     &                 & 12.75$\pm$0.04 & 11.59$\pm$0.04 & 10.10$\pm$0.03 & 2 & V & .0132 & 13 &  55 &  8.70$\pm$0.03 &  8.07$\pm$0.04 &  7.81$\pm$0.03 & M3.5Ve & (2) & 14.23$\pm$2.22 & 12 & * \\
GJ2122AB      & HD 150848       &  9.68$\pm$0.03 &  8.73$\pm$0.03 &  7.69$\pm$0.03 & 3 & V & .0122 & 40 & 215 &  6.57$\pm$0.02 &  5.94$\pm$0.03 &  5.72$\pm$0.03 & M1.0V  & (2) &  9.81$\pm$1.57 & 12 & \\
UPM1710-5300AB &                & 11.75$\pm$0.03 & 10.65$\pm$0.03 &  9.31$\pm$0.03 & 2 & V & .0149 & 10 &  53 &  8.00$\pm$0.03 &  7.41$\pm$0.02 &  7.16$\pm$0.02 &        &     & 13.11$\pm$2.03 & 12 & \\
SIP1809-7613  &                 & 15.11$\pm$0.04 & 13.62$\pm$0.03 & 11.71$\pm$0.03 & 2 & I & .0090 & 13 &  67 &  9.82$\pm$0.02 &  9.28$\pm$0.02 &  8.99$\pm$0.02 & M4.5Ve & (2) & 12.00$\pm$1.91 & 12 & * \\
SCR1816-5844  &                 & 12.78$\pm$0.05 & 11.62$\pm$0.06 & 10.08$\pm$0.04 & 4 & V & .0675 & 15 &  62 &  8.60$\pm$0.02 &  7.96$\pm$0.06 &  7.70$\pm$0.02 & M3.5Ve & (2) & 12.20$\pm$1.97 & 12 & * \\
DEN1956-3207B &                 & 13.25$\pm$0.03 & 12.01$\pm$0.04 & 10.45$\pm$0.03 & 2 & V & .0196 &  9 &  45 &  8.96$\pm$0.03 &  8.34$\pm$0.04 &  8.11$\pm$0.03 & M4     & (3) & 14.18$\pm$2.22 & 12 & \\
DEN1956-3207A & TYC 7443-1102-1 & 11.54$\pm$0.03 & 10.64$\pm$0.04 &  9.74$\pm$0.03 & 2 & V & .0307 &  9 &  45 &  8.71$\pm$0.03 &  8.03$\pm$0.04 &  7.85$\pm$0.02 &        &     & 30.21$\pm$5.14 & 12 & \\
BD-13-06424   &                 & 10.51$\pm$0.04 &  9.58$\pm$0.04 &  8.59$\pm$0.03 & 3 & V & .0280 & 15 &  77 &  7.45$\pm$0.02 &  6.77$\pm$0.04 &  6.57$\pm$0.02 & M0Ve   & (6) & 14.50$\pm$2.67 & 12 & \\
\hline
\enddata

\tablecomments{Photometry data collected on the sample. Asterisks in the notes column indicate TINYMO stars identified in the TINYMO survey itself. $VRI$ photometry and variability are original, $JHK$ is reprinted from the 2MASS Point Source Catalog \citep{Cutri2003}.}
\tablenotetext{a}{Astrometric results and relative photometry use new $V$ filter data.}
\tablenotetext{b}{References: (1) \citet{Leggett2001}; (2) This paper; (3) \citet{Riaz2006}; (4) \citet{Schmidt2007}; (5) \citet{Shkolnik2009}; (6) \citet{Torres2006}. ``J'' indicates joint spectral types from unresolved multiples.}

\end{deluxetable*}
\end{rotatetable*}

\subsection{Spectroscopy}
\label{sec:spectroscopy}

For the stars of interest, we obtained high-SNR low resolution long-slit spectroscopy. The primary purpose of this spectroscopy was to identify (and remove) giant stars from our astrometric sample. The data were also intended for spectral typing and used for measuring gravity-sensitive spectral features. 

\subsubsection{CTIO 1.5m/RCSpec}
Most of the spectroscopy were collected on the CTIO/SMARTS 1.5m telescope with the Ritchie-Cr{\'e}tchien spectrograph using the 32/Ia first-order grating (15.13$^{\circ}$ tilt, 5994\AA--9600\AA, R=500, OG570 blocking filter), and a 2\arcsec~slit to maximize the stellar flux. The RC spectrograph uses a relatively old 1200x800 Loral CCD with few bad columns and no backthinning, which minimizes fringing in the red end of the spectrum. Two distinct epochs of observations were conducted, from 2003-2006 for some of the additional CTIOPI targets now being presented here, and from 2009-2011 specifically for the TINYMO survey targets. In both cases, the regular operation was two exposures of the target object, followed by one Neon-Argon (NeAr) lamp exposure for wavelength solution, with one flux standard taken per night.

From 2003-2006, observing was done in person on nine user runs. From 2009-2011, observing was done in SMARTS queue mode. At that time, the 32/Ia setting was no longer a common setup, so for the most part data for TINYMO stars were also collected in single-night blocks. The flux standard was chosen by the queue manager from a small subset of stars, all of which are in IRAF's standard {\it onedstds\$iidscal/ctionewcal} directory. Spectra were reduced using standard IRAF {\it onedspec}, {\it ccdred}, and {\it ctioslit} packages.
  
\subsubsection{Lowell 1.8m/DeVeny}

Additional spectra were gathered at Lowell Observatory's Perkins 1.8m telescope with the DeVeny spectrograph and its 400 g/mm grating tilted at 17 degrees, with the OG570 blocking filter, for coverage from 5800-9200\AA~at a spectral resolution of roughly R=1500. Spectra were obtained on five runs from 2009 -- 2010. Owing to the observatory's northern latitude, only targets north of DEC=$-$36\arcdeg were observed from Lowell.

The process of obtaining spectra changed considerably over the course of the project, partly owing to the fact that the DeVeny was not regularly used and rarely in the red end of the spectral range. For the first run (Feb 2009), only one spectrum was taken of each target and standard IRAF flux standard, with Neon-Argon calibration lamp spectra taken at four different times throughout the night. Subsequent runs (May 2009 and Dec 2009) included lamps taken after each exposure and a large catalog of flatfields, and finally (Mar 2010 and May 2010) flat lamps were taken after every exposure.  Spectra were reduced using standard IRAF {\it onedspec}, {\it ccdred} and {\it kpnoslit} packages.

\subsubsection{CTIO 4.0m/RCSpec}

Ten objects were observed with the CTIO 4.0m RCSpec on 18 Sep 2008 and 19 Sep 2008, using the KPGLF-1 grating (632 g/mm) and an unknown blocking filter (S. Kafka, private communication).  The spectra are higher resolution than our CTIO 1.5m spectra ($\Delta\lambda = $1.90\AA, R$\approx$ 3000), and cover 4900\AA -- 8050\AA.  These spectra do not have the Na~I doublet nor Ca~II triplet used for gravity and luminosity class detection, but do contain H$\alpha$ and the K~I doublet.  For some stars this is the only spectrum available.

\subsubsection{CFHT/ESPaDONs}

\object{LP 780-032} was observed with ESPaDONs on CFHT on 28 Jan 2016. This spectrum is much higher resolution than our CTIO 1.5m spectra and covers 3730\AA-10290\AA~at a resolving power of R=75000. Spectra were processed and flat-fielded through standard methods, and barycentric velocity was removed. LP 780-032 was determined to have a radial velocity of $-$7 km s$^{-1}$ and rotational velocity of 2 km s$^{-1}$.

\subsection{Astrometry}

\subsubsection{CTIOPI}

TINYMO targets that were spectroscopically identified as dwarfs and were within 15 pc according to $VRIJHK$ photometric distance estimates were placed on the CTIOPI astrometric program.

The RECONS group has been conducting the Cerro Tololo Inter-american Observatory Parallax Investigation at the CTIO 0.9m since 1999, until 2003 as an NOAO survey, and 2003-present through the SMARTS Consortium. CTIOPI uses the facility Tek \#2 $VRI$ filters for observations. For a period of time between 2005 and 2009, the Tek \#1 $V$ filter was used instead (see \citealt{Subasavage2007} for more information). The filter had different astrometric (though not photometric) properties, and all results incorporating data taken in that filter are marked as such in Table \ref{tab:photometry} and Table \ref{tab:astrometry}.

For astrometric observations, target fields are observed usually three times a year within two hours of transit for at least two years in a single filter, chosen out of the $VRI$ set to provide the optimal balance between exposure time and brightness of the reference field. Photometric frames in the appropriate filter that meet image quality and hour angle requirements may be used for astrometry.  Data are reduced using the pipeline described in \citet{Jao2005}, and as used in all subsequent CTIOPI publications\footnote{See http://www.recons.org for a list of publications}. 

The parallax results (Table \ref{tab:astrometry}) indicate that 15 of the 26 systems presented here are between 25 and 50 pc away, counter to the expectations of the TINYMO selection process, while 11 systems were within the expected 25 parsecs.

\begin{deluxetable*}{llrcrclcccrrrcccl}
\setlength{\tabcolsep}{0.02in}
\tablewidth{0pt}
\tabletypesize{\tiny}
\tablecaption{Astrometric Results for 26 Selected Star Systems}
\label{tab:astrometry}
\tablehead{\colhead{}              &
	   \colhead{R.A.}          &
	   \colhead{Decl.}         &
 	   \colhead{}              &
	   \colhead{}              &
	   \colhead{}              &
	   \colhead{}              &
	   \colhead{}              &
	   \colhead{}              &
	   \colhead{$\pi$(Rel)}    &
	   \colhead{$\pi$(Corr)}   &
	   \colhead{$\pi$(Abs)}    &
	   \colhead{$\mu$}         &
	   \colhead{P.A.}          &
	   \colhead{V$_{tan}$}     &
	   \colhead{}             \\
           \colhead{Name}          &
           \colhead{(J2000)}       &
	   \colhead{(J2000)}       &
 	   \colhead{Filter}        &
	   \colhead{$N_{sea}$}     &
	   \colhead{$N_{frm}$}     &
	   \colhead{Coverage\tablenotemark{a}}      &
	   \colhead{Years}         &
 	   \colhead{$N_{ref}$}     &
 	   \colhead{(mas)}         &
	   \colhead{(mas)}         &
	   \colhead{(mas)}         &
	   \colhead{(mas yr$^{-1}$)}&
	   \colhead{(deg)}         &
	   \colhead{(km s$^{-1}$)} &
	   \colhead{Notes}        \\
           \colhead{(1)}           &         
           \colhead{(2)}           &
           \colhead{(3)}           &
           \colhead{(4)}           &
           \colhead{(5)}           &
           \colhead{(6)}           &
           \colhead{(7)}           &
           \colhead{(8)}           &
           \colhead{(9)}           &
           \colhead{(10)}          &
           \colhead{(11)}          &
           \colhead{(12)}          &
           \colhead{(13)}          &
           \colhead{(14)}          &
           \colhead{(15)}          &
           \colhead{(16)}          }
\startdata
NLTT01261      & 00 24 24.63 & -01 58 20.0 & I & 7s &  29 & 2008.70-2015.82 &  7.12 &  6 & 81.58$\pm$2.22 & 0.85$\pm$0.07 & 82.43$\pm$2.22 & 157.1$\pm$0.8 & 336.4$\pm$0.54 &  9.0 & \\
GIC0050        & 00 32 53.14 & -04 34 07.0 & R & 7s &  85 & 2007.82-2015.83 &  8.01 &  7 & 51.89$\pm$1.05 & 0.72$\pm$0.07 & 52.61$\pm$1.05 & 167.2$\pm$0.5 & 156.4$\pm$0.32 & 15.1 & * \\
2MA0112+1703   & 01 12 35.06 & +17 03 55.5 & I & 3s &  36 & 2013.67-2015.96 &  2.29 &  6 & 18.75$\pm$2.15 & 1.85$\pm$0.18 & 20.60$\pm$2.16 & 134.1$\pm$2.6 & 135.1$\pm$2.19 & 30.8 & \\
2MA0123-6921   & 01 23 11.27 & -69 21 38.0 & I & 7s &  53 & 2008.70-2014.92 &  6.22 & 10 & 22.14$\pm$1.37 & 0.78$\pm$0.07 & 22.92$\pm$1.37 &  87.4$\pm$0.7 & 107.4$\pm$0.83 & 18.1 & \\
SCR0128-1458   & 01 28 39.53 & -14 58 04.2 & V & 7s &  70 & 2009.93-2015.97 &  6.04 &  5 & 71.84$\pm$1.35 & 3.01$\pm$0.23 & 74.85$\pm$1.37 &  71.9$\pm$0.8 & 226.0$\pm$1.30 &  4.6 & * \\
BAR161-012     & 01 35 13.94 & -07 12 51.8 & R & 6s &  70 & 2009.94-2014.93 &  4.99 &  7 & 26.11$\pm$1.78 & 1.56$\pm$0.27 & 27.67$\pm$1.80 &  93.3$\pm$1.2 & 114.1$\pm$1.38 & 16.0 & * \\
SCR0143-0602   & 01 43 45.13 & -06 02 40.1 & V & 6s &  61 & 2009.74-2014.91 &  5.18 &  7 & 49.59$\pm$1.48 & 0.93$\pm$0.13 & 50.52$\pm$1.49 &  47.3$\pm$0.9 & 104.5$\pm$1.96 &  4.4 & * \\
SIP0152-6329   & 01 52 55.35 & -63 29 30.2 & R & 8s &  56 & 2007.82-2014.93 &  7.11 &  7 & 25.14$\pm$1.17 & 1.39$\pm$0.23 & 26.53$\pm$1.19 & 127.0$\pm$0.5 &  95.7$\pm$0.38 & 22.7 & * \\
SCR0222-6022   & 02 22 44.17 & -60 22 47.6 & V & 6s &  52 & 2009.75-2014.65 &  4.90 &  8 & 31.76$\pm$1.66 & 0.94$\pm$0.13 & 32.70$\pm$1.67 & 126.2$\pm$1.3 &  98.0$\pm$0.92 & 18.3 & * \\
2MA0236-5203   & 02 36 51.71 & -52 03 03.7 & V & 6s &  63 & 2009.92-2014.93 &  5.00 &  6 & 25.80$\pm$1.24 & 1.90$\pm$0.25 & 27.70$\pm$1.26 &  80.4$\pm$0.8 &  96.5$\pm$0.93 & 13.8 & \\
2MA0254-5108A  & 02 54 33.17 & -51 08 31.4 & V & 6c &  70 & 2009.92-2014.91 &  4.99 &  6 & 25.10$\pm$1.64 & 1.95$\pm$0.24 & 27.05$\pm$1.66 &  85.9$\pm$1.0 &  94.6$\pm$1.04 & 15.0 & \\
2MA0254-5108B  & 02 54 34.77 & -51 08 28.8 & V & 6c &  70 & 2009.92-2014.91 &  4.99 &  6 & 20.52$\pm$2.04 & 1.95$\pm$0.25 & 22.47$\pm$2.06 &  88.2$\pm$1.3 &  92.1$\pm$1.20 & 18.6 & \\
SCR0336-2619   & 03 36 31.46 & -26 19 57.9 & I & 7s &  68 & 2008.70-2015.08 &  6.38 &  9 & 21.12$\pm$1.09 & 0.68$\pm$0.07 & 21.80$\pm$1.09 &  76.4$\pm$0.5 & 107.9$\pm$0.73 & 16.6 & * \\
RX0413-0139    & 04 13 26.64 & -01 39 21.2 & V & 6s &  58 & 2009.94-2015.08 &  5.14 & 10 & 35.68$\pm$1.90 & 0.74$\pm$0.17 & 36.42$\pm$1.91 & 127.0$\pm$1.5 &  93.2$\pm$0.99 & 16.5 & \\
2MA0446-1116AB & 04 46 51.74 & -11 16 47.7 & V & 5c &  71 & 2011.73-2016.04 &  4.31 &  8 & 70.77$\pm$3.41 & 0.91$\pm$0.23 & 71.68$\pm$3.42 & 149.1$\pm$2.1 & 249.2$\pm$1.48 &  9.9 & * \\
HD271076       & 05 10 09.69 & -72 36 27.9 & V\tablenotemark{b} & 5c &  40 & 2007.81-2011.74 &  3.93 &  7 & 46.73$\pm$2.75 & 2.78$\pm$0.46 & 49.51$\pm$2.79 & 130.7$\pm$2.6 &  80.3$\pm$1.81 & 12.5 & * \\
SCR0533-4257AB & 05 33 28.03 & -42 57 20.5 & R & 9c & 141 & 2007.81-2016.05 &  8.24 &  9 & 95.46$\pm$1.32 & 0.98$\pm$0.21 & 96.44$\pm$1.34 &  38.8$\pm$0.5 & 328.8$\pm$1.48 &  1.9 & * \\
LP780-032      & 06 39 37.41 & -21 01 33.3 & V\tablenotemark{b} & 8c & 111 & 2008.70-2016.04 &  7.34 & 13 & 62.26$\pm$0.58 & 1.17$\pm$0.11 & 63.43$\pm$0.59 & 179.2$\pm$0.3 & 294.9$\pm$0.17 & 13.4 & * \\
2MA0936-2610AC & 09 36 57.83 & -26 10 11.2 & V & 4c &  46 & 2010.16-2013.38 &  3.22 &  9 & 52.74$\pm$1.41 & 1.01$\pm$0.15 & 53.75$\pm$1.42 &  44.1$\pm$1.2 & 137.7$\pm$2.99 &  3.9 & * \\
SIP1110-3731AC & 11 10 27.88 & -37 31 52.0 & V & 6s &  63 & 2009.32-2014.17 &  4.85 &  8 & 28.41$\pm$3.97 & 0.98$\pm$0.12 & 29.39$\pm$3.97 &  91.8$\pm$2.5 & 263.8$\pm$2.42 & 14.8 & * \\
SIP1110-3731B  & 11 10 27.88 & -37 31 52.0 & V & 6s &  46 & 2009.32-2014.17 &  4.85 &  8 & 30.31$\pm$6.82 & 0.98$\pm$0.12 & 31.29$\pm$6.82 & 114.6$\pm$4.4 & 246.9$\pm$4.06 & 17.4 & * \\
STEPH0164      & 12 06 22.15 & -13 14 56.1 & V & 5c &  55 & 2010.20-2014.44 &  4.23 &  5 & 31.63$\pm$2.36 & 0.58$\pm$0.15 & 32.21$\pm$2.36 & 113.5$\pm$1.7 & 128.7$\pm$1.68 & 16.7 & * \\
GJ2122AB       & 16 45 16.97 & -38 48 33.3 & V\tablenotemark{b} &16s & 215 & 2000.58-2016.21 & 15.63 &  8 & 75.69$\pm$1.57 & 1.50$\pm$0.50\tablenotemark{c} & 77.19$\pm$1.64 &  60.9$\pm$0.4 & 203.8$\pm$0.62 &  3.6 & \\
UPM1710-5300AB & 17 10 44.31 & -53 00 25.1 & V & 5c &  53 & 2010.50-2014.27 &  3.78 & 10 & 61.88$\pm$2.94 & 2.67$\pm$0.31 & 64.55$\pm$2.96 & 169.1$\pm$2.1 & 195.9$\pm$1.31 & 12.4 & \\
SIP1809-7613   & 18 09 06.94 & -76 13 23.9 & I & 5c &  67 & 2010.40-2014.28 &  3.88 & 10 & 36.12$\pm$1.50 & 1.69$\pm$0.21 & 37.81$\pm$1.51 & 143.8$\pm$1.2 & 178.2$\pm$0.70 & 18.0 & * \\
SCR1816-5844   & 18 16 12.37 & -58 44 05.6 & V & 6c &  62 & 2010.50-2015.29 &  4.79 &  9 & 33.61$\pm$1.22 & 0.86$\pm$0.15 & 34.47$\pm$1.23 & 139.8$\pm$0.8 & 172.8$\pm$0.48 & 19.2 & * \\
DEN1956-3207B  & 19 56 02.94 & -32 07 18.7 & V & 4s &  45 & 2012.83-2015.68 &  2.85 &  6 & 21.66$\pm$1.83 & 1.09$\pm$0.20 & 22.75$\pm$1.84 &  64.1$\pm$1.9 & 149.3$\pm$3.30 & 13.3 & \\
DEN1956-3207A  & 19 56 04.38 & -32 07 37.7 & V & 4s &  45 & 2012.83-2015.68 &  2.85 &  6 & 20.93$\pm$1.77 & 1.09$\pm$0.20 & 22.02$\pm$1.78 &  62.9$\pm$1.8 & 148.5$\pm$3.24 & 13.5 & \\
BD-13-06424    & 23 32 30.87 & -12 15 51.4 & V & 5s &  77 & 2010.73-2015.56 &  4.83 &  5 & 34.19$\pm$1.84 & 0.58$\pm$0.05 & 34.77$\pm$1.84 & 156.5$\pm$1.6 & 108.7$\pm$1.04 & 21.3 & \\
\hline
\enddata

\tablecomments{Astrometric results derived for the sample. Asterisks in the notes column indicate TINYMO stars identified in the TINYMO survey itself.}

\tablenotetext{a}{'c' indicates continuous coverage, at least two epochs per observing season. 's' indicates scattered observations, with years missing.}

\tablenotetext{b}{Astrometric results and relative photometry use new $V$ filter data.}

\tablenotetext{c}{Generic correction to absolute parallax was used because the reference starfield appears to be reddened by the nearby dust cloud \object{[DB2002b] G344.85+4.27}}

\end{deluxetable*}

\subsubsection{FGS}

One additional opportunity occurred in 2008 when the Hubble Space Telescope's data bus developed a fault. As a result, the only available instruments for Cycle 16B were the Fine Guidance Sensors (FGS), which communicate via the telemetry subsystems. 

The FGS system on HST can be used as an interferometer, where two of the three onboard Koesters Prisms are used, with one fixed on the target and another scanning around the source to sample the interference pattern, while the third maintains observatory pointing. The output of the interference is two S-shaped curves along orthogonal axes, from which binary stars with separations on the order of tens of milliarcseconds can be resolved by either visually identifying a second overlapping S-curve or, for close-in objects, deviations from the S-curves of a single star.

We took advantage of this opportunity to observe 66 stars from our X-ray bright sample as part of HST program  \#11943/11944 ``Binaries at the Extremes of the H-R Diagram'', PI Douglas Gies. Roughly half of the intended list was observed, and results of newly discovered binaries are mentioned where appropriate in Section \ref{sec:systemnotes}.

\section{The Complete Catalog (1215 sources)}
\label{sec:catalog}

The catalog is divided into our five subsamples of descending quality, as described in Section \ref{sec:5_quality}:
\begin{enumerate}
\item Good targets with X-ray detections in the ROSAT All Sky Survey (RASS) -- 88 stars, of which 68 have less than 0.18\arcsec yr$^{-1}$ (tiny) proper motion.
\item Good targets without X-ray detection -- 563 stars, of which 394 have tiny proper motion.
\item Probable giants ($J-K>1.2, |R_1-R_2|>1$) -- 222 stars, of which all are tiny proper motion.
\item Known giants (from SIMBAD, the General Catalog of Variable Stars \citep[][GCVS,]{Samus2010}, and the Catalog of Galactic Carbon Stars \citep[][CGCS]{Alksnis2001}) -- 223 stars, of which all are tiny proper motion.
\item Discarded objects not within 25 pc or not within the color-selection boxes. These were the ``flyers'' or objects found by eye, but are included for completeness -- 119 stars, of which 109 are tiny proper motion.
\end{enumerate}

The final catalog is presented in Table \ref{tab:tinymo}. All told, 114 of the stars in the catalog of 1215 targets now have published parallaxes (66 from CTIOPI efforts; 48 from \citealt{van-Leeuwen2007} and \citealt{gaia2016}). 251 stars have new $VRI$ photometry, and 229 have new spectral types from red-optical spectra.

\begin{deluxetable}{lll}
\tablecaption{TINYMO Catalog Headers\label{tab:tinymo}}
\tablehead{
\colhead{Number} &
\colhead{Column} &
\colhead{Unit} }
\startdata
 1 & Sample Type\tablenotemark{a} & \\
 2 & Name & \\
\hline
\multicolumn{3}{c}{CTIOPI Astrometry} \\
\hline
 3 & RA & h:m:s \\
 4 & DEC & d:m:s \\
 5 & pm & arcsec \\
 6 & P.A. & deg \\
 7 & pi & mas \\
 8 & e\_pi & mas \\
 9 & r\_pi & \\
\hline
\multicolumn{3}{c}{SuperCOSMOS photometry} \\
\hline
10 & B$_j$ & mag \\
11 & R1 & mag \\
12 & R2 & mag \\
13 & I$_{59F}$ & mag \\
14 & Blend & \\
\hline
\multicolumn{3}{c}{CTIOPI photometry} \\
\hline
15 & V & mag \\
16 & V {Blend} & \\
17 & e\_V & mag \\
18 & R & mag \\
19 & R {Blend} & \\
20 & e\_R & mag \\
21 & I & mag \\
22 & I {Blend} & \\
23 & e\_I & mag \\
24 & n\_phot & \\
\hline
\multicolumn{3}{c}{2MASS photometry} \\
\hline
25 & J & mag \\
26 & J {Blend} & \\
27 & e\_J & mag \\
28 & H & mag \\
29 & H {Blend} & \\
30 & e\_H & mag \\
31 & K & mag \\
32 & K {Blend} & \\
33 & e\_K & mag \\
\hline
\multicolumn{3}{c}{Spectra} \\
\hline
34 & SpType & \\
35 & SpType Ref & \\
36 & ewHa & \AA \\
37 & NaI Index & \\
38 & ewKI7699 & \AA \\
39 & ewNaI & \AA \\
\hline
\multicolumn{3}{c}{Distance Estimates} \\
\hline
40 & plate relations & \\
41 & plate distance & pc \\
42 & e\_plate distance & pc \\
43 & CCD relations & \\
44 & CCD distance & pc \\
45 & e\_CCD distance & pc \\
\hline
\multicolumn{3}{c}{HR Diagram Values} \\
\hline
46 & Mv & mag \\
47 & V-K & mag \\
48 & v-K & mag \\
49 & J-K & mag \\
50 & R1-R2 & mag \\
\hline
\multicolumn{3}{c}{ROSAT X-ray Data} \\
\hline 
51 & X-ray flux & cnts sec$^{-1}$ \\
52 & X-ray flux blend & \\
53 & e\_X ray flux & cnts sec$^{-1}$ \\
54 & HR1 Hardness Ratio & \\
55 & HR1 blend & \\
\enddata
\tablecomments{The full catalog is available electronically.}
\tablenotetext{a}{The Samples referred to are 1) X-ray bright stars, 2) Good stars, 3) Very Red/Probable Giants, 4) Known giants, 5) Discarded objects, as per Section \ref{sec:catalog}}
\end{deluxetable}

\section{Survey Discussion}
\label{sec:discussion}

\subsection{Analysis of the photometric cuts}

\begin{figure*}
\centering
\includegraphics[angle=0,width=\textwidth]{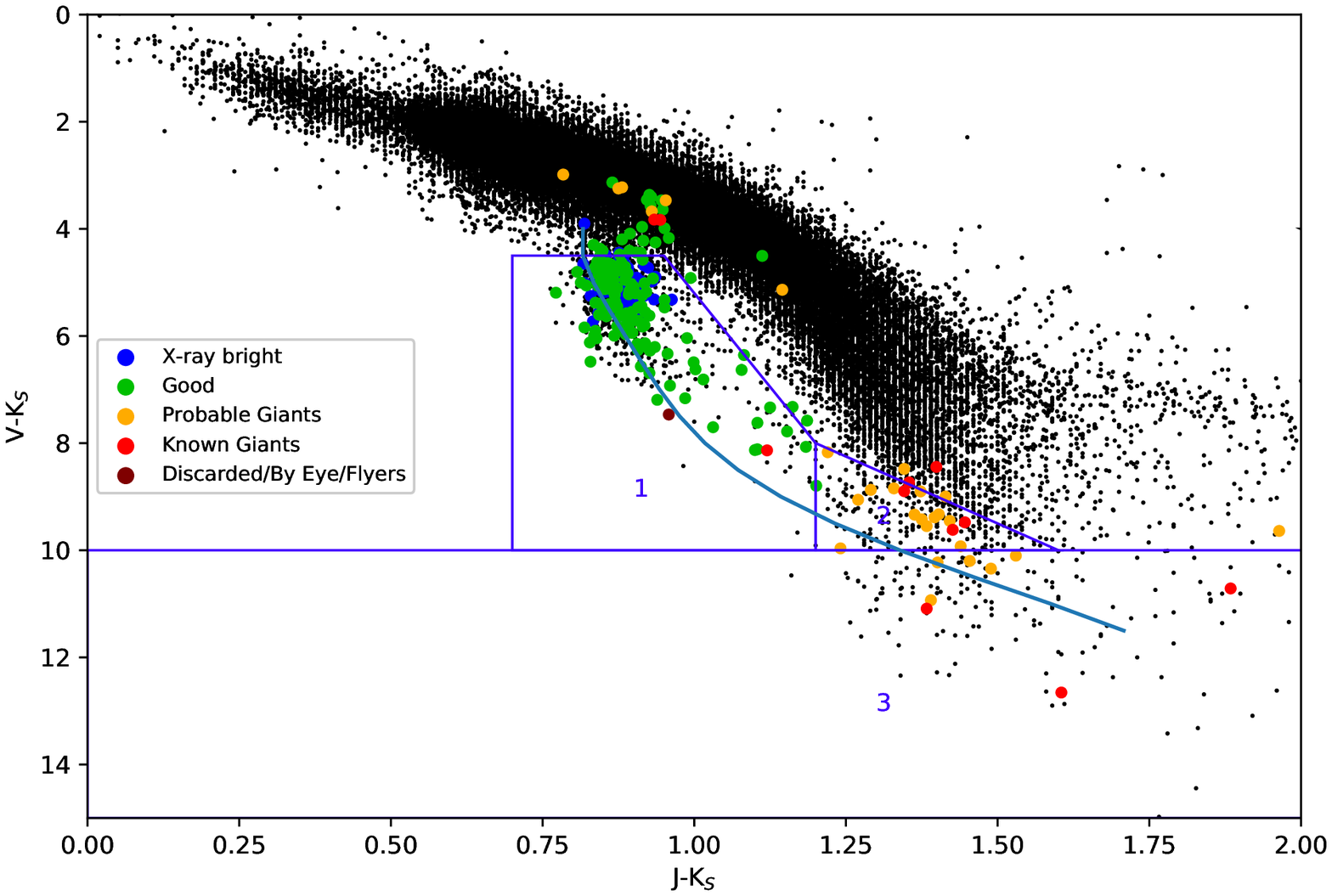}
\caption{The final state of the TINYMO sample's subset of stars observed for CCD photometry, plotted on top of the 88,586 stars from the photometric distance cut from Figure \ref{fig:TINYMO_boxes} (black points still use simulated $v$-$K$) for comparison. There has been some vertical shifting in the plotted positions of our photometric sample due to the differences between our simulated $v$ and actual Johnson $V$. 163 of the stars with CCD photometry are low proper motion ($<$0.18\arcsec~yr$^{-1}$), the remaining 103 stars are high proper motion stars observed for other reasons (other CTIOPI targets recovered by TINYMO).}
\label{fig:superTINYMOvk9}
\end{figure*}

Figure \ref{fig:superTINYMOvk9} shows the true $V-K$ colors for all targets actually observed for $VRI$ photometry by CTIOPI. This displaces the targets from where they appeared in Figure \ref{fig:TINYMO_boxes}, which was based on the simulated $v-K$ colors. It is apparent from Figure \ref{fig:superTINYMOvk9} that not all of our ``good'' (green) targets are actually dwarfs; some of them now lie in the giant locus (which is still drawn with $v-K$ color as in Figure \ref{fig:TINYMO_boxes}). This is unsurprising, as we arrived at our ``good'' sample by process of elimination, and we did not have the resources to completely vet the sample.

\begin{figure}
\centering
\includegraphics[angle=0,width=0.5\textwidth]{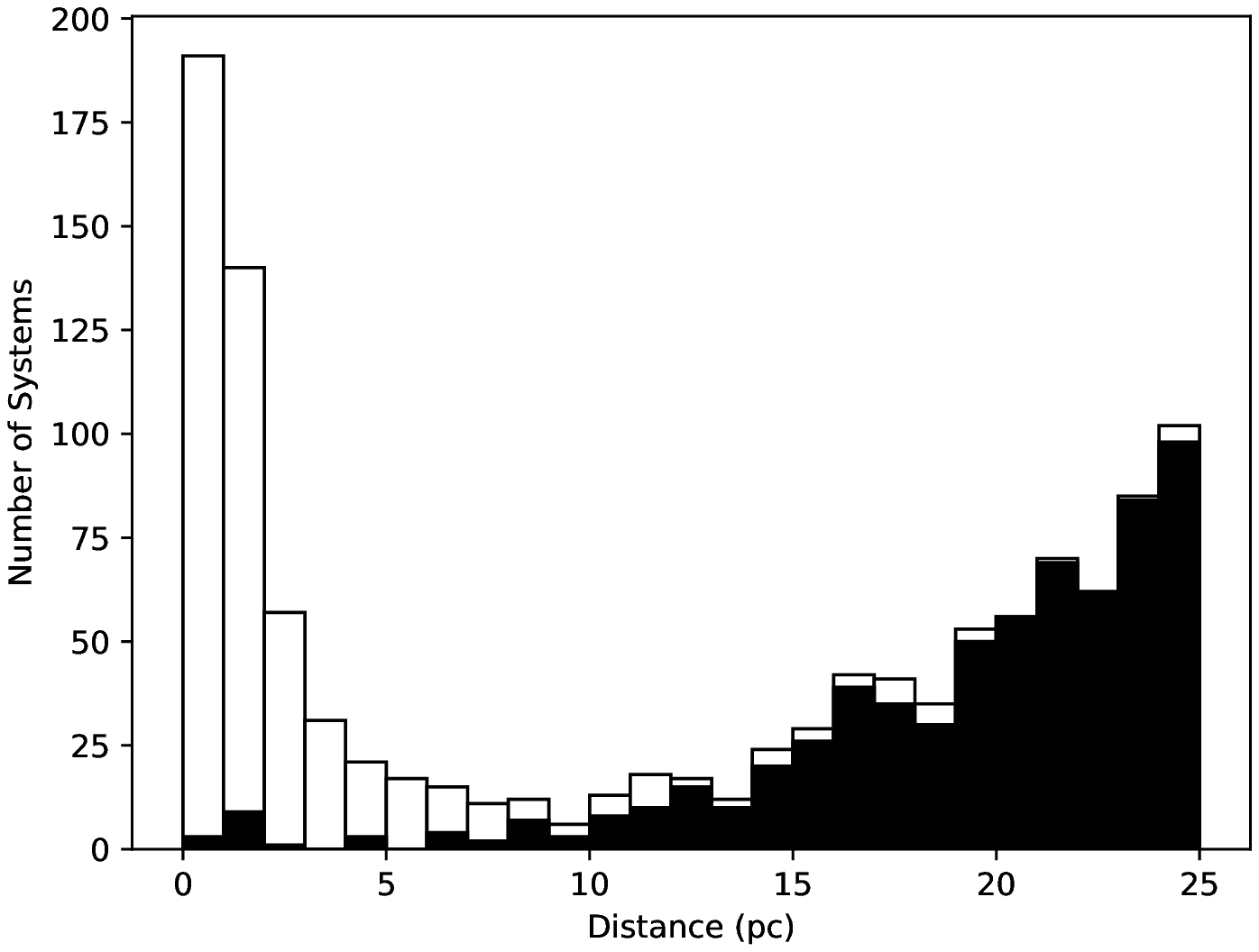}
\caption{Plate photometric distances for (white) the entire sample of 1215 objects (including later additions found by eye) and (black) the X-ray bright and good candidate samples. The trend of photometric distances is clearly bimodal, although applying our additional photometric cuts has weeded out an immense number of giants that only appeared to be nearby.}
\label{fig:TINYMO_distance_histogram}
\end{figure}

Plotting a histogram of distance estimates from plate $BRI$ and 2MASS $JHK$ photometry (Figure \ref{fig:TINYMO_distance_histogram}) shows that the original sample was bimodal, with peaks at 25 pc and 1 pc. The giant-sensitive photometric cuts remove most of the stars with predicted distances less than 2 pc. As expected, all potential nearby stars with distances less than 2 parsecs were confirmed with spectroscopy to be giants. All of the 462 nearby low proper motion stars (Categories 1 and 2) can be found among the rest of the sample in the complete Table described in Section \ref{sec:catalog}.


\subsection{Completion of TINYMO sample}

\begin{figure}
\centering
\includegraphics[angle=0,width=0.5\textwidth]{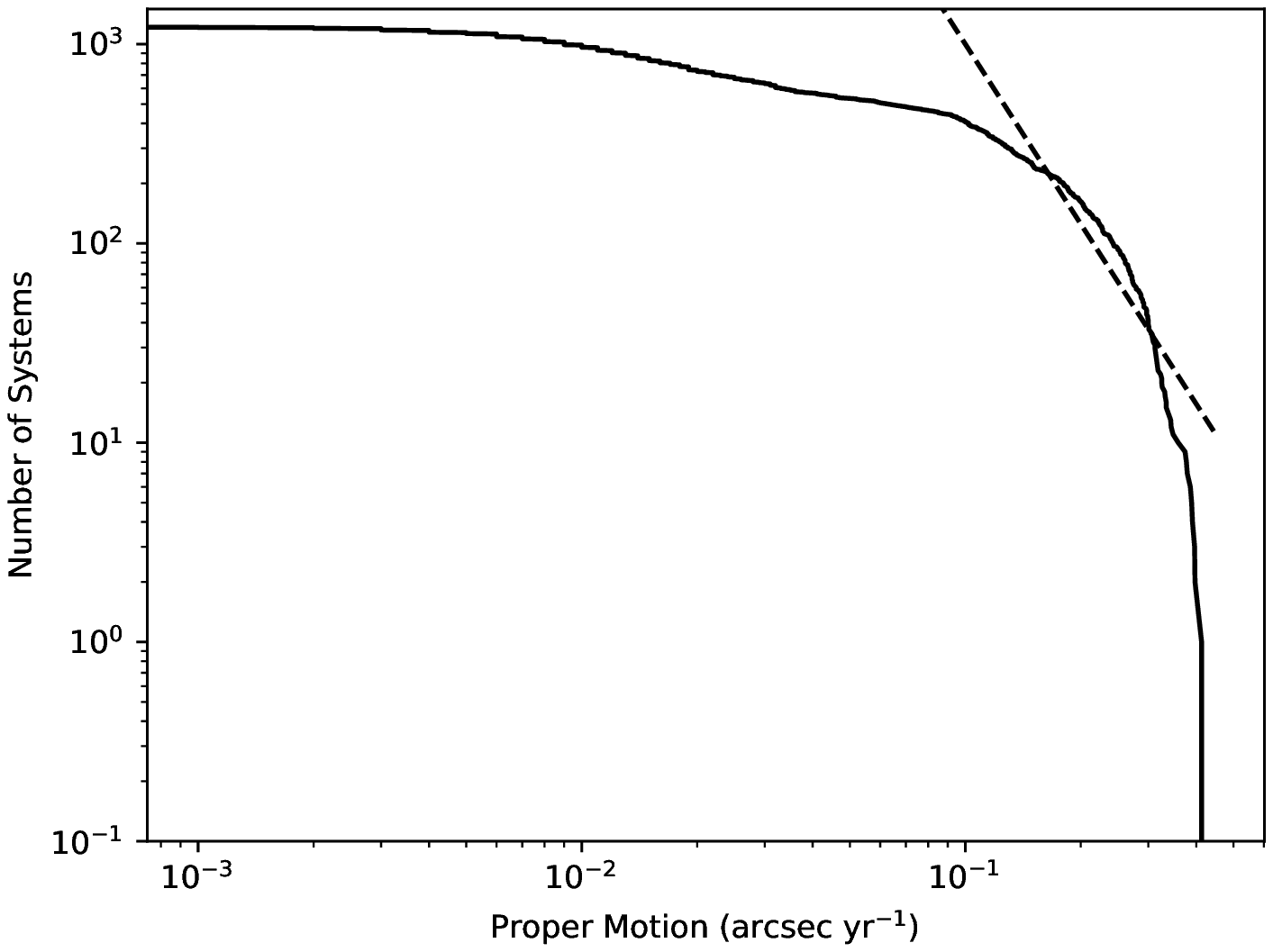}
\caption{A diagram of the completeness of stars, as in Figure 1 of \citet{Lepine2005b}.  The $\mu^{-3}$ curve (dotted line) follows from the assumption that the number density of stars varies as $n \propto d^3$, and proper motion varies as $\mu \propto d^{-1}$. The TINYMO sample is not complete for its proper motions.}
\label{fig:pmcompleteness}
\end{figure}

\begin{figure}
\centering
\includegraphics[angle=0,width=0.5\textwidth]{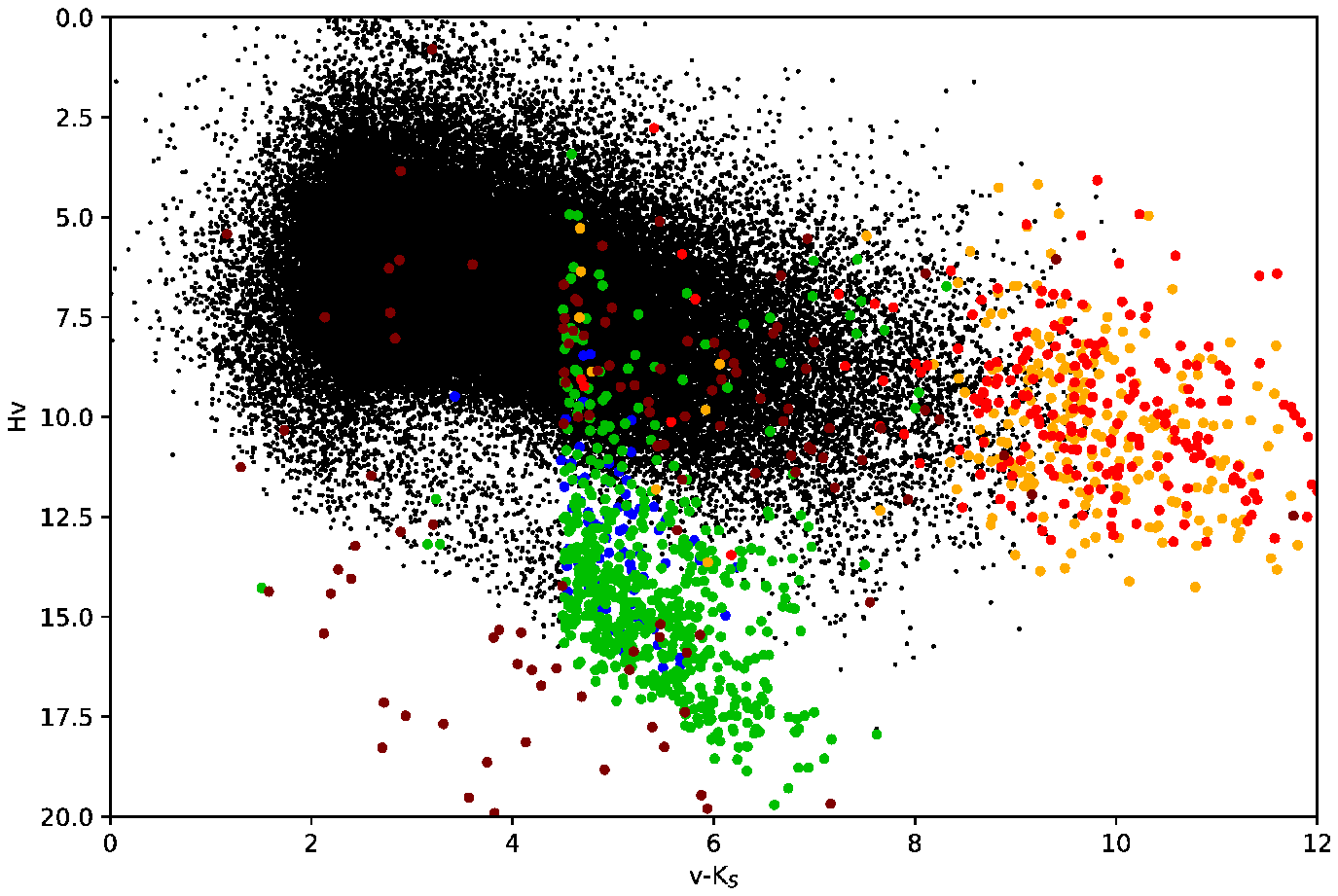}
\caption{Reduced Proper Motion diagram for the TINYMO sample, where the subsamples are colorized in the same manner as Figure \ref{fig:superTINYMOvk9}.  Reduced Proper Motion (H) is $H-v=5 log(\frac{1}{\mu}) - 5$ (i.e., $\mu$ replaces $\pi$ in the distance modulus equation). As can be seen above, there are clearly two locii, one (top) for giants, and one (bottom) for dwarfs; while most of the green/blue ``good'' sample of stars obey those trends, there are clearly green/blue ``good'' stars in the giant locus.}
\label{fig:reducedpm}
\end{figure}

TINYMO is not complete in terms of proper motions (Figure \ref{fig:pmcompleteness}), but this is not surprising as TINYMO uses photometric cuts with no lower proper motion limit, and is not a traditional proper motion survey. TINYMO is probing the range of proper motions more common for giants, which means we also cannot make use of reduced proper motion diagrams that operate under the assumption that lower proper motion objects are farther away; we are specifically looking for nearby stars that move like distant giants, and as Figure \ref{fig:reducedpm} shows, the survey contains several such targets.

\subsection{The Limit of Meaningful Proper Motion}
\label{sec:pm_lower_limit}

\begin{figure}
\centering
\includegraphics[angle=0,width=0.5\textwidth]{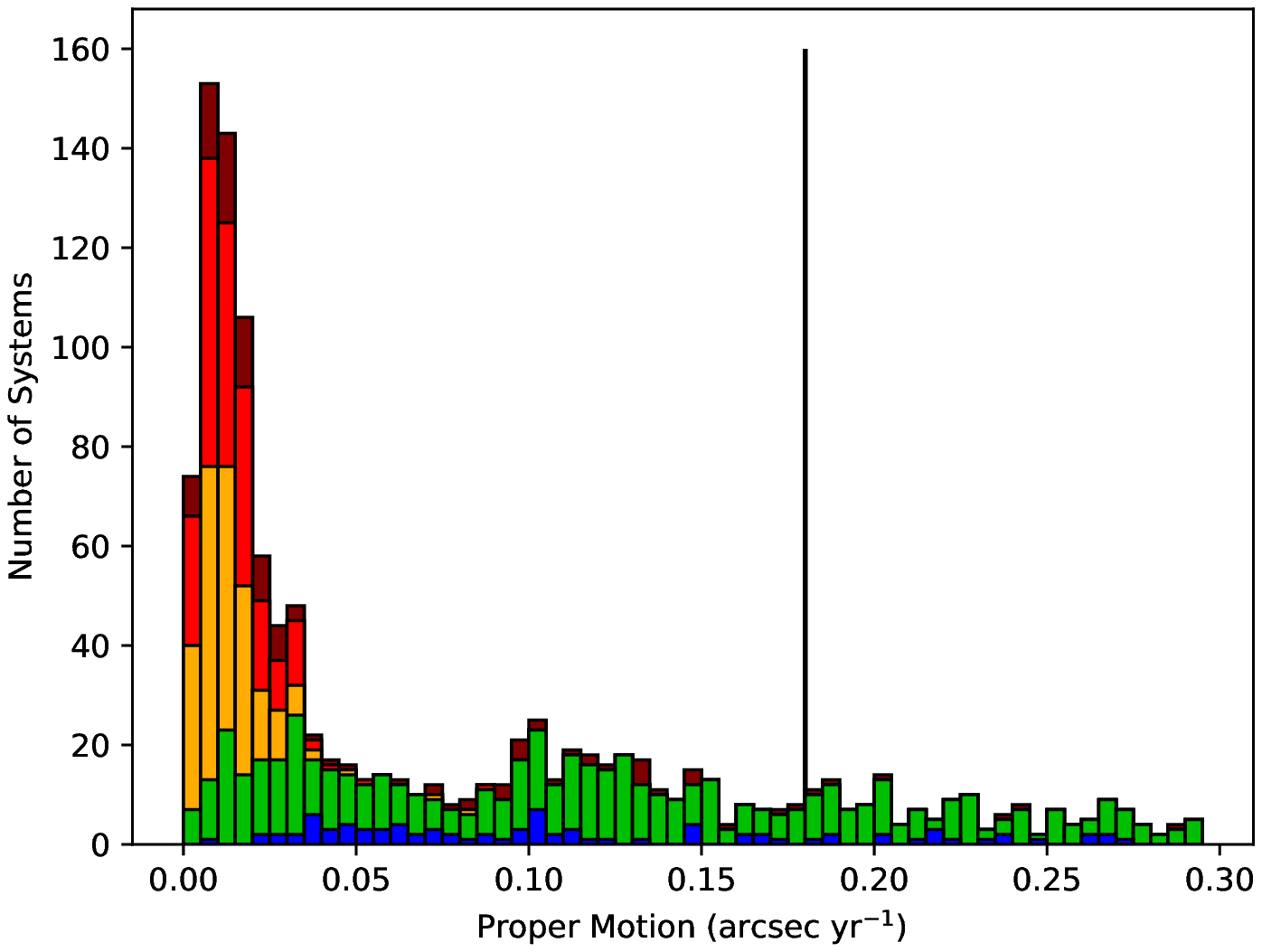}
\caption{Proper motions of the entire survey (including all higher-proper-motion stars discovered within), in 0.005\arcsec~yr$^{-1}$ bins. Color follows Figure \ref{fig:superTINYMOvk9}. Below 0.035\arcsec~yr$^{-1}$, the sample is dominated by giants and suspected giants, while above that, it is dominated by X-ray bright and regular stars.  The vertical line is at 0.18\arcsec~yr$^{-1}$, and divides TINYMO and higher proper motion targets found in the survey.}
\label{fig:propermotions}
\end{figure}

TINYMO offers a rough idea of the point at which a proper motion search (even if the proper motions are accurate) will be overwhelmed by giants. This limit (seen in Figure \ref{fig:propermotions}) appears to be around 0.035\arcsec~yr$^{-1}$, which is not coincidentally near the lower limit of Lepine's SUPERBLINK surveys \citep{Lepine2013}, 0.04\arcsec~yr$^{-1}$.

\subsection{Why so many young stars?}

\begin{figure}
\center
\includegraphics[angle=0,width=.5\textwidth]{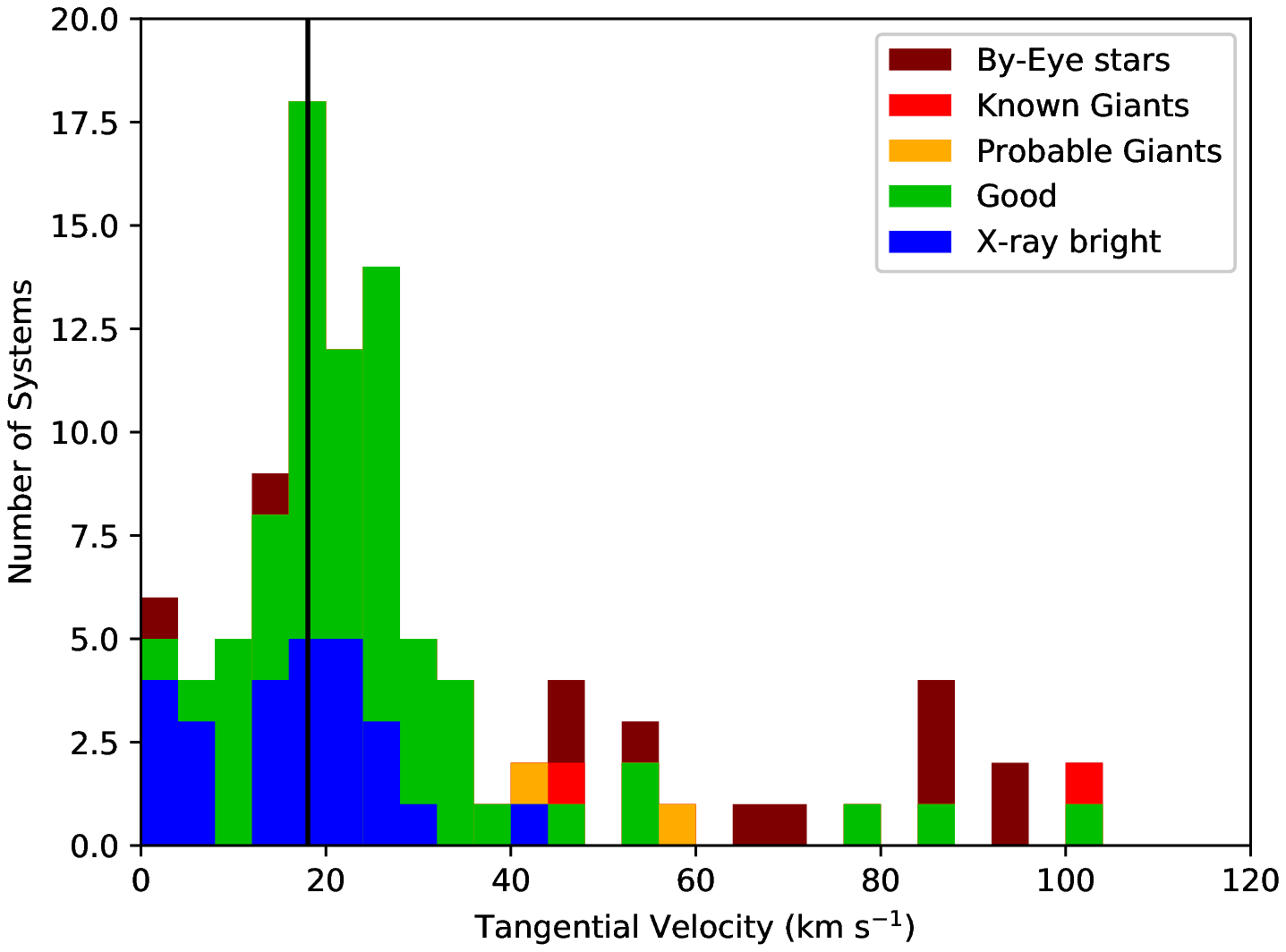}
\caption{The tangential velocity distribution for the stars in the TINYMO sample with parallaxes. Despite having no overall constraint on $V_{tan}$, the distribution peaks at 15-20 km s$^{-1}$, near the Local Standard of Rest (vertical black line).}
\label{fig:tangential_velocities}
\end{figure}

\begin{figure}
\centering
\includegraphics[angle=0,width=0.5\textwidth]{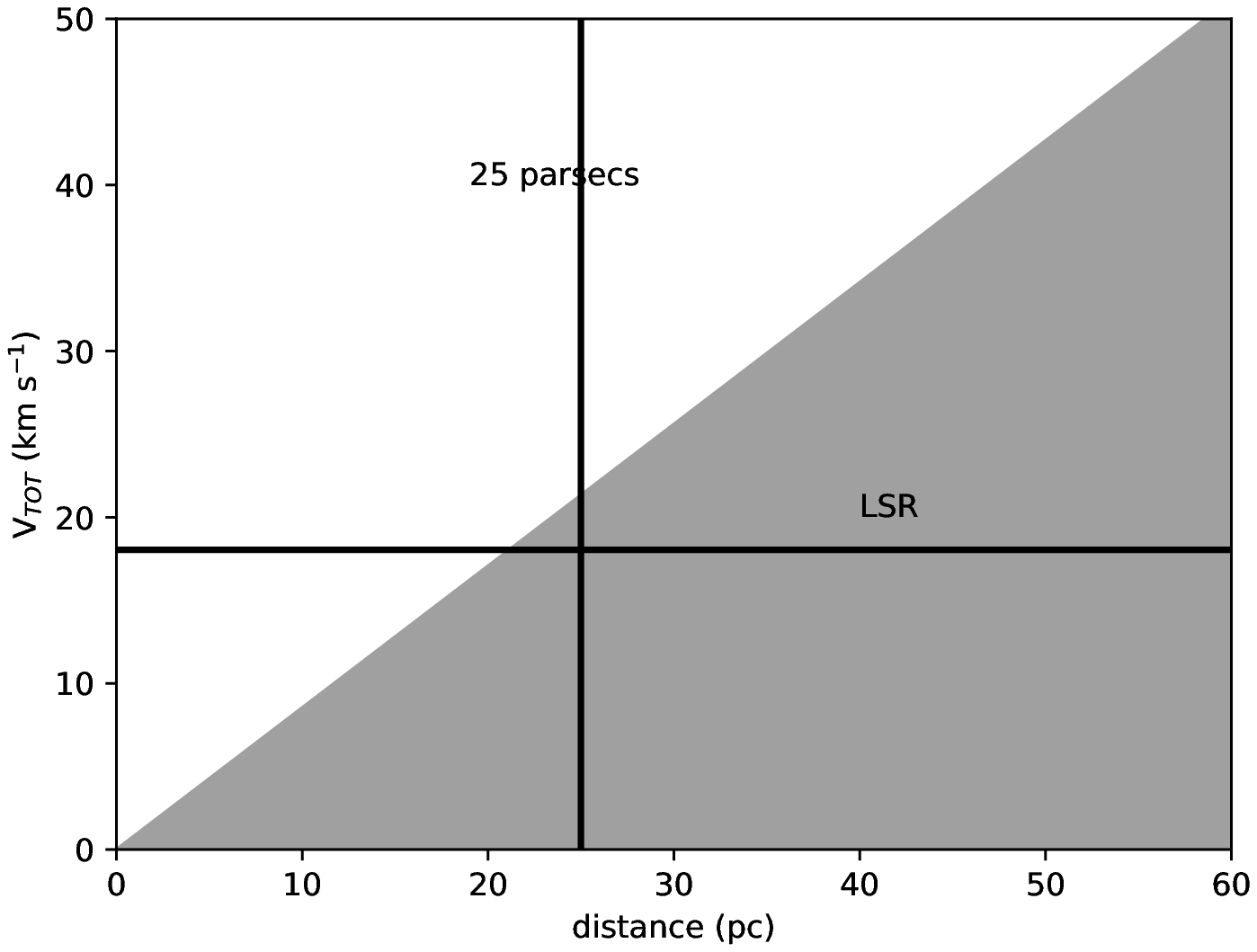}
\caption{Objects whose distance and tangential velocity fall within the white region will move more than 0.18\arcsec~yr$^{-1}$; stars within the gray region will have lower proper motions. Stars moving at the velocity of the Local Standard of Rest (LSR, 18 km s$^{-1}$) move slower than 0.18\arcsec~yr$^{-1}$ if they are more than 21.1 pc away.}
\label{fig:ellipsevtan}
\end{figure}

The TINYMO survey contains a large number of nearby young stars, (55, counting \citealt{Riedel2014}, \citealt{Riedel2017b}, and this paper) where they make up perhaps 4\% of all stars \citep{Riedel2017a}. There are two primary reasons for this. First, the TINYMO search was carried out using photometric distance estimates, which assumed every star was a single main-sequence star. Pre-main-sequence M dwarfs are brighter and therefore appear closer when estimating distances photometrically, and thus preferentially appear in the sample. Second, the space velocities of nearby stars are clustered around the local standard of rest (Figure \ref{fig:tangential_velocities}) because they are still largely following the paths of the gas clouds from which they formed, and the velocity of the LSR falls below 0.18\arcsec yr$^{-1}$ beyond 21 parsecs (Figure \ref{fig:ellipsevtan}).

The velocity peak at 15-20 km s$^{-1}$ is only partially a result of the 0.18\arcsec~yr$^{-1}$ proper motion limit. While it is true that stars moving at 0.18\arcsec~yr$^{-1}$ could have at most 21 km s$^{-1}$ tangential velocities if they were within 25 pc, nearly half of the sample of low proper motion stars were {\it not} within 25 pc, and thus their $V_{\rm{tan}}$ was not constrained to 21 km s$^{-1}$.

\subsection{Close passes to the Solar System}

Without radial velocities, it is difficult to determine which, if any, of our objects have made close passes to the Solar System.  As an educated guess, however, we can take the stars with the lowest $V_{tan}$ velocities as being the most likely to have purely radial motion. The most obvious contender is SCR~0613-2742AB, the $\beta$ Pic member published in \citet{Malo2013} and \citet{Riedel2014}. It does have a published radial velocity, ($+$22.54$\pm$1.16 km s$^{-1}$, \citealt{Riedel2014}), which places its closest approach to the Solar System (using an epicyclic approximation to Galactic motion, \citealt{Riedel2017a}) as 1.2 Myr ago, at a distance of 6.1 pc. SCR~0533-4257 may be a more likely target, but without a radial velocity it is hard to identify.

\citet{Bobylev2010} lists no less than six stars predicted (via a more rigorous Galactic potential analysis) to come closer than SCR~0613-2742AB: GJ~710 (0.21 pc), GJ~551=Proxima Centauri (0.89 pc), GJ~559A=$\alpha$ Centauri A (0.91 pc), GJ~559B=$\alpha$ Centauri B (0.91 pc), GJ~445 (1.06 pc), and GJ~699=Barnard's Star (1.15 pc).  Of those stars, the most remarkable is GJ~710, with proper motion vectors ($\mu_{RA}=$1.15$\pm$1.66 mas yr$^{-1}$, $\mu_{DEC}=$1.99$\pm$1.22 mas yr$^{-1}$), far smaller than any stars in the TINYMO survey.





\section{Results}
\label{sec:results}

\subsection{Nearby Stars}
\label{sec:nearby_stars}

Although the majority of the stars followed up by the TINYMO survey were not within the 15 pc limit for which they were selected, there are 11 new stars within 25 pc in this sample. Most notable among them are SCR~0533-4257AB, a binary almost within 10 pc of the Sun, and HD 271076, which sits in front of the Large Magellanic Cloud and was at one time mistaken for a supergiant member of that satellite galaxy. More details of these two stars as well as other highlighted nearby stars are given in Section \ref{sec:systemnotes}.

\subsection{Spectral types}
Initial classification was done by eye using the techniques from \citet{Henry2002}, \citet{Kirkpatrick1991}, \citet{Boeshaar1976}, and \citet{Keenan1976}, which solely focused on identifying dwarfs and giants by Na~I, Ca~II and K~I line features.  Many giants were identified this way, as well as two carbon stars. Stars confirmed as dwarfs were placed on the CTIOPI astrometric observing program.

Spectral types (given in Table \ref{tab:photometry}) were determined using the MATCHSTAR code \citep{Riedel2014}, a template matching code which operates by comparing input red optical spectra to a series of spectral standard star spectra (\citealt{Kirkpatrick1991}, \citealt{Henry1994}). The code selects the portions of the spectrum held in common between both target and standard star, masks out the atmospheric bands and H$\alpha$ emission line, divides the spectra by the templates, and takes the lowest standard deviation of a match as the correct spectral type. In this way, the code is able to type K0-K9 stars in whole types and M0.0-M9.0 in half types with a 0.5 spectral type uncertainty, as determined from fitting spectra of stars taken on different dates and with different instruments. 

The code also measures, through simple numerical integration, the H$\alpha$ line equivalent width at 6563\AA, the K I doublet line at 7699\AA~(the 7665\AA~line is masked out as part of the atmospheric A band), and the \citet{Lyo2004a} Na{\sc I} 8200\AA~doublet index. Emission is reported as negative equivalent widths.

\subsection{They Might Be Giants}
\label{sec:giants}

Table \ref{tab:giants} contains a list of the new giants (confirmed by spectroscopy) discovered in the TINYMO search, a list distinct from the stars described in Table \ref{tab:photometry} and Table \ref{tab:astrometry}. The spectral types given in the table were assigned by matching to M dwarf spectra, and identified as giants by Na~I index measures of less than 1.02. Accordingly, not much stock should be placed in the actual spectral types of the giants in Table \ref{tab:giants}, as M dwarf types do not correspond directly to giant or supergiant classifications; we also do not provide luminosity classes. The H$\alpha$ emission (denoted by ``e'' in Table \ref{tab:giants}) reported for three stars does appear to be genuine. \citet{Samus2010} mentions ``characteristic late-type emission spectra'' in its description of Mira variables, which implies this is a known phenomenon in at least Mira-type giants.

This sample contains thirteen new large-amplitude photometric variables (denoted by ``var'' in Table \ref{tab:giants}) based on either much larger than typical uncertainties on their CCD photometry ($>0.1$ mag mean uncertainty, which matches that of known Miras observed by CTIOPI), or $>1$ mag discrepancy between their $R$ magnitudes (SuperCOSMOS and our CCD photometry). These may be Mira variables, but we lack sufficient evidence of periodicity or the required 2.5 magnitude amplitude for the formal definition of Miras. Photometry and other details for these stars can be found in Table \ref{tab:tinymo}.

\startlongtable
\begin{deluxetable}{lccl}
\setlength{\tabcolsep}{0.02in}
\tablewidth{0pt}
\tablecaption{New Giants and Supergiants in the TINYMO sample}
\label{tab:giants}
\tablehead{
\colhead{Name} &
\colhead{RA} &
\colhead{DEC} &
\colhead{SpType\tablenotemark{a}} \\
\colhead{(1)} &
\colhead{(2)} &
\colhead{(3)} &
\colhead{(4)}
}
\startdata
\object{HD270965}      & 05 00 40.38 & -71 57 52.9 & K7.0var\tablenotemark{b}\\
\object{SCR0659-5954}  & 06 59 10.94 & -59 54 58.6 & M6.5var,e\tablenotemark{c}\\
\object{SCR0703-3507}  & 07 03 49.64 & -35 07 44.3 & M6.5     \\
\object{SCR0705-3534}  & 07 05 47.36 & -35 34 25.8 & M6.5     \\
\object{SCR0711-3600}  & 07 11 03.53 & -36 00 59.7 & CARBON   \\
\object{SCR0747-5412}  & 07 47 14.27 & -54 12 02.5 & CARBON   \\
\object{SCR0747-6355}  & 07 47 25.60 & -63 55 42.3 & K7.0     \\
\object{SCR0749-6502}  & 07 49 05.69 & -65 02 40.0 & K8.0     \\
\object{SCR0753-5150}  & 07 53 24.57 & -51 50 22.0 & M9.0     \\
\object{SCR0753-6641}  & 07 53 49.77 & -66 41 38.3 & K9.0     \\
\object{SCR0805-0743}  & 08 05 52.81 & -07 43 05.7 & M9.0     \\
\object{STEPH0097}     & 08 14 24.82 & -13 02 22.6 & M6.5     \\
\object{SCR0833-6107}  & 08 33 27.67 & -61 07 58.4 & M4.5var\tablenotemark{b}\\
\object{SCR0857-6734}  & 08 57 38.21 & -67 34 10.5 & M5.0     \\
\object{IRA08583-2531} & 09 00 32.06 & -25 43 14.1 & M8.0     \\
\object{SCR0902-7823}  & 09 02 35.97 & -78 23 14.7 & M7.5     \\
\object{SCR0910-7214}  & 09 10 57.71 & -72 14 52.9 & M5.0     \\
\object{SCR0927-8105}  & 09 27 04.18 & -81 05 00.7 & M4.5var,e\tablenotemark{c}\\
\object{SCR0932-2806}  & 09 32 03.32 & -28 06 27.0 & M9.0     \\ 
\object{SCR0938-3748}  & 09 38 20.24 & -37 48 44.6 & M6.5     \\
\object{SCR0945-3430}  & 09 45 43.54 & -34 30 18.1 & M4.5var\tablenotemark{c}\\
\object{SCR1044-7543}  & 10 44 06.77 & -75 43 42.2 & M2.5     \\
\object{SCR1044-4330}  & 10 44 40.73 & -43 30 44.2 & M6.5     \\
\object{SCR1048-7739}  & 10 48 26.67 & -77 39 19.1 & M0.0     \\
\object{SCR1058-4218}  & 10 58 44.39 & -42 18 12.3 & M6.5     \\
\object{SCR1111-4856}  & 11 11 28.25 & -48 56 14.3 & M2.5     \\
\object{SCR1138-4338}  & 11 38 13.34 & -43 38 04.6 & M9.0     \\
\object{SCR1228-4949}  & 12 28 06.16 & -49 49 34.5 & M5.0     \\
\object{STEPH0172}     & 12 34 41.61 & -00 14 14.1 & M9.0     \\
\object{SCR1306-4745}  & 13 06 42.81 & -47 45 25.7 & M6.5var\tablenotemark{c}\\
\object{SCR1316-5206}  & 13 16 42.18 & -52 06 38.3 & M6.5     \\
\object{SCR1317-4643}  & 13 17 56.50 & -46 43 54.0 & M9.0var\tablenotemark{d}\\
\object{SCR1321-4913}  & 13 21 31.72 & -49 13 09.6 & M7.5     \\
\object{SCR1349-7417}  & 13 49 16.98 & -74 17 15.4 & M9.0     \\
\object{SCR1358-4910}  & 13 58 43.58 & -49 10 52.0 & M7.0     \\
\object{SCR1408-3506}  & 14 08 36.51 & -35 06 02.3 & M6.5     \\
\object{SCR1424-4427}  & 14 24 36.78 & -44 27 05.6 & M7.5     \\
\object{SCR1427-4731}  & 14 27 43.90 & -47 31 13.2 & M4.0     \\
\object{SCR1431-4823}  & 14 31 28.46 & -48 23 12.1 & M7.0     \\
\object{SCR1439-4506}  & 14 39 33.26 & -45 06 42.3 & M4.5     \\   
\object{SCR1440-7837}  & 14 40 37.43 & -78 37 11.4 & K8.0     \\
\object{SCR1458-4102}  & 14 58 23.80 & -41 02 27.9 & M7.0var\tablenotemark{e} \\
\object{CD-81-00572}   & 15 32 44.68 & -81 43 53.0 & K8.0     \\
\object{SCR1534-7237}  & 15 34 02.51 & -72 37 11.1 & M6.5     \\
\object{SCR1544-1805}  & 15 44 44.97 & -18 05 07.1 & M9.0     \\
\object{SCR1551-8047}  & 15 51 10.25 & -80 47 51.5 & M3.0     \\   
\object{STEPH0257}     & 15 58 20.04 & -06 03 37.4 & M7.0     \\
\object{SCR1604-7009}  & 16 04 23.14 & -70 09 03.1 & M5.0     \\
\object{SCR1612-6858}  & 16 12 30.09 & -68 58 52.7 & M6.5     \\
\object{SCR1621-6843}  & 16 21 18.53 & -68 43 58.4 & M6.5     \\
\object{SCR1647-6436}  & 16 47 48.35 & -64 36 43.6 & M5.0     \\
\object{SCR1654-0055}  & 16 54 08.17 & -00 55 04.9 & M9.0     \\     
\object{SCR1658-6350}  & 16 58 12.94 & -63 50 49.3 & M7.5     \\
\object{SCR1706-6426}  & 17 06 39.02 & -64 26 23.3 & M7.5var\tablenotemark{c} \\
\object{SCR1719-6151}  & 17 19 09.42 & -61 51 55.7 & M9.0     \\
\object{SCR1738-6844}  & 17 38 14.51 & -68 44 52.8 & M5.0var\tablenotemark{c} \\
\object{SCR1743-4959}  & 17 43 35.28 & -49 59 10.6 & M9.0     \\
\object{SCR1803-7807}  & 18 03 30.88 & -78 07 21.7 & M4.5     \\
\object{SCR1807-5839}  & 18 07 22.90 & -58 39 59.9 & M4.5     \\   
\object{SCR1919-2943}  & 19 19 23.11 & -29 43 15.0 & M9.0     \\
\object{CD-35-13495}   & 19 27 08.18 & -35 15 09.6 & M7.5     \\
\object{SCR1943-0138}  & 19 43 43.06 & -01 38 31.6 & M6.5     \\
\object{SCR1944-3414}  & 19 44 45.52 & -34 14 41.2 & M9.0     \\
\object{CD-45-13476}   & 19 53 08.97 & -45 15 15.5 & M7.5     \\
\object{SCR1959-1639}  & 19 59 35.79 & -16 39 20.3 & M8.0     \\
\object{SCR2000-0837}  & 20 00 58.33 & -08 37 27.5 & M9.0e    \\
\object{SCR2024-2500}  & 20 24 15.40 & -25 00 56.8 & M6.5     \\
\object{SCR2038-0409}  & 20 38 45.49 & -04 09 27.0 & M5.0     \\
\object{SCR2107-5734}  & 21 07 58.01 & -57 34 17.5 & M7.0var\tablenotemark{d} \\
\object{SCR2138-4308}  & 21 38 15.11 & -43 08 40.6 & M6.5var\tablenotemark{f} \\
\object{CD-24-17228}   & 22 34 29.69 & -24 15 17.7 & M6.5     \\
\object{SCR2305-3054}  & 23 05 14.88 & -30 54 37.1 & M5.0var\tablenotemark{g} \\
\enddata
\tablenotetext{a}{Spectral Types derived from comparisons to dwarfs, and may not be accurate.}
\tablenotemark{c}{Variable status inferred from $>$1 magnitude R2 and $R_{kc}$ magnitude mismatch.}
\tablenotemark{c}{Variable status inferred from $>$1 magnitude R1 and R2 plate magnitude mismatch.}
\tablenotemark{d}{Variable status inferred from $>$1 magnitude R1 and $R_{kc}$ magnitude mismatch.}
\tablenotemark{e}{$R_{kc}$ filter variability 0.18 mag.}
\tablenotemark{f}{$R_{kc}$ filter variability 0.31 mag.}
\tablenotemark{g}{$R_{kc}$ filter variability 0.54 mag.}

\end{deluxetable}

\subsection{Carbon Stars}
\label{sec:carbon}

\begin{figure}
\centering
\includegraphics[angle=90,width=.4\textwidth]{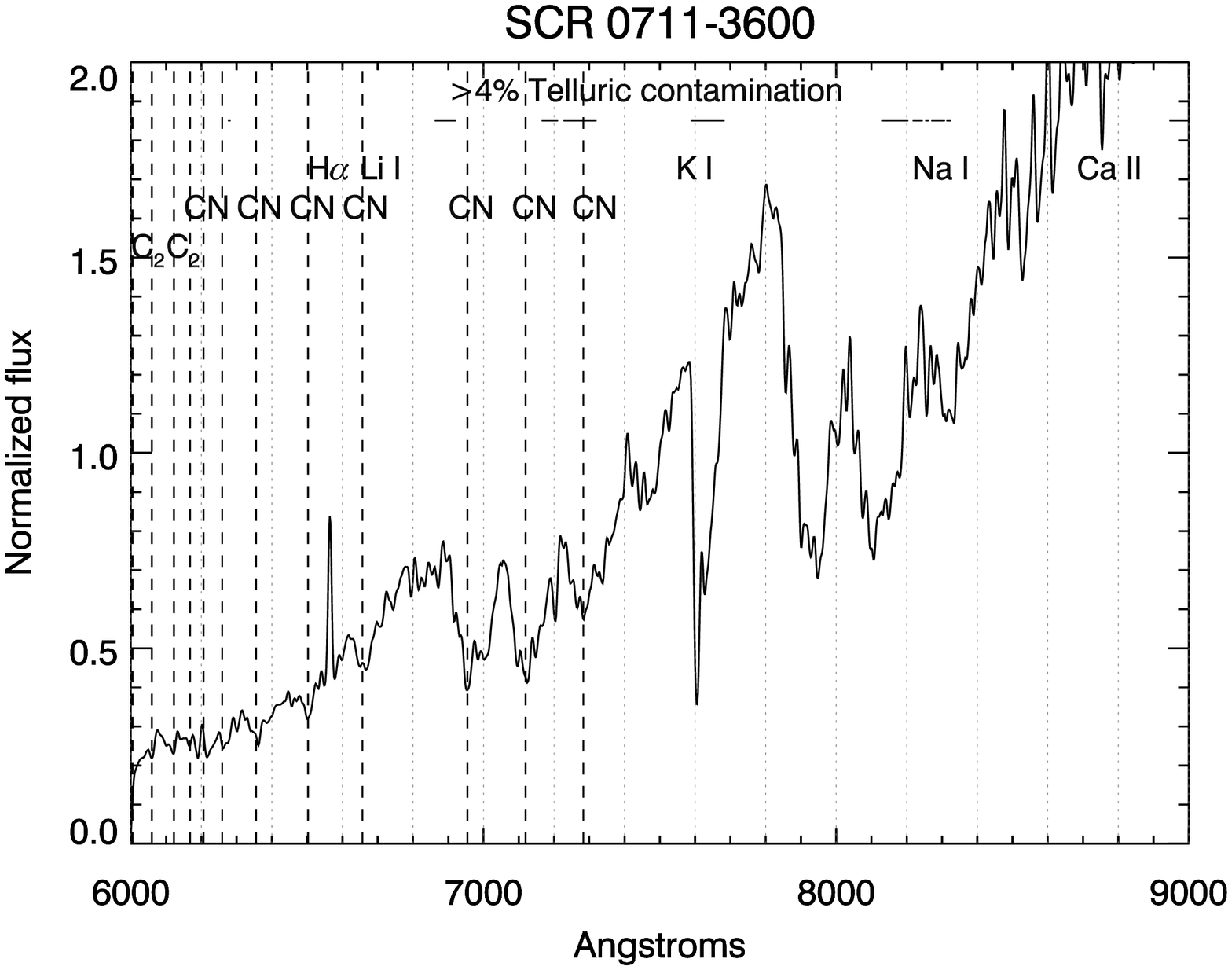}
\includegraphics[angle=90,width=.4\textwidth]{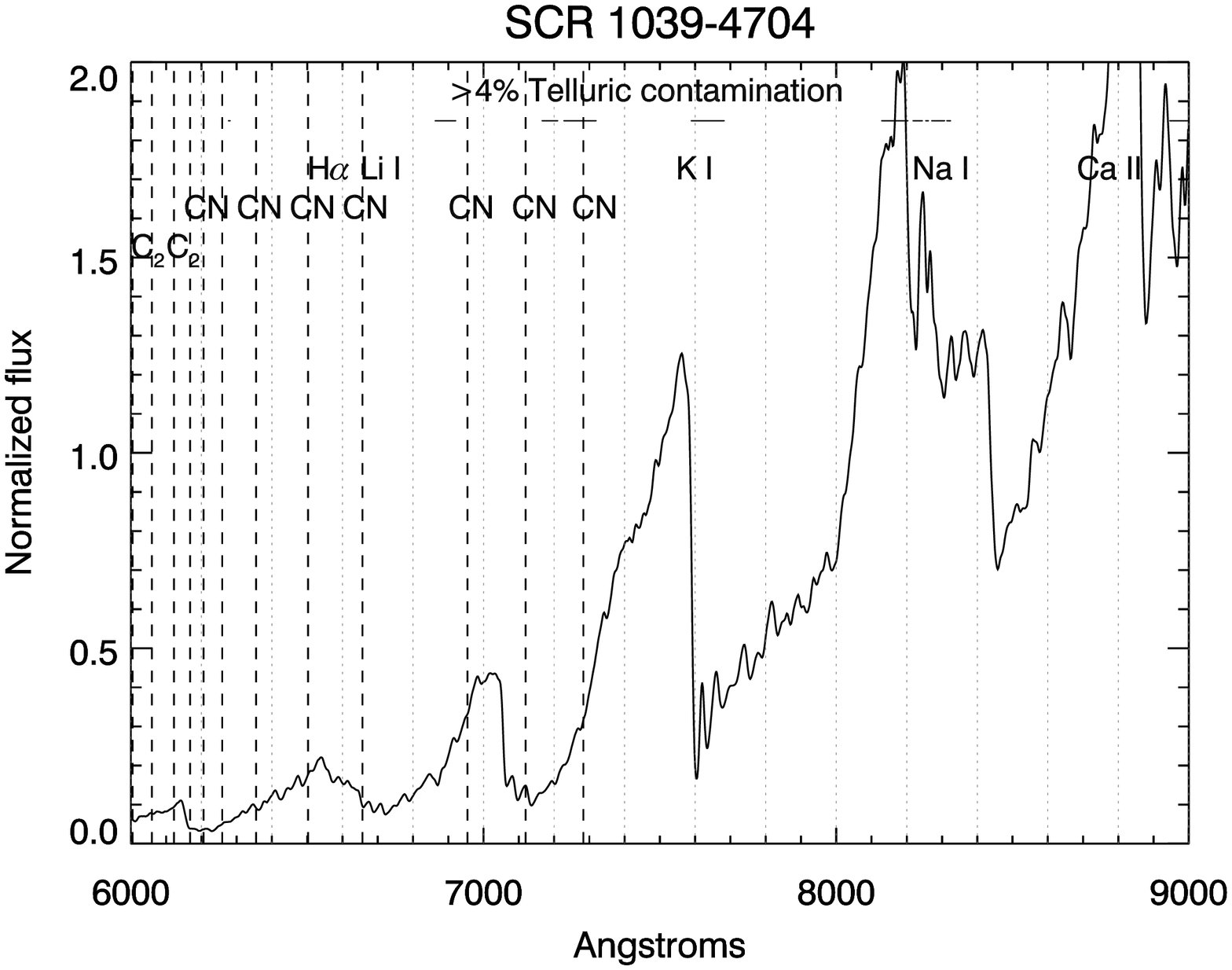}
\includegraphics[angle=90,width=.4\textwidth]{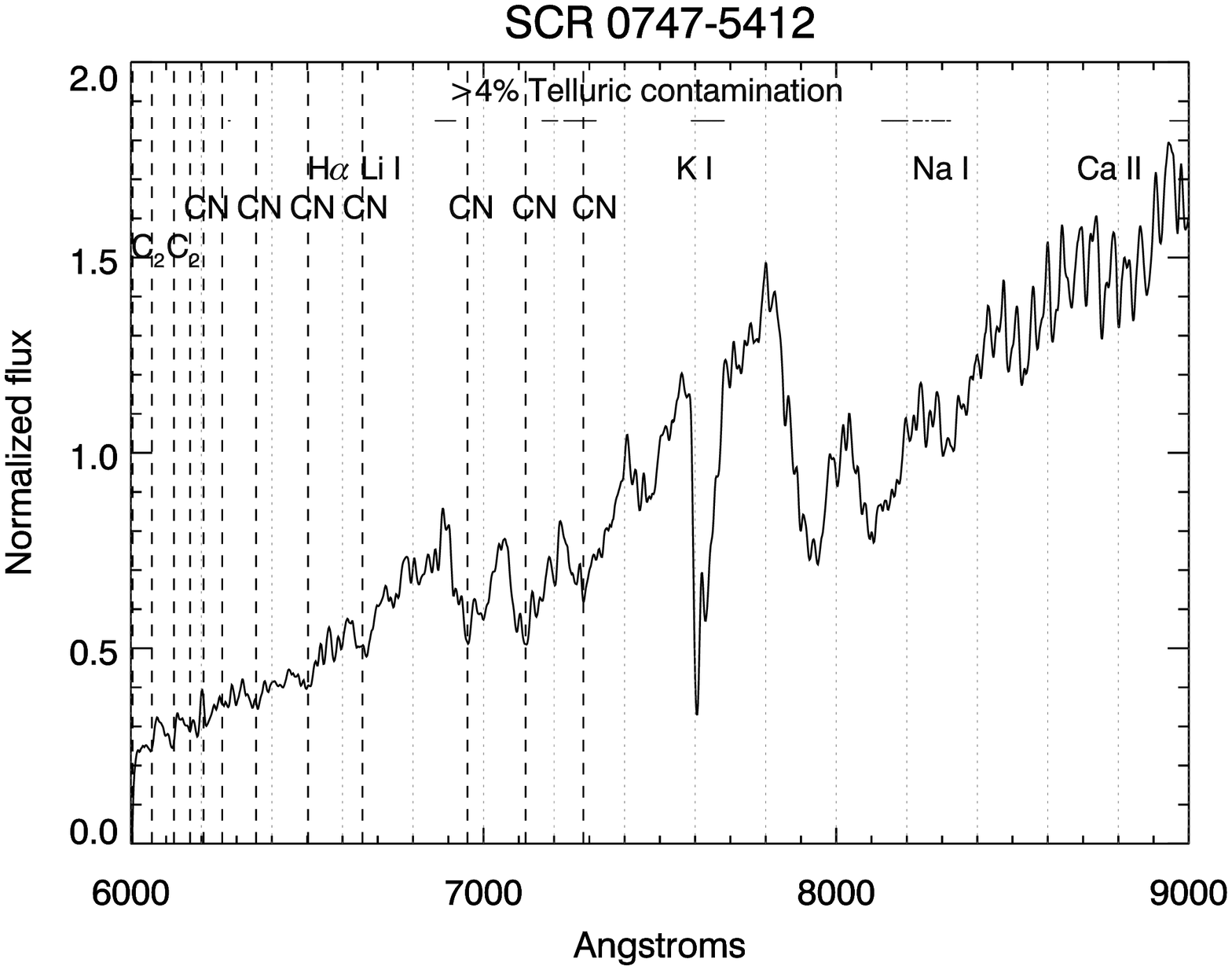}
\caption{CTIO 1.5m spectra of the two new carbon stars, SCR~0711-3600 (top), and SCR~0747-5412 (bottom), both from 2009 SEP 16, along with the approximately M7.5III giant SCR~1039-4704 (middle, from 2010 Dec 20) for comparison. The spectrum of a carbon star is unlike an M dwarf or M giant, and contains unusual concentrations of carbon molecules (here, CN bands known as Swan bands) rather than the typical TiO or VO bands of an M dwarf (compare, for instance, the spectral morphology at 7100\AA).}
\label{fig:carbon}
\end{figure}

Three carbon stars were observed during data collection. One, IY~Hya, was observed as a comparison object; the other two are new discoveries. Figure \ref{fig:carbon} shows the spectra of the new stars and a normal M giant for comparison. Based on comparisons with spectra in \citet{Turnshek1985}, they appear to be genuine C-type stars with CN bands at 6900, 7100, 7500, 7900, and 8100\AA.

\subsection{Reddened Stars}
\label{sec:reddened}
\begin{figure}
\centering
\includegraphics[angle=0,width=.5\textwidth]{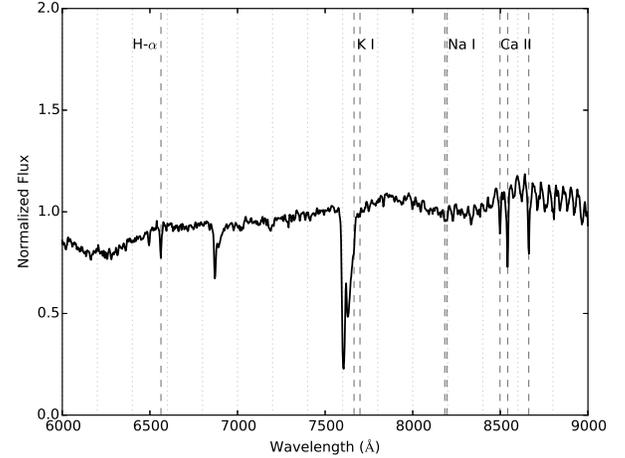}
\caption{CTIO 1.5m spectrum of the reddened star BD-19-04371 on 17 Sep 2010. The star is an apparent member of the Upper Scorpius star forming region. Its colors are of an M dwarf, but it is missing the strong TiO bands of M dwarfs.}
\label{fig:BD-19-04371}
\end{figure}

Several reddened stars were picked up in TINYMO; these mostly appear to be members of various subsets of the Sco-Cen star forming region. BD-19~04371 (16:26:23.37 -19:31:35.7), SCR~1627-1925 (16:27:14.03 -19:25:46.7), and SCR~1627-1924 (16:27:14.79 -19:24:16.3) are all in the region of the sky with the Upper Scorpius star forming region, and all appear to be reddened stars of hotter spectral types (Figure \ref{fig:BD-19-04371}).

A few other stars were pre-identified in SIMBAD as members of the Cham\ae leontis I dark cloud (CHXR11, 11:03:11.61 -77:21:04.2), or $\epsilon$ Cham\ae leontis association. The only truly unusual set of reddened potential nearby stars were a quartet of reddened objects:
\begin{itemize}
\item CD-58~07828 20:39:19.56  -58:02:29.4   $\mu=$0.036 @ 099.3$^{\circ}" yr^{-1}$
\item CD-61~06505 20:54:02.76  -61:28:25.4   $\mu=$0.002 @ 010.5$^{\circ}" yr^{-1}$
\item SCR~2055-6001 20:55:43.94  -60:01:46.1 $\mu=$0.018 @ 010.4$^{\circ}" yr^{-1}$
\item SCR~2116-5825 21:16:44.72  -58:25:25.2 $\mu=$0.014 @ 218.1$^{\circ}" yr^{-1}$ 
\end{itemize}
There is no known cloud in this location (as per WEBDA), which is at a high Galactic latitude. It may be that these stars are truly unrelated (their proper motion vectors from SuperCOSMOS appear different, but statistically consistent with $\mu=0$) and all just happen to be reddened, but they are the only concentration of reddened objects that cannot be immediately explained.
   
\subsection{New Young Stars}

\clearpage
\startlongtable
\begin{deluxetable*}{llccccclc}
\tabletypesize{\small}
\setlength{\tabcolsep}{0.02in}
\tablecaption{Youth Criteria of Parallax targets}
\label{tab:spectroscopic}
\tablehead{
\colhead{Name} &
\colhead{LACEwING} &
\colhead{} &
\colhead{Kinematic} &
\colhead{H-$\alpha$} &
\colhead{Na Index} &
\colhead{KI EW} &
\colhead{Youth} &
\colhead{Note} \\
\colhead{} &
\colhead{Group} &
\colhead{Prob. (\%)} &
\colhead{RV (km s$^{-1}$)} &
\colhead{\AA} &
\colhead{idx.} &
\colhead{\AA} &
\colhead{Flags\tablenotemark{a}} &
\colhead{} \\
\colhead{(1)} &
\colhead{(2)} &
\colhead{(3)} &
\colhead{(4)} &
\colhead{(5)} &
\colhead{(6)} &
\colhead{(7)} &
\colhead{(8)} &
\colhead{(9)} }
\startdata
NLTT 1261       & (None)  &    &                 &         &      &      & \ldots & \\
GIC 50          & (None)  &    &                 & $-$1.83 & 1.22 & 2.05 &        & * \\
2MA 0112+1703   & AB Dor  & 73 &  $-$1.4$\pm$1.9 &         &      &      & \ldots & \\
2MA 0123-6921   & Tuc-Hor & 85 &  $+$9.9$\pm$3.4 &         &      &      & \ldots & \\
SCR 0128-1458   & (None)  &    &                 & $-$3.49 & 1.22 & 2.23 &        & * \\
BAR 161-012     & (None)  &    &                 &$-$10.51 & 1.15 & 1.43 & h N K  & * \\
SCR 0143-0602   & (None)  &    &                 & $-$5.32 & 1.20 & 1.76 &        & * \\
SIPS 0152-6329  & Tuc-Hor & 80 & $+$10.4$\pm$3.4 & $-$9.72 & 1.19 & 2.52 & ~~~N   & * \\ 
SCR 0222-6022   & Tuc-Hor & 88 & $+$11.5$\pm$3.3 &$-$11.56 & 1.18 & 1.07 & h N K  & * \\
2MA 0236-5203   & Tuc-Hor & 86 & $+$11.7$\pm$3.1 & $-$5.53 & 1.09 & 0.54 & ~~~N   & \\
2MA 0254-5108A  & Tuc-Hor & 33 & $+$12.7$\pm$3.1 & $-$2.08 & 1.09 & 0.66 &        & \\
2MA 0254-5108B  & Tuc-Hor & 63 & $+$12.6$\pm$3.1 &         &      &      & \ldots & \\
SCR 0336-2619   & Tuc-Hor & 67 & $+$13.5$\pm$2.6 &$-$10.19 & 1.25 & 2.47 & h N K  & * \\
RX 0413-0139    & (None)  &    &                 &$-$10.54 & 1.18 & 1.11 & h N K  & \\
2MA 0446-1116AB & (None)  &    &                 &         &      &      & \ldots & \\
HD 271076       & (None)  &    &                 & $+$0.20 & 1.11 & 0.82 &        & * \\
SCR 0533-4257AB & (None)  &    &                 & $-$4.63 & 1.20 & 1.95 &        & * \\
LP 780-032      & Argus   & 38 & $+$23.8$\pm$1.8 & $-$0.26 & 1.21 & 1.69 &        & \\
2MA 0936-2610AC & (None)  &    &                 & $-$2.36 & 1.26 & 2.45 &        & * \\
SIPS 1110-3731AC& TW Hya  & 72 & $+$12.7$\pm$2.2 & $-$9.21 & 1.10 & 0.51 & ~~~N K & * \\
SIPS 1110-3731B & TW Hya  & 62 & $+$12.7$\pm$2.2 & $-$9.21 & 1.10 & 0.51 & ~~~N K & * \\
STEPH 164       & (None)  &    &                 & $-$4.25 & 1.15 & 1.38 &        & * \\
GJ 2122AB       & (None)  &    &                 & $+$0.28 & 1.08 & 0.82 &        & \\
UPM 1710-5300AB & (None)  &    &                 &         &      &      & \ldots & \\
SIPS 1809-7613  &$\beta$ Pic&31&  $+$6.4$\pm$2.6 & $-$8.31 & 1.17 & 1.98 & ~~~N K & * \\
SCR 1816-5844   & Argus   & 69 & $-$13.0$\pm$1.9 & $-$6.50 & 1.15 & 1.09 & ~~~N   & * \\
DEN 1956-3207B  & (None)  &    &                 &         &      &      & \ldots & \\
DEN 1956-3207A  & (None)  &    &                 &         &      &      & \ldots & \\
BD-13 6424      &$\beta$ Pic&30&  $+$0.9$\pm$1.6 &         &      &      & \ldots & \\
\enddata

\tablecomments{Youth properties are of stars in Table \ref{tab:photometry} and Table \ref{tab:astrometry}.}
\tablenotetext{a}{Youth flags are: ``h'': H-$\alpha$ stronger than $-$10\AA; ``N'': Low surface gravity by sodium index; ``K'': Low surface gravity by potassium EW.}
\end{deluxetable*}

\begin{figure}
\centering
\includegraphics[angle=0,width=.5\textwidth]{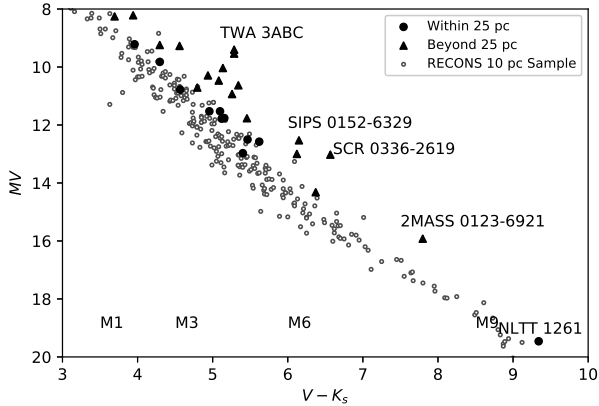}
\caption{Color-Magnitude Diagram for the stars with follow-up astrometry and photometry from this paper, plotted against the RECONS 10 pc sample. A substantial number of stars lie beyond 25 pc of the Sun, and are more than a magnitude overluminous compared to the mostly main-sequence stars in the RECONS 10 pc sample.}
\label{fig:CMD}
\end{figure}

A substantial number of targets found in the TINYMO survey were found to be young (Figure \ref{fig:CMD}). In red spectra (6000\AA~-- 9000\AA) there are three useful spectroscopic features that distinguish dwarfs from giants. Ca~II is strong in giants and weak in dwarfs; Na~I and K~I are weak in giants and strong in dwarfs; the general principle, as outlined in \citet{Allers2007}, is that neutral alkali species are stronger in dwarfs, while singly-ionized species are stronger in giants. The Ca II triplet is almost completely absent in mid-M dwarfs, but prominent in M giants, which makes it an easy diagnostic to use in luminosity classifying.

\begin{figure}
\centering
\includegraphics[angle=0,width=.5\textwidth]{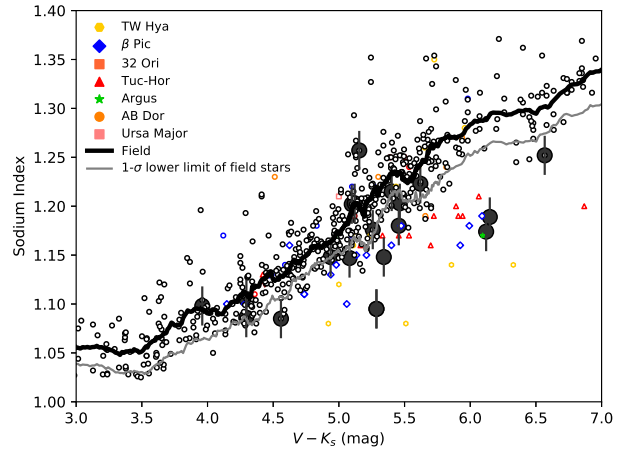}
\caption{The NaI Index from \citet{Lyo2004a} versus $V-K_s$. TINYMO objects measured spectroscopically are shown as large black points with error bars; the smaller points are young stars from \citet{Riedel2017a} and other RECONS spectral holdings measured with MATCHSTAR. A 25-point moving average (thick black line) and standard deviation (gray line) show the boundaries of inactive stars, and demonstrate that many of our targets have lower surface gravities than typical field stars, though many are consistent with or even higher than the field locus.}
\label{fig:NaI}
\end{figure}

The Na~I index is particularly useful for determining the relative surface gravities of mid and cool M dwarfs \citep{Schlieder2012b}. For our purposes, we use the \citet{Lyo2004a} index constructed from a 24\AA~wide region redward of the Na~I 8200\AA~doublet divided by a 24\AA~wide region containing the Na~I 8200\AA~doublet, as used in \citet{Murphy2010},\citet{Riedel2011}, \citet{Murphy2013}, \citet{Rodriguez2013}, and \citet{Riedel2014}. Empirically, we have found that an index of 1.02 or less indicates a giant (see Figure \ref{fig:NaI}), and intermediate index values between dwarfs (which increase to lower temperatures) and giants (which remain flat at 1.02) indicate a low-surface-gravity pre-main-sequence star. The results for this sample of stars are shown in Figure \ref{fig:NaI}.

Unfortunately, giants and dwarfs overlap at colors bluer than $V-K_s=5$. Alkali metal lines such as Na~I can also be affected by stellar activity, where emission fills in the absorption line cores, leading to lower EWs \citep{Reid1999}. \citet{Slesnick2006b} notes that the Na~I doublet can be affected by telluric absorption over the region 8161--8282~\AA, leading to artificially low Na~I index values for stars observed at large airmasses. Our results have large systematic errors because of this uncorrected telluric absorption. 

\begin{figure}
\centering
\includegraphics[angle=0,width=.5\textwidth]{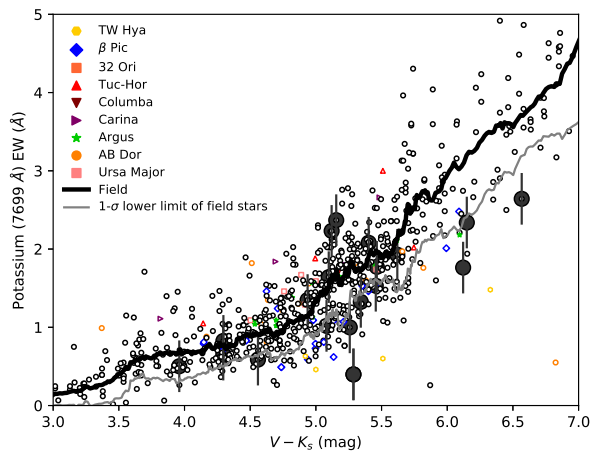}
\caption{The same as Figure \ref{fig:NaI}, except showing the K~I equivalent width.}
\label{fig:KI}
\end{figure}

We also use the K~I 7699\AA~doublet line (though not its companion at 7665\AA, because that portion of the spectrum is contaminated by the atmospheric A band) equivalent width as an independent indicator of surface gravity. As with the Na~I index, the K~I values for giants and dwarfs overlap at colors bluer than $V-K_s=5$. Those results are shown in Figure \ref{fig:KI}.
 
\begin{figure}
\centering
\includegraphics[angle=0,width=.5\textwidth]{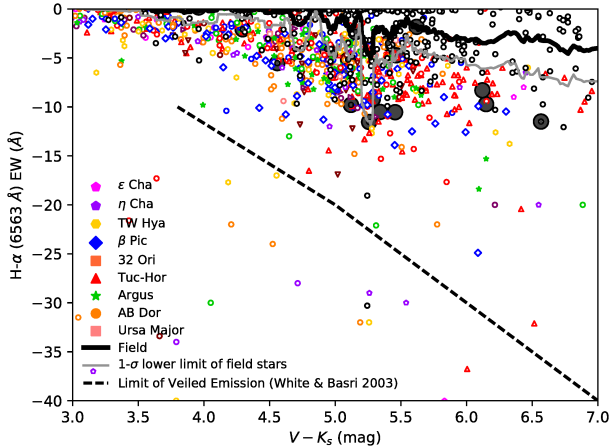}
\caption{The same as Figure \ref{fig:NaI}, except showing the H$\alpha$ EW, and the \citet{White2003} limit (dotted black line) below which veiling from accretion is a probable concern.}
\label{fig:H-alpha}
\end{figure}

Many of the stars in this sample have H$\alpha$ in emission (see Figure \ref{fig:H-alpha}). As noted by \citet{West2008} and \citet{Zuckerman2004}, H$\alpha$ activity persists in M dwarfs for long periods of time, which means H$\alpha$ itself is not a suitable source of youth. Strong H$\alpha$ emission has been linked to accretion and T Tauri status, but none of these stars comes close to the \citet{White2003} limit.

The other available means for judging the youth of the stars studied here is kinematics, through which stars can be matched to nearby young moving groups (NYMGs) like $\beta$ Pictoris \citep{Song2002} and Tucana-Horologium \citep{Zuckerman2001}. CTIOPI astrometry provides accurate values of five of the six kinematic elements (RA, DEC,$\mu_{RA}$, $\mu_{DEC}$, and Parallax; missing only RV) necessary to fully describe a star's position and motion. The LACEwING code \citep{Riedel2017a} can accommodate partial information and calculate the probabilities of membership in 13 nearby young moving groups, and three nearby open clusters.

LACEwING has two modes of operation: Field and Young star mode. In Field star mode, the probabilities are calculated allowing for the possibility that the star is a field star with coincidentally similar space velocities to a young moving group (where field stars outnumber moving group members 50:1). In Young star mode, the probabilities are calculated assuming that the star is known to be spectroscopically or photometrically young, and young field stars are evenly matched with young moving group members, 1:1. We consider LACEwING membership probabilities of 20-50\% to be low, 50-75\% to be medium, and 75\%-100\% to be high probability memberships.

For objects with low surface gravity (below the gray lines in either Figure \ref{fig:NaI} or Figure \ref{fig:KI}), we have used LACEwING's young star mode. For all other objects, we have used field star mode. We are accordingly biased against identifying members of AB Doradus (125 Myr) and older groups, where M dwarf surface gravities are indistinguishable from field stars. The results of this study, and spectroscopic measurements, are given in Table \ref{tab:spectroscopic}. 

\section{System Notes}
\label{sec:systemnotes}

\begin{deluxetable*}{llrccccccrrrcccl}
\setlength{\tabcolsep}{0.02in}
\tablewidth{0pt}
\tablecaption{Multiple Star Results}
\label{tab:multiples}
\tablehead{\colhead{Name}            &
	\colhead{Binary}             &
	\colhead{Type}		     &
	\colhead{Separation}	     &
	\colhead{Position Angle}     &
	\colhead{$\Delta$ Mag.}	     &
	\colhead{Filter}  	     &
	\colhead{Ref.}		     \\
	\colhead{}		&
	\colhead{}		&
	\colhead{}		&
	\colhead{(arcsec)}	&
	\colhead{(deg)}		&
	\colhead{(mag)}		&
	\colhead{}		&
	\colhead{}		\\
    \colhead{(1)}           &
    \colhead{(2)}           &
    \colhead{(3)}           &
    \colhead{(4)}           &
    \colhead{(5)}           &
    \colhead{(6)}           &
    \colhead{(7)}           &
    \colhead{(8)}
}
\startdata
GIC 50 & AB & VB & 0.51 & 184 &  & $z'$ & \citet{Janson2014} \\
GIC 50 & AB & VB & 0.21 & 17 &  & $z'$ & \citet{Janson2014} \\
2MASS 0123-6921 &  &  &  &  &  &  & \\
2MASS 0254-5103 & AB & VB & 15.3 & 80.2 & 5.48 & $V$ & \\
2MASS 0446-1116 & AB & VB & $\sim$1.0 & $\sim$285 & $\sim$0.9 & $V$ & \\
SCR 0533-4257 & AB & IB & 0.056 & ? & 0.7 & $F583W$ & \\
2MASS 0936-2610 & AB & VB & 41 & 314 & 3.25 & $K$ & \\
2MASS 0936-2610 & AC & VB & 0.39 & 284 & 0.5 & ? & (B. Mason, Priv. Comm.) \\
SIPS 1110-3731 & AB & VB & $\sim$1.16 & $\sim$209 & $\sim$0.38 & $V$ & \\
SIPS 1110-3731 & AC & SB &  &  &  &  & \citet{Webb1999} \\
Stephenson 164 &  &  &  &  &  &  & \\
GJ 2122 & AB & VB & 0.59 & 255 & 2 & $V$ & \citet{Heintz1987} \\
UPM 1710-5300 & AB & VB & $\sim$0.77 & $\sim$343 & $\sim$0.69 & $V$ & \\
DENIS 1956-3207 & AB & VB & 26.37 & 43.9 & 1.71 & $V$ & \\
\hline
\enddata
\tablecomments{Measurements are this work unless otherwise noted. AB= Astrometric Binary, IB= Interferometric Binary, SB = Spectroscopic Binary, VB = Visual Binary. Approximate measurements were determined by eye.} 

\end{deluxetable*}

Here we describe each of the 26 systems for which parallaxes are
published in this paper in Table \ref{tab:astrometry}. See also Table
\ref{tab:multiples} for details on the various multiple systems.

\paragraph{(0024-0158) NLTT~1261} BRI~0021-0214 ($M_V=19.46, V-K_s=9.34$) has
$V$ = 19.88, making it the faintest star in the optical $VRI$
bandpasses in this survey. Our parallax (82.4$\pm$2.2 mas) is consistent
with that of \citet{Tinney1995} (86.6$\pm$4.0 mas), and represents a
factor of two improvement in the uncertainty.

\begin{figure}
\centering
\includegraphics[angle=0,width=0.5\textwidth]{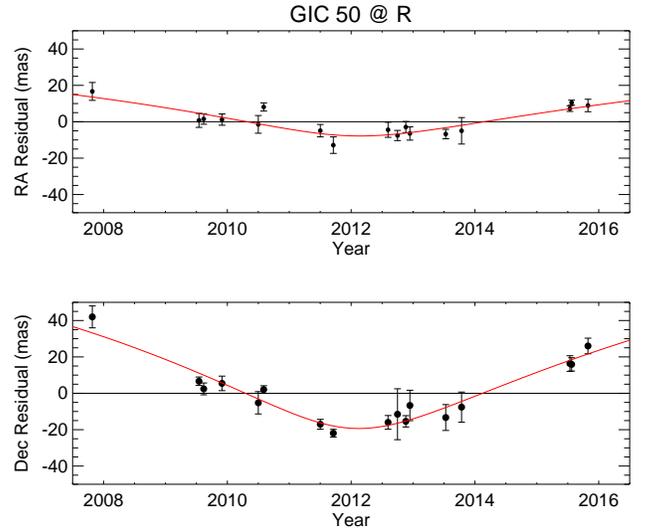}
\caption{The perturbations in the astrometric residuals of 
GIC~50 (curve) due to its otherwise-unseen companion have been removed 
from the astrometry before the final parallax was fit.}
\label{fig:GIC0050_PB}
\end{figure}

\paragraph{(0032-0434) GIC~50} ($M_V=12.75, V-K_s=5.62$) exhibits a
perturbation due to an unseen companion spanning the full 8 years of
our data, as shown in Figure \ref{fig:GIC0050_PB}.  A fit has been made 
to the data and the perturbation removed to derive the astrometry 
results given in Table \ref{tab:astrometry}. The system was resolved with AstraLux by \citet{Janson2014}, who found it to be a triple with companions at 0.508\arcsec and 0.213\arcsec. 

\paragraph{(0112+1703) 2MA~0112+1703} ($M_V=10.71, V-K_s=4.79$) was 
identified by \citet{Malo2013} as a potential member of AB Dor, but the 
identification was less certain because the star had no measured parallax 
or radial velocity. While \citet{Malo2014a} furnished a radial velocity, 
this is the first parallax. With all available information, the system is 
still a high probability member of AB Dor.
We do not have a spectrum of 2MA~0112+1703, but AB Dor members are too 
old to distinguish from field stars by Na~I or K~I surface gravity, so a spectrum
would not be expected to show any of the signs of youth we are looking for.

\paragraph{(0123-6921) 2MA~0123-6921} ($M_V=15.92, V-K_s=7.80$) is a 
color-magnitude diagram match (we have no spectrum to measure its 
H$\alpha$ or gravity features) for the TW Hydra association, but like 
SCR~0103-5515 and SCR~0336-2610 in \citet{Riedel2014}, this system is 
on the wrong side of the sky from all known members. It is kinematically 
consistent with Tuc-Hor and to a lesser extent AB Dor, but would have to 
be a higher-order multiple to align with other members of those groups on 
the HR diagram. We conclude that the system must be young and assign it 
to Tuc-Hor, and note that it is likely to be a triple or quadruple in that
case, but find no evidence in our astrometric data.

\begin{figure}
\centering
\includegraphics[angle=0,width=0.5\textwidth]{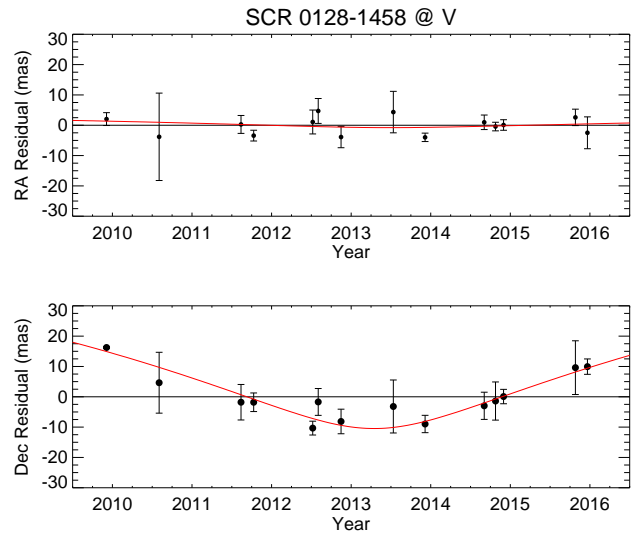}
\caption{Same as Figure \ref{fig:GIC0050_PB}, but for the star SCR0128-1458. There is no visible perturbation on the RA axis.
\label{fig:SCR0128-1458_PB}}
\end{figure}

\paragraph{(0128-1458) SCR~0128-1458} ($M_V=12.97, V-K_s=5.40$) shows a
possible perturbation in the 6 years of data available, but only in
the DEC axis (Figure \ref{fig:SCR0128-1458_PB}) More data are required 
before the companion can be confirmed. Regardless, a fit has been made 
to the data and the slight possible perturbation removed to derive the 
astrometry results given in Table \ref{tab:astrometry}.

\begin{figure}
\centering
\includegraphics[angle=0,width=0.5\textwidth]{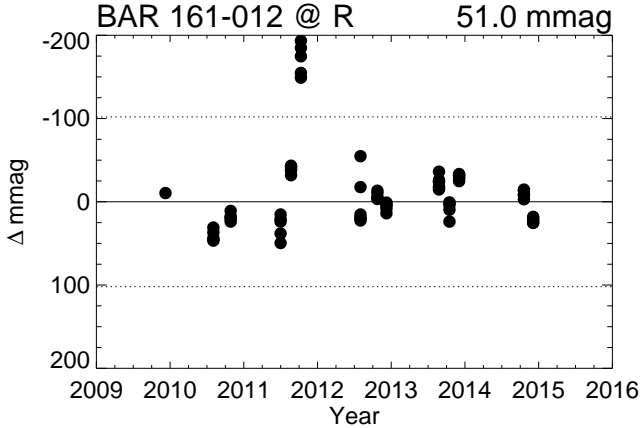}
\caption{Relative $R$-band photometry of BAR~161-012 from 2009-2015. A flare can be seen on 11 Oct 2011, as well as generally high photometric variability ($>$20 mmag).
\label{fig:BAR161-012_LC}}
\end{figure}

\paragraph{(0135-0712) BAR~161-012} ($M_V=10.63, V-K_s=5.34$) flared to a
maximum of 193 mmag in $R$ on UT 11 Oct 2011, and shows a variability
of 51 mmag over the 5 year dataset, as shown in Figure \ref{fig:BAR161-012_LC}.
This photometric variability and a discrepancy between a photometric
distance estimate of 12.3 pc and a trigonometric distance of 36.1 pc
indicates that this star is likely young. While several sources, most 
notably \citet{Shkolnik2012}, place this star in $\beta$ Pic, we find that 
this star's kinematics are inconsistent with any known NYMG. This star is
yet another young star without membership in a NYMG. 

\paragraph{(0143-0602) SCR~0143-0602} ($M_V=11.53, V-K_s=5.10$) was observed
to be 178 mmag in $V$ above its baseline brightness on UT 18 Oct 2014,
although the peak may have been higher because this offset was
measured in the first frame taken that night.  The star's photometric
variability is 37 mmag over 5 years, indicating it might be young.
The discrepancy between the photometric distance estimate of 13.3 pc
and trigonometric distance of 19.8 pc also hints that the star might
be young, or alternately, an unresolved multiple.

\begin{figure}
\centering
\includegraphics[angle=0,width=0.5\textwidth]{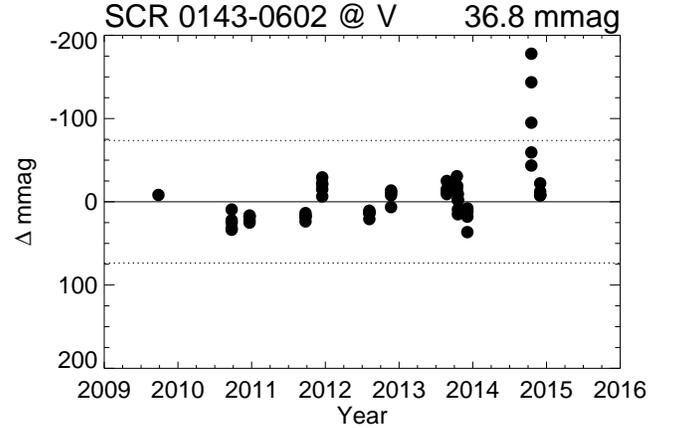}
\caption{Same as Figure \ref{fig:BAR161-012_LC}, but for SCR~0143-0602. A flare can be
seen on UT 17 Oct 2014.
\label{fig:SCR0143-0602_LC}}
\end{figure}

\paragraph{(0152-6329) SIP~0152-6329} ($M_V=12.53, V-K_s=6.15$) is a new Tuc-Hor member as identified by kinematics and the sodium gravity test, with a LACEwING-derived membership probability of 80\%. \citet{Gagne2015} found it to be a member of $\beta$ Pic, but we find only an 11\% probability of this. We do not find the star to be significantly photometrically variable in $R$ over 7 years.

\begin{figure}
\centering
\includegraphics[angle=0,width=0.5\textwidth]{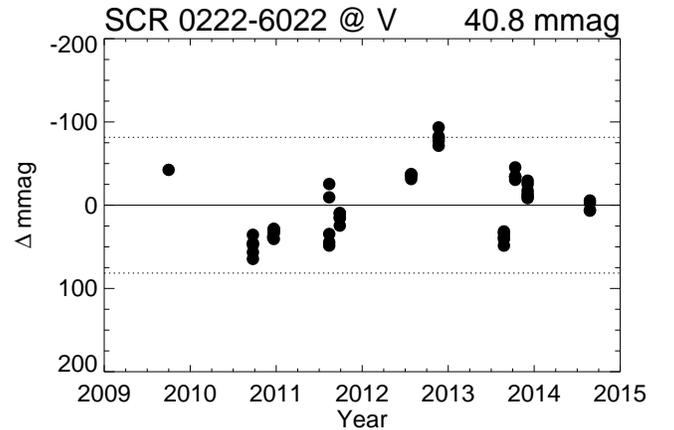}
\caption{Same as Figure \ref{fig:BAR161-012_LC}, but for SCR~0222-6022.
\label{fig:SCR0222-6022_LC}}
\end{figure}

\paragraph{(0222-6022) SCR~0222-6022} ($M_V=10.93, V-K_s=5.26$) was 
first identified as a member of Tuc-Hor by \citet{Rodriguez2013}. It 
is confirmed as a member based on kinematics (with 88\% probability in LACEwING) 
and both gravity tests. The star varies by 41 mmag over 5 years, as shown in Figure 
\ref{fig:SCR0222-6022_LC}, supporting the premise that it is young.

\paragraph{(0236-5203) 2MA~0236-5203} ($M_V=9.27, V-K_s=4.56$) has a proper
motion of 80 mas/yr at position angle 97$^{\circ}$, similar to the 2MA
0254-5108AB system (discussed next) with 87 mas yr$^{-1}$ at 93$^{\circ}$.
2MA~0236-5203 was identified in \citet{Zuckerman2004} as
a Tuc-Hor member, and we find it to be a member of that group via 
kinematics.

\begin{figure}
\centering
\includegraphics[angle=0,width=0.5\textwidth]{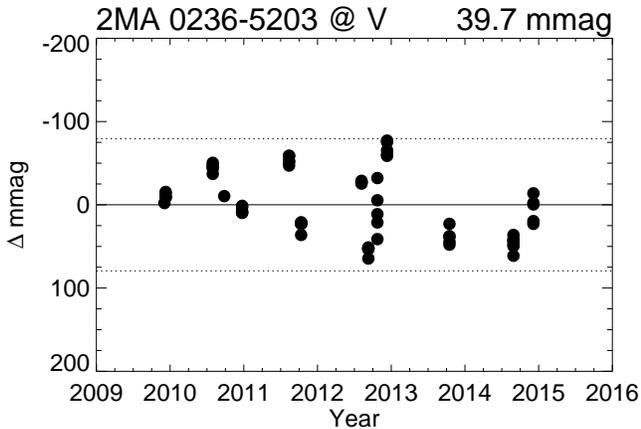}
\caption{Same as Figure \ref{fig:BAR161-012_LC}, but for 2MA~0236-5203.
\label{fig:2MA0236-5203_LC}}
\end{figure}

This star does show a photometric variability of 40 mmag as
shown in Figure \ref{fig:2MA0236-5203_LC}, indicative of youth, which corroborates the low
surface gravity measurement from the sodium index.

\paragraph{(0254-5108) 2MA~0254-5108AB} ($M_V=9.24, V-K_s=4.29 (A), M_V=14.32,
V-K_s=6.37 (B)$) is a binary with a separation of 15.3\arcsec at a position angle of
80.2$^{\circ}$. 
We do not see
any indication of orbital motion of the components in the 5 year
timespan of our observations.  The two components have the largest
$\Delta V$ (5.48) of any resolved system under consideration here; the
astrometry for the B component suffers due to its low SNR because
images were taken based on the brightness of the A component.  The
trigonometric parallaxes differ by 1.9$\sigma$, which may be caused by
the low signal on B, or (alternately) taken as evidence that these are
two separate members of Tuc-Hor serendipitously aligned on the sky.
Assuming they are a bound system (and with the weighted mean system
parallax), the A component is only marginally consistent with Tuc-Hor
membership and would need to be an equal-luminosity binary to fit the
Tuc-Hor isochrone.  Currently, the agreement with Tuc-Hor (for both
components) actually improves if the parallaxes are {\it not}
combined, and the two components are treated as separate star systems.
The single ROSAT X-ray detection is likely for the A component.

\begin{figure}
\centering
\includegraphics[angle=0,width=0.5\textwidth]{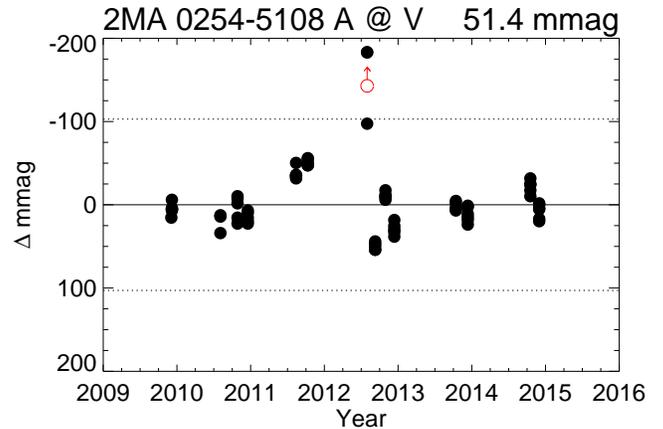}
\caption{Same as Figure \ref{fig:BAR161-012_LC} but for 2MA~0254-5108A. A flare is seen on 31 Jul 2012. One photometric point (indicated in red) from that sequence is outside the scale of the figure.
\label{fig:2MA0254-5108A_LC}}
\end{figure}

Both of the stars are variable, by 51 mmag (A) and 46 mmag (B) in $V$
over 5 years, although the variability for B is suspect given its low
level of counts throughout the observing sequence.  Component A flared
to a maximum of 255 mmag above its mean value on UT 31 Jul 2012 (Figure 
\ref{fig:2MA0254-5108A_LC}).  Both
stars are almost certainly young, supported by the parallaxes that
place them well above the main sequence --- photometric estimates
place the stars at 21 pc (A) and 31 pc (B), whereas the trigonometric distances
are 37 pc and 45 pc, respectively.

\begin{figure}
\centering
\includegraphics[angle=0,width=0.5\textwidth]{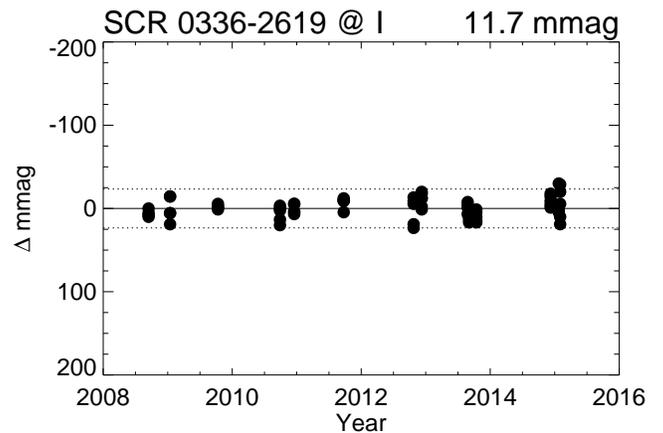}
\caption{Same as Figure \ref{fig:BAR161-012_LC} but for SCR~0336-2619.
\label{fig:SCR0336-2619_LC}}
\end{figure}

\paragraph{(0336-2619) SCR~0336-2619} ($M_V=13.02, V-K_s=6.57$) is a member of Tuc-Hor, as first suggested by \citet{Gagne2015}. Despite its clear spectroscopic signatures of gravity and chromospheric activity, it has an exceptionally flat lightcurve, shown in Figure \ref{fig:SCR0336-2619_LC}. 

\begin{figure}
\centering
\includegraphics[angle=0,width=0.5\textwidth]{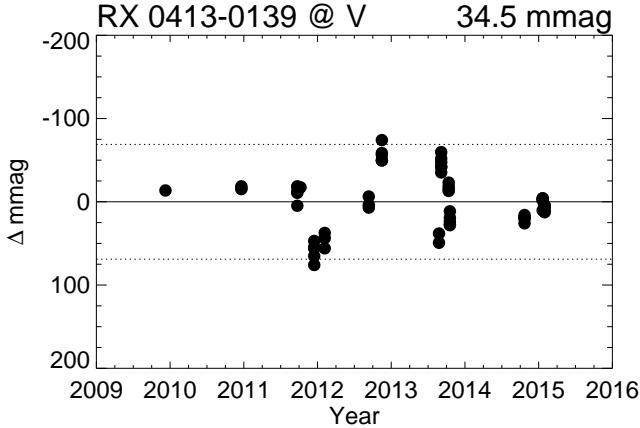}
\caption{Same as Figure \ref{fig:BAR161-012_LC} but for RX~0413-0139.
\label{fig:RX0413-0139_LC}}
\end{figure}

\paragraph{(0413-0139) RX~0413-0139} ($M_V=11.77, V-K_s=5.46$) exhibits 
photometric variability at the 35 mmag level over 5 years, as 
shown in Figure \ref{fig:RX0413-0139_LC}.  The astrometric
residuals are poor ($\sim$9.5 mas) in the DEC axis due to few
reference stars in the southern portion of the field.

RX~0413-0139 was reported to be a member of the Argus association by \citet{Malo2013}, but we find no support for that identification using 
our new parallax. This star and BAR~161-012 are new examples of nearby young 
stars without membership in any known group.

\paragraph{(0446-1116) 2MA~0446-1116AB} ($M_V=11.53, V-K_s=4.96$) appears
elongated in our images, with two components separated by
$\sim$1\farcs0. This companion is too close to be properly centroided
or even distinguished from the primary, which results in a parallax with
a relatively high error of 3.4 mas that is unlikely to improve with
additional data from our observing program.

\paragraph{(0510-7236) HD~271076} ($M_V=9.82, V-K_s=4.30$) was reasonably
suspected to be a supergiant in the LMC by \citet{Westerlund1981}
given its location in the sky.  However, at a distance of 20.2$\pm$1.1
pc it is clearly a foreground M2.0V star and a member of the Solar
Neighborhood.  Much of the error in the parallax (2.8 mas) can be
attributed to a faint reference field.

\paragraph{(0533-4257) SCR~0533-4257AB} ($M_V=12.50, V-K_s=5.46$) is the
closest system in our sample, at a distance of only 10.4$\pm$0.1 pc.
It has a very low proper motion, only 39 mas yr$^{-1}$.  The relatively large
position angle error (1.5 deg) in Table \ref{tab:astrometry} is
due to this small proper motion; the proper motion errors themselves
are no worse than for other stars on the program.  

\begin{figure}
\centering
\includegraphics[angle=0,width=.23\textwidth]{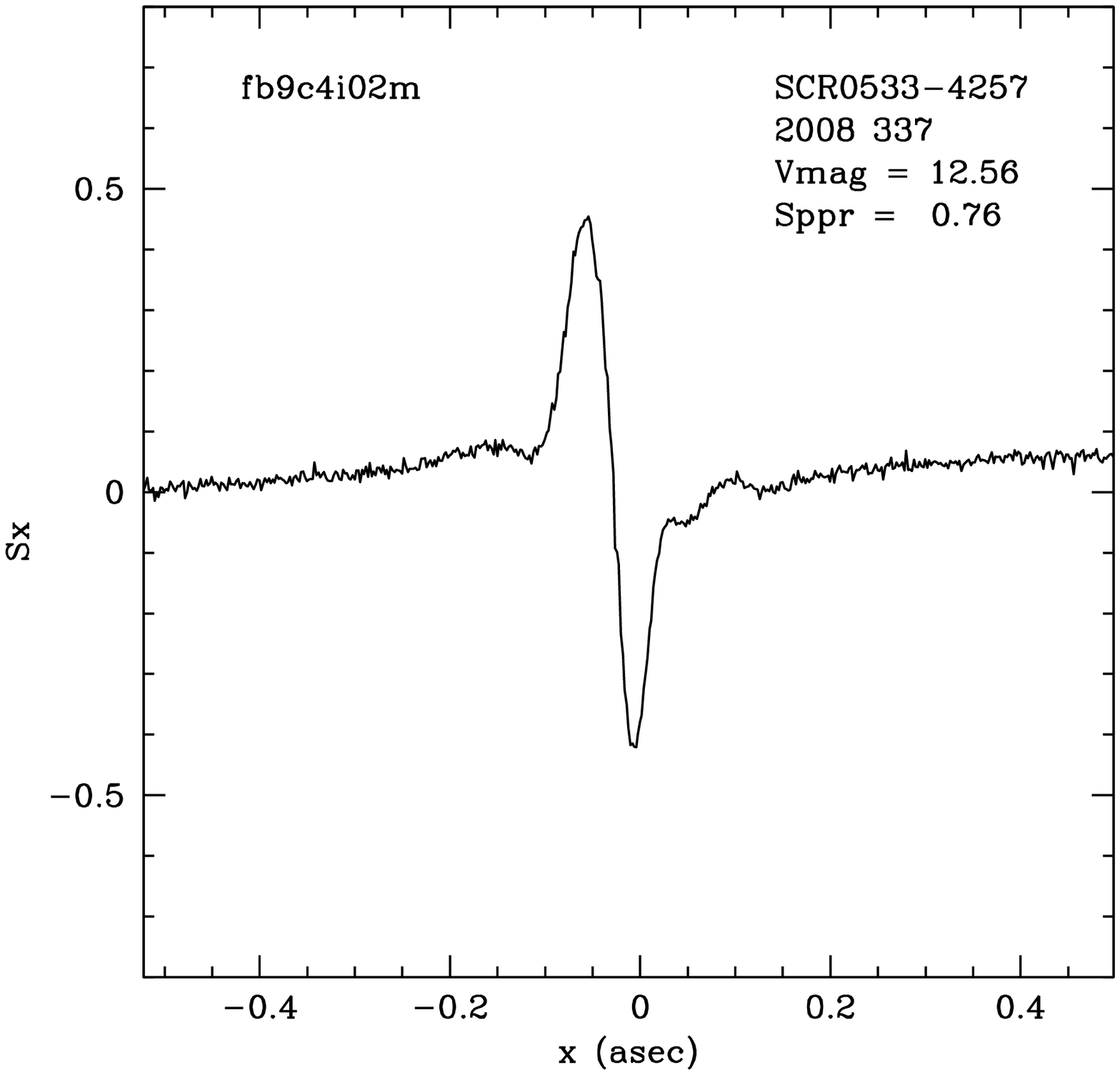}
\includegraphics[angle=0,width=.23\textwidth]{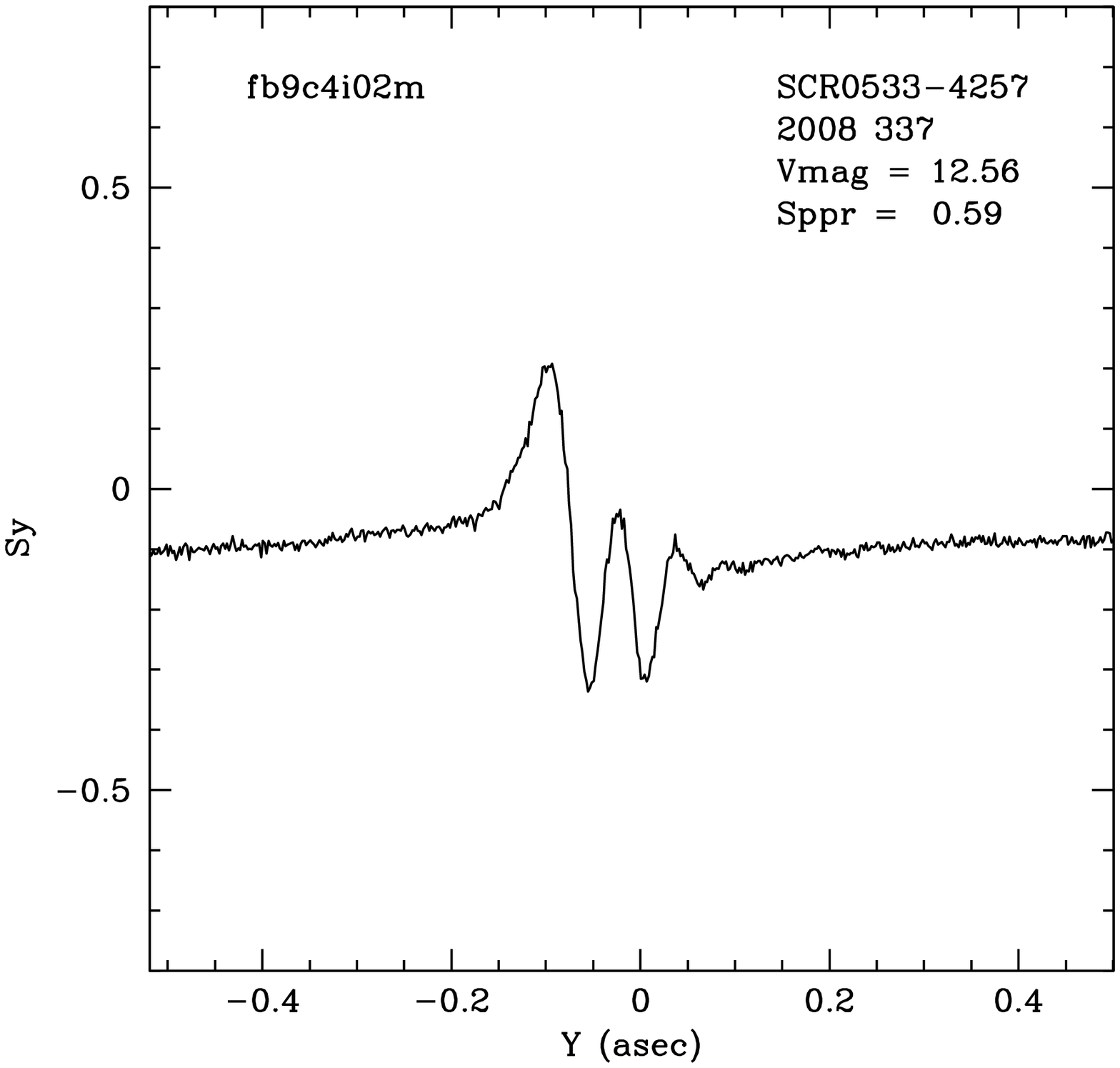}
\caption{The X-axis (left) and Y-axis (right)
Hubble Space Telescope Fine Guidance Sensor preliminary results for
SCR~0533-4257AB. The Y-axis ``S-curve'' of the Fine Guidance Sensor
shows a second dip to the right of the main one, revealing a
companion.  The companion can also be identified as a deformation in the in the X-axis S
curve, compared to a single star, though it is not visibly apparent.  Figure by Ed Nelan.
\label{fig:SCR0533-4257AB_FGS}}
\end{figure}

The system was identified in \citet{Riaz2006} to emit X-rays, so we
observed it during our HST-FGS Cycle 16B campaign.  It was resolved
(Figure \ref{fig:SCR0533-4257AB_FGS}) into a close binary 
with a 56 mas separation and $\Delta
F583W$=0.7 mag on 2 Dec 2008. Thus, the stars have a projected separation of roughly 0.56
AU, and the system's X-ray flux is unlikely to be due to the components'
interactions.  A periodogram of our astrometric data shows a peak with
a period near 9 months, consistent with the projected separation for
the companion in a Keplerian orbit.  However, the amplitude of the
perturbation is small and evident in the RA direction only, so we
present an uncorrected parallax in Table \ref{tab:astrometry}.

Given the system's X-ray emission, it is probably at least somewhat youthful, 
but based on its lack of low-gravity features and color-magnitude diagram 
position (corrected for its multiplicity), the system is over 120 Myr old. 
The kinematics of the system do not place it in any of the nearby young 
moving groups.

%

%

\paragraph{(0639-2101) LP~780-32} ($M_V=11.78, V-K_s=5.12$) has a photometric
distance estimate of 11.6 pc, which places the system closer than the parallax
(15.8 pc). Kinematic analysis shows it to be a low-probability member of Argus.

\paragraph{(0936-2610) 2MA~0936-2610ABC} ($M_V=11.76, V-K=5.15, M_K=6.61 (A),
M_K=9.86 (B)$) is a likely common proper motion binary separated by
41\arcsec~at position angle 314$^{\circ}$.  While comparing images
from multiple epochs (SuperCOSMOS, 2MASS, and WISE images) using
Aladin, we discovered the possible secondary, which is not in SIMBAD.
We are unable to determine a reliable proper motion or parallax for
the companion using our existing data because it is 3.2 mag fainter
than the primary at $K$, and not exposed well enough in our images at
$V$ for reliable astrometry.  With $VRI$ = 13.11, 11.86, 10.31 and $JHK$ =
8.86, 8.29, 7.96, we estimate a photometric distance of
17.0$\pm$2.7 pc for the possible companion, which is consistent with
the trigonometric distance of 18.6 pc for A.
The A component's photometric distance is closer, 13.6$\pm$2.2 pc. Speckle 
observations from 2010 (B. Mason, private communication) indicate the A 
component may be a close binary at 0\farcs39 at 284 degrees with a delta mag 
of 0.5. No sources were near the star at either the Palomar Deep Sky Survey 
(DSS) 1 red plate epoch (1955) or DSS2 red plate epoch (1995), making it likely 
that the speckle source is a co-moving companion and the system as a whole is 
a triple.

\paragraph{(1110-3731) SIP~1110-3731ABC=TWA~3ABC} ($M_V=9.40, V-K_s=5.29$),
is one of the first known members of the TW Hya
association \citep{de-la-Reza1989}, and has for some time been
considered the closest genuine member (TWA 22AB, at 17.5$\pm$0.2 pc, 
is now widely believed to be a member of $\beta$ Pic instead, 
\citealt{Mamajek2005,Teixeira2009}.) \citet{Webb1999} and 
\citet{Zuckerman2004} claim that this system is a triple (see Table 
\ref{tab:multiples}) where A is a spectroscopic binary, presumably 
SB2; with no other information given we assume $\Delta V = \Delta K = 0$, 
which makes B the actual brightest component.

\begin{figure}
\centering
\includegraphics[angle=0,width=.45\textwidth]{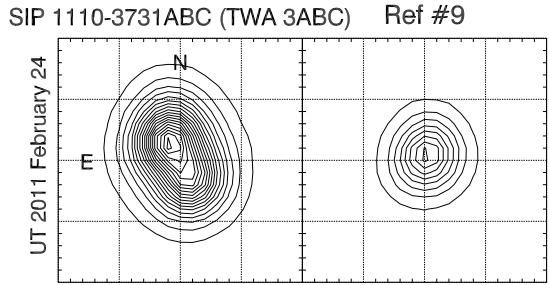}
\caption{Positions of TWA 3 AC (NE) and B (SW) on 24 Feb 2011.
  As seen here, TWA 3B often appears as a mere elongation of the TWA 3AC
  PSF, making parallax reduction difficult.  Reference star \#9 is shown on the right
  as an example single-star PSF, with the same contour
  intervals. Grid lines are 2.05\arcsec~apart, or 5 pixels at the
  CTIO 0.9m.
\label{fig:TWA003AB}}
\end{figure}

As shown in Figure \ref{fig:TWA003AB}, we detect two nearly equal
magnitude sources separated by 1\farcs16 at a position angle of 210.2\arcdeg, which 
implies a projected separation of 38 AU.
During reductions, many frames were
thrown out and most of the remaining SExtractor output had to be
manually edited to correctly identify the B component.  The resulting
parallax precision for the two components in Table
\ref{tab:astrometry} is poor, with errors of 4.0 and 6.8 mas,
and the measured variability is unreliable.  Nevertheless, the
combined weighted mean result of 33 pc is close to the expected
distance to the system (42 pc) based on kinematics in
\citet{Zuckerman2004}.  The astrometry also confirms that each star is
a potential member of the TW Hydra association, though only TWA 3B has
a published radial velocity consistent with the predicted best-fit
value.

We see minimal evidence for orbital motion in the form of different proper motions for the two components. We find the difference to be marginally significant: $\Delta
\mu_{RA\cos{DEC}}$=20$\pm$11 mas yr$^{-1}$, $\Delta
\mu_{DEC}$=35$\pm$12 mas yr$^{-1}$.  The separation of the two
components in our images on 02 Apr 2012 is 1.16\arcsec~at position
angle 210.2$^{\circ}$, but aperture photometry for each component is
not possible with our data.  Instead, a PSF fit of the photometry data
for each component is used to apply the DCR correction that is the
same for both components, consistent with their similar $V-I$ colors.

%

\paragraph{(1206-1314) STEPH~164} ($M_V=10.29, V-K_s=4.94$) exhibits a 
possible perturbation in the somewhat limited set of
data we have spanning 4 years.  The discrepancy between photometric
(14.2 pc) and trigonometric distance (31.0 pc) also implies the object
might be a young star and/or an unresolved multiple star. As neither kinematics
nor spectroscopy identify it as a young object, we suspect this star 
is a multiple star system.

\begin{figure}
\centering
\includegraphics[angle=0,width=0.5\textwidth]{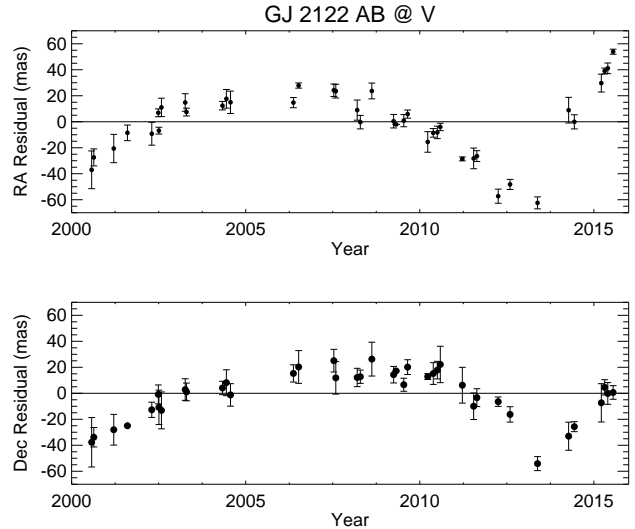}
\caption{Same as Figure \ref{fig:GIC0050_PB}, but for the star GJ2122AB. Large astrometric perturbations are seen in both axes.
\label{fig:GJ2122AB_PB}}
\end{figure}

\paragraph{(1645-3848) GJ~2122AB} ($M_V=9.29, V-K_s=3.96$) was found by
\citet{Heintz1987} to be a binary with separation 0.59\arcsec~and an 
estimated by-eye delta magnitude of 2.0. (see Table \ref{tab:multiples}) 
We see a single source in our images spanning 16 years, but find it to 
be an obvious astrometric binary, as shown in the nightly mean residuals 
of the positions after the parallax and proper motion fit (Figure 
\ref{fig:GJ2122AB_PB}).  Although the orbital period remains
uncertain, the parallax given in Table \ref{tab:astrometry} has been 
derived after removing our best approximation to the perturbation 
measured to date. Our calculated correction to absolute is an unrealistic 
5.08$\pm$1.34 mas because of a reddened reference field, so we have adopted a 
generic correction of 1.5$\pm$0.5 mas, and find the system to be at a distance 
of 12.4 pc.

The binary is also known as HIP~82021, but a bad position (off by
19\arcsec, more than the scale of the astrometer grating) in the
Hipparcos Input Catalog \citep{Turon1993} leads to an enormous
parallax error, and it was omitted from both official {\it HIPPARCOS}
catalogs.  \citet{Fabricius2000} re-reduced the {\it HIPPARCOS} data
and found a parallax of poor quality (71.3$\pm$14.8 mas, 14.0$\pm$2.9
pc) and again blame the pointing error.  Our parallax result
(77.2$\pm$2.1 mas, 13.0$\pm$0.3 pc) is consistent with the
\citet{Fabricius2000} result. (There is as yet no result from Gaia
for this star, likely for the same reason.)

\paragraph{(1710-5300) UPM~1710-5300AB} ($M_V=10.78, V-K_s=4.57$) is a binary
for which we estimate a separation of 0\farcs8 (see Table \ref{tab:multiples}).
The system appears as one elongated source in our data, and we determine 
a single parallax for the combined system with relatively high error (3.0 mas).


\paragraph{(1809-7613) SIPS~1809-7613} ($M_V=13.00, V-K_s=6.12$) is a 
possible member of $\beta$ Pictoris, as determined by a low-probability 
LACEwING membership of 31\%, supported by low-gravity features in both 
sodium and potassium.
The photometric variability is measured to be very low at 9 mmag, but
frames were taken in the $I$ filter, where variability is lower than
in $V$ or $R$.

\begin{figure}
\centering
\includegraphics[angle=0,width=0.5\textwidth]{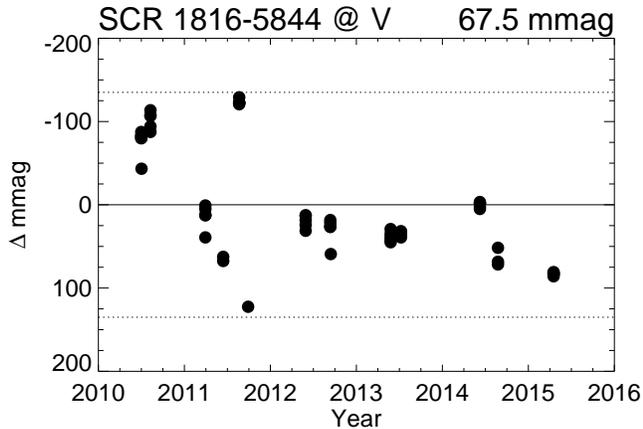}
\caption{Same as Figure \ref{fig:BAR161-012_LC} but for SCR1816-5844.
\label{fig:SCR1816-5844_LC}}
\end{figure}

\paragraph{(1816-5844) SCR~1816-5844} ($M_V=10.47, V-K_s=5.08$) is a 
new member of Argus, according to kinematics, corroborated by the 
sodium and potassium line strengths indicative of low surface gravity.
This star exhibits the highest level of photometric variability of any 
star reported here --- 68 mmag at $V$ over the 5 years, as shown in Figure 
\ref{fig:SCR1816-5844_LC}.  This youth indicator is supported by the 
discrepancy between a photometric distance of 12.2 pc and a trigonometric 
distance of 29.0 pc, placing the star well above the main sequence in the 
HR diagram.

\paragraph{(1956-3207) DEN~1956-3207AB} ($M_V=8.254, V-K_s=3.69 (A),
M_V=10.04,V-K_s=5.14 (B)$) is a binary separated by 26\arcsec~for
which we determine separate but entirely consistent parallaxes placing
the system at 45 pc. We do not have spectroscopy of either member of this 
system with which to comment on its age, but kinematic analysis with 
LACEwING shows no probability of membership in any known NYMG. 
The A component shows significant photometric variability of 31 mmag
at $V$, while B's variability of 20 mmag is more muted. This system 
does not seem to be young.

\paragraph{(2332-1215) BD-13~06424} ($M_V=8.22, V-K_s=3.94$) is a known member
of $\beta$ Pic \citep{Torres2006}, which we confirm with our
astrometric results (Table \ref{tab:astrometry}). 
We find the star to be photometrically variable at $V$ at a level of 28 
mmag during the 5 years of data in hand.

 
\section{Conclusions}
The effort described in this paper was an experiment to determine whether or not we might reveal nearby stars via a completely photometric search, rather than via the traditional route of assuming high proper motions. The combination of optical photometry from SuperCOSMOS and infrared photometry from 2MASS proved to be a powerful method to find nearby stars with minimal proper motions.  In this paper, we report:

$\bullet$ 29 parallaxes for 26 stellar systems, including 11 systems within 25 pc and 15 between 25 and 50 pc.  The closest three systems are between 10 and 12.5 pc distant.  All of the systems have $\mu$ = 38--179 mas yr$^{-1}$, which is slower moving than the 180 mas yr$^{-1}$ threshold used by Luyten for his compendia of proper motion stars.  Thus, the experiment to find nearby, slow-moving stars, was successful.

$\bullet$ 12 pre-main sequence stellar systems that are identified to be part of the AB Doradus (two systems), Argus (three system), Tucana-Horologium (5), $\beta$ Pictoris (1), and TW Hydra (1) moving groups, plus two additional stars that do not appear to be associated with any group.  The unassociated stars (along with other young non-members identified in \citealt{Riedel2014} and \citealt{Riedel2017b}) hint at a complex outcome to the star formation process that yields relatively young stars that cannot be straightforwardly linked to known assocations or moving groups.


$\bullet$ Among those stars is LP~780-032, another possible new member of the Argus moving group at a distance of only 15 pc. This system would rank as the fifth closest young star system.

\acknowledgments{
The RECONS effort has been supported by the National Science
Foundation through grants AST 05-07711, AST 09-08402, and
AST 14-12026. We also thank the members of the SMARTS
Consortium and the CTIO staff, who enable the operations of
the small telescopes at CTIO. This research has made use of
results from the SAO/NASA Astrophysics Data System
Bibliographic Services, as well as the SIMBAD and VizieR
databases operated at CDS, Strasbourg, France, and the Two
Micron All Sky Survey, which is a joint project of the
University of Massachusetts and the Infrared Processing and
Analysis Center, funded by NASA and NSF. The authors would like to thank the HST and Lowell Observatory staff, and Ed Nelan, for their assistance with data collection and processing.
}

\facility{CTIO:0.9m, CTIO:1.5m (RCSpec), Blanco (RCSpec), Perkins (DeVeny Spectrograph), CFHT (ESPaDONs), HST (FGS)}
\software{Astropy \citep{Astropy2013}, Numpy \citep{Walt2011}, Matplotlib \citep{Hunter2007}, IRAF, MATCHSTAR \citep{Riedel2014}, LACEwING \citep{Riedel2017a}}

\bibliographystyle{aasjournal} \bibliography{riedel_a}

\end{document}